\documentclass[epj]{svjour}
\usepackage{graphics}
\usepackage{hepunits2007}
\usepackage{lhc_defs}
\usepackage{calc}
\usepackage{bm}
\usepackage{makeidx}
\usepackage{subfigure}
\usepackage{xspace}
\usepackage[noprefix]{nomencl}
\usepackage{marvosym}
\usepackage{wasysym}
\usepackage{delarray}
\usepackage{footmisc}
\usepackage{graphicx}
\usepackage[sectionbib,numbers,sort,compress]{natbib}
\usepackage{color}
\usepackage[hyperfootnotes=false]{hyperref}
\begin{document}

\title{Top-Quark Physics at the LHC} 
\author{Kevin Kr{\"o}ninger$^{\dagger}$, Andreas B.~Meyer$^{\ddagger}$ and Peter Uwer$^{\amalg}$} 
\institute{$^{\dagger}$Technische Universit\"at Dortmund, Fakult\"at Physik, Otto-Hahn-Str.~4, 44227~Dortmund, Germany\\$^{\ddagger}$Deutsches Elektronen-Synchrotron (DESY), Notkestr.~85, 22607~Hamburg, Germany\\$^{\amalg}$Humboldt-Universit\"at zu Berlin, Institut f\"ur Physik,
Newtonstra{\ss}e~15, 12489~Berlin, Germany}

\abstract{The top quark
        is the heaviest of all known elementary particles. It was
        discovered in 1995 by the CDF and \Dzero experiments at
        the \Tevatron. With the start of the LHC in 2009, an
        unprecedented wealth of measurements of the top quark's
        production mechanisms and properties have been performed by
        the ATLAS and CMS collaborations, most of these resulting in
        smaller uncertainties than those achieved previously. At the
        same time, huge progress was made on the theoretical side
        yielding significantly improved predictions up to 
        next-to-next-to-leading order in perturbative QCD. Due to the vast amount
        of events containing top quarks, a variety of new measurements
        became feasible and opened a new window to precisions tests of
        the Standard Model and to contributions of new physics. In this review, originally written for a recent book on the results of LHC Run~1~\cite{thisbook}, 
        top-quark measurements obtained so far from the LHC Run~1 are summarised 
		and put in context with the current understanding of the 
		Standard Model.}
		
\authorrunning{K.\,Kr\"oninger, A.B.\,Meyer, P.\,Uwer}
\titlerunning{Top-Quark Physics at the LHC}
\maketitle
\tableofcontents
\blfootnote{Originally published in ``The Large Hadron Collider --- Harvest of Run~1'', edited by T.~Sch\"orner-Sadenius, Springer, 2015, pp. 259--300~\cite{thisbook}.}


\section{Introduction} \label{sec:top:intro}

Top quarks have been a subject of scientific research ever since Kobayashi's and Maskawa's 
speculations about a third family of quarks in the context of solving the problem of weak \CP violation 
in the early 1970s~\cite{Kobayashi:1973fv}. After a two-decade long period of searches at 
various colliders and experiments, the top quark was finally discovered\index{top quark!discovery} in 1995 by the CDF 
and \Dzero experiments at Fermilab's \Tevatron,
a proton-antiproton collider, operated at the time at 
a centre-of-mass energy of $\sqrt{s}=\unit{1.80}{\TeV}$~\cite{Abe:1995hr,Abachi:1995iq}.\index{Fermilab}\index{Fermilab!Tevatron@\Tevatron}\index{CDF experiment}\index{D0 experiment@\Dzero experiment} Since then, 
and in particular after the upgrade of the \Tevatron to a centre-of-mass energy 
of \unit{1.96}{\TeV}, pioneering precision measurements were performed at this machine.
Prime examples are the measurements of the total cross section for single-top 
and top-quark pair (\ttbar) production and the measurements of the top-quark mass 
(see Refs.~\cite{Aaltonen:2013wca,Tevatron:2014cka} and references therein). 
Currently, both CDF and \Dzero are in the process of publishing their legacy measurements based on 
data sets corresponding to an integrated luminosity of about \unit{10}{\invfb} collected by each experiment 
during \Tevatron Run~2---and these measurements set a standard for the LHC.

A new era in experimental top-quark physics was marked by the start of the LHC in 2009. 
At the LHC, top quarks are produced abundantly due to the high centre-of-mass energy, 
the resulting large rise of the parton luminosities, 
and the large instantaneous luminosity of the accelerator. 
During LHC's Run~1 more than 5 million top-quark events were produced 
at the collision points in ATLAS and CMS each. Based on these huge data samples, most of 
the measurements performed at the \Tevatron have already been improved and/or complementary studies have been performed. This has only been made possible by inheriting a wealth of analysis techniques from the \Tevatron experiments where they were pioneered and brought to perfection.
At the LHC, top quarks have become tools for searches for new physics, e.g.\ 
the search for rare decays, for measurements of top-quark couplings to gauge bosons, or 
for the investigation of the proton structure. 

In this chapter the experimental findings obtained during LHC Run~1 are summarised 
and put in context with our current understanding of the Standard Model (SM). 
Emphasis is placed on three aspects: 
i) the description of the precision measurements in comparison with results from the \Tevatron and 
with theory predictions, e.g.\ production cross sections or top-quark properties; 
ii) the presentation of new measurements that were not performed at the \Tevatron and that 
either improve the current understanding of already measured quantities or enable measurements of yet unexplored processes and quantities, e.g.\ small couplings or associated production processes; 
iii) the discussion of the physics lessons and insights gained from the LHC Run~1 results. 
These results do not only impact on
our understanding of the Standard Model, but also help us 
prepare for future investigations at the LHC and beyond. 
Several important measurements based on the 
data from LHC Run~1 are still being performed. This chapter can thus not be complete, and only a 
selection of the results is presented.

\section{Top-Quark Pair Production}\label{sec:top:pairproduction}

\begin{figure}[htbp]
  \begin{center}
\includegraphics[width=\textwidth]{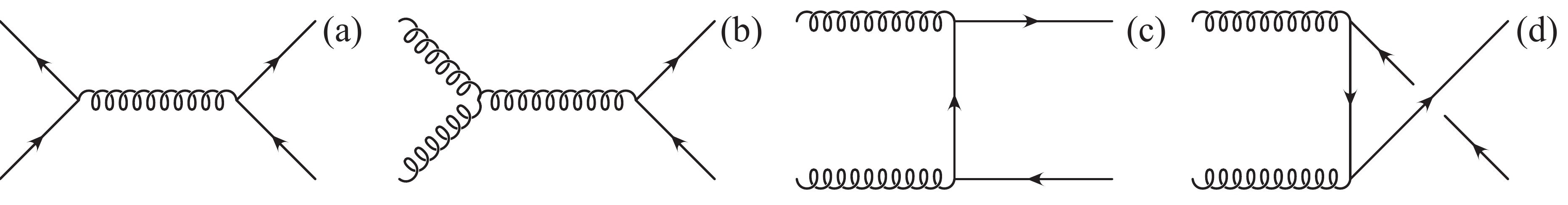}
    \caption{Leading-order Feynman diagrams contributing to top-quark pair production 
      in hadronic collisions.
    \label{fig:top:FeynmanDiagramsPairProduction}}
  \end{center}
\end{figure}

In hadronic collisions, top-quark pairs are dominantly produced
through the strong interaction. At the parton level, the production
mechanisms are quark-antiquark annihilation and gluon fusion.
Figure~\ref{fig:top:FeynmanDiagramsPairProduction} shows the
corresponding leading-order (LO) Feynman diagrams.  The differential
cross section for quark-antiquark annihilation can easily be
obtained from $\epem\to\mpmm$ by replacing the QED coupling \alpQED 
with the QCD coupling \alpS and introducing an
appropriate colour factor (2/9) (see also Chap.~5 of Ref.~\cite{thisbook}):
%
\begin{equation*}
  {\dif{\hat \sigma_{\qqbar \to \ttbar}}\over \dif{z}} = 
  {\pi\alpS^2\over 9s} \beta \left(2 - \left(1-z^2\right)
  \beta\right)\, .
\end{equation*}
%
Here $s$ denotes the center-of-mass energy squared, $\beta =
\sqrt{1-4\Mt^2/s}$ is the velocity of the top quark in the partonic
centre-of-mass system, $\Mt$ is the top-quark mass, and
$z=\cos(\theta)$ is the cosine of the scattering angle defined as the
angle between the incoming quark and the outgoing top quark. The
corresponding result for gluon fusion reads
%
\begin{equation*}
  {\dif{\hat \sigma_{gg \to \ttbar}}\over \dif{z}}
  ={\pi\alpS^2\over 96 s}\beta {7+9z^2\beta^2\over
    \left(1-z^2\beta^2\right)^2}
  \left(1+2\beta^2-2z^2\beta^2-2\beta^4+2z^2\beta^4-z^4\beta^4\right) \, .
\end{equation*}
%
The hadronic cross section is obtained from the partonic cross section
through a convolution with the parton distribution functions (PDFs)
$f_{i/H}(x,\mufs)$ which, roughly speaking, describe the probability to
find a parton $i$ inside a hadron $H$ with a momentum fraction between
$x$ and $x+\dif{x}$ of the mother hadron. 
The factorisation scale $\mufs$
denotes the scale at which, in higher-order calculations, the initial-state 
singularities are factorised into the parton distribution
functions.
The final formula for the hadronic cross section is thus given by
\begin{eqnarray}
  \label{eq:top:Factorisation}
  \dif{\sigma_{H_1H_2\to \ttbar +X}} &=& \sum_{i,j}\int \dif{x_1}\dif{x_2}
  f_{i/{H_1}}(x_1,\mufs)f_{j/{H_2}}(x_2,\mufs)\nonumber \\ 
  &\times& \dif{\hat \sigma_{ij\to \ttbar
  +X}(x_1P_1,x_2P_2,k_{\tq},k_{\tbarq},\alpS(\murs),\mufs)}\, ,
\end{eqnarray}
where $P_1$, $P_2$ are the momenta of the incoming hadrons, while
$k_{\tq}$, $k_{\tbarq}$ are the momenta of the outgoing top quarks. The
sum is over all possible partons. The coupling constant $\alpS$ is
evaluated at the renormalisation scale $\murs$. Note that in higher order, the
partonic cross section $\dif{\hat \sigma}$ also depends on the
factorisation scale $\mufs$ such that the factorisation scale
dependence of the parton distribution functions is cancelled 
order by
order and the hadronic cross section becomes independent of the
non-physical scale $\mufs$ at fixed order in perturbation
theory. For more details we refer to Chap.~5 of Ref.~\cite{thisbook}.
Because the QCD coupling constant is not small
($\alpS\approx 0.1$), higher-order corrections can give sizeable
contributions and need to be taken into account in most cases.  Since
the top-quark mass sets a large energy scale, non-perturbative effects
are essentially cut off and QCD perturbation theory
is believed to give reliable predictions.  
The next-to-leading order
(NLO) QCD corrections
for \ttbar production were
calculated a long time ago~\cite{Nason:1987xz,Beenakker:1988bq}. Further improvements were
obtained by resumming soft-gluon corrections, which lead to a
logarithmic enhancement of the cross section, to next-to-leading
logarithmic accuracy~\cite{Catani:1996dj,Bonciani:1998vc,Kidonakis:2001nj}.  Soft-gluon
resummation
has been extended recently to the next-to-next-to-leading
logarithmic (NNLL) 
accuracy~\cite{Kidonakis:2003qe,Moch:2008qy,Czakon:2008cx,Cacciari:2011hy,Beneke:2011mq}. 
Despite the fact that top quarks do not form bound states (because of their short lifetime),
binding effects still lead to minor corrections of the cross section close
to the $\ttbar$-pair production threshold. 
In principle, such a would-be bound state
could be observed
as a narrow peak in the $\ttbar$ invariant-mass spectrum, just below
the production threshold. However, the energy resolution of the LHC
experiments is not sufficient to resolve this effect.\footnote{At
  a future \epem collider, operating at the top-quark production
  threshold, the effect would be visible thanks to the high energy resolution 
  and could be used for very precise measurements of the top-quark mass.} 
The corresponding corrections to the inclusive cross section are small and have been studied in 
detail~\cite{Beneke:2011mq,Hagiwara:2008df,Kiyo:2008bv}. 
Electroweak corrections have also been 
investigated~\cite{Beenakker:1993yr,Kuhn:2005it,Bernreuther:2005is,Kuhn:2006vh,Bernreuther:2006vg,Hollik:2007sw,Bernreuther:2008md}. 
For the inclusive cross section of \ttbar production at the LHC operating 
at \unit{14}{\TeV}, they are negative and at the percent level.  On the other hand,
weak Sudakov logarithms can result in suppression of the cross section
at the level of 10--20\% for differential distributions at large
momentum transfer (see also Chap.~4 of Ref.~\cite{thisbook}).\index{Sudakov}
Since this region is precisely the one where new heavy resonances could
lead to an increase of the cross section, neglecting weak
corrections could potentially hide signs of new physics.  

Very recently, full next-to-next-to-leading order (NNLO)
QCD corrections have been
presented for the inclusive cross section~\cite{Czakon:2013goa}.  For
a centre-of-mass energy of \unit{8}{\TeV}, the result---including soft-gluon resummation at NNLL accuracy
(assuming $\Mt=\unit{173.3}{\GeV}$ and the MSTW2008nnlo68cl PDF 
set~\cite{Martin:2009iq})---reads
\begin{equation}
  \label{eq:top:NNLOxsection}
  \sigma_{\ttbar}(\sqrt{s}=\unit{8}{\TeV}) =
  245.8^{+6.2}_{-8.4}~(\text{scale}) ^{+6.2}_{-6.4}~(\text{PDF})~\pico\barn\, .
\end{equation}
Note that the result does not include the aforementioned weak
corrections. The first uncertainty is due to the residual\footnote{We
  call the remaining scale dependence ``residual'' because it is formally
  of higher order.} scale dependence which is used as an estimate of the
unknown higher-order contributions. It has been determined by a
variation of the renormalisation and factorisation scales in the range
$\Mt/2\ldots 2\Mt$.  
The second uncertainty is due to the incomplete
knowledge of the parton distribution functions.
From \eqn{\eqref{eq:top:NNLOxsection}} we conclude that the total cross
section is known with a precision better than 5\%. Moving to higher
collider energies will slightly improve the PDF uncertainties since
less weight is put to the large $x$ region where the PDFs are 
less precisely known. 

So far, only very few NNLO results exist for differential distributions. 
Most predictions are currently restricted to NLO accuracy
(extended in some cases by the resummation of soft-gluon corrections). 
Fixed-order NLO corrections
are available, for example, through the parton-level Monte Carlo (MC)
program \MCFM~\cite{Campbell:2012uf}. Combining parton-level NLO
calculations with parton-shower simulations is in general not straight
forward. A naive combination would count real emission processes
twice, since real emission is simulated through the parton
shower, but, on the other hand, is also explicitly taken into
account in the real-emission processes contributing at NLO. A
consistent matching that avoids double-counting has been developed in
the past~\cite{Frixione:2002ik, Nason:2004rx}. 
Differential distributions for top-quark pair production at NLO 
including parton shower effects are given in ~\cite{Frixione:2003ei,Frixione:2007nw}.

In all calculations mentioned before, the production of
stable top quarks is assumed, which is then followed by an on-shell decay. This
corresponds to the ``narrow-width'' or ``double-pole approximation''.
The naive expectation is
that corrections to this approximation should be suppressed like
$\Gamma_t/\Mt$ or even $\Gamma_t^2/\Mt^2$, where $\Gamma_t$ denotes the
top-quark decay width (see also Ref.~\cite{Melnikov:1995fx}). Obviously, the
naive expectation can only be true if the observable under
consideration is not directly related to off-shell effects. A counter-example 
is the invariant-mass distribution of the top-quark
decay products.  Off-shell effects for \ttbar production have
been investigated in detail in Refs.~\cite{Denner:2010jp,Denner:2012yc,Bevilacqua:2010qb}, where the
QCD NLO corrections for the process $\pp \to \WW \bbbar$ have been
calculated. Indeed, the effects are typically small, unless
observables of the type mentioned before are studied.

As can be seen from the vast amount of different theoretical studies,
\ttbar production is well understood in the Standard Model. A
variety of different corrections have been considered in the past, and
precise theoretical predictions for inclusive as well as differential
quantities are available. Top-quark pair production is thus an ideal
laboratory to test the consistency of the SM and to search for possible
deviations. In the following sections we describe the measurements of
inclusive and exclusive cross sections for top-quark pair production.

\subsection{Inclusive \ttbar Cross Section}

First measurements of the inclusive \ttbar cross section were published by the ATLAS and CMS 
collaborations already in the year 2010~\cite{Aad:2010ey,Khachatryan:2010ez}. These are based 
on data corresponding to an integrated luminosity of about \unit{3}{\invpb}, 
a fraction of the data collected in 2010.
More precise measurements, based on the full 2010 data sample corresponding to an integrated luminosity of \unit{35}{\invpb} became available shortly after~\cite{Aad:2012qf, 
                   Aad:2011yb, 
                   Chatrchyan:2011nb, 
                   Chatrchyan:2011ew, 
                   Chatrchyan:2011yy}. 

The amount of data collected in the years 2011 ($\sqrt{s}=\unit{7}{\TeV}$) and 2012 ($\sqrt{s}=\unit{8}{\TeV}$) correspond to integrated luminosities of about \unit{5 and 20}{\invfb}, respectively---altogether more than a factor 500 more than that
of 2010. With this wealth of data, top-quark physics entered a completely new realm of precision and detail.
With the 2011 data samples, measurements of the inclusive top-quark pair cross section were performed in all decay channels of the \ttbar system (except the one with two $\tau$ leptons in the final state), reaching an unprecedented level of statistical and systematic precision~\cite{
       Aad:2012vip, 
       Aad:2012mza, 
       ATLAS:2012aa, 
       Chatrchyan:2012ria, 
       Chatrchyan:2012bra, 
       Chatrchyan:2013ual, 
       Chatrchyan:2013kff, 
       Chatrchyan:2012vs, 
       Khachatryan:2014loa, 
       Aad:2014jra, 
       Aad:2014kva}. 

\begin{figure} [h] 
                \begin{center}
                        \includegraphics[width=0.95\textwidth]{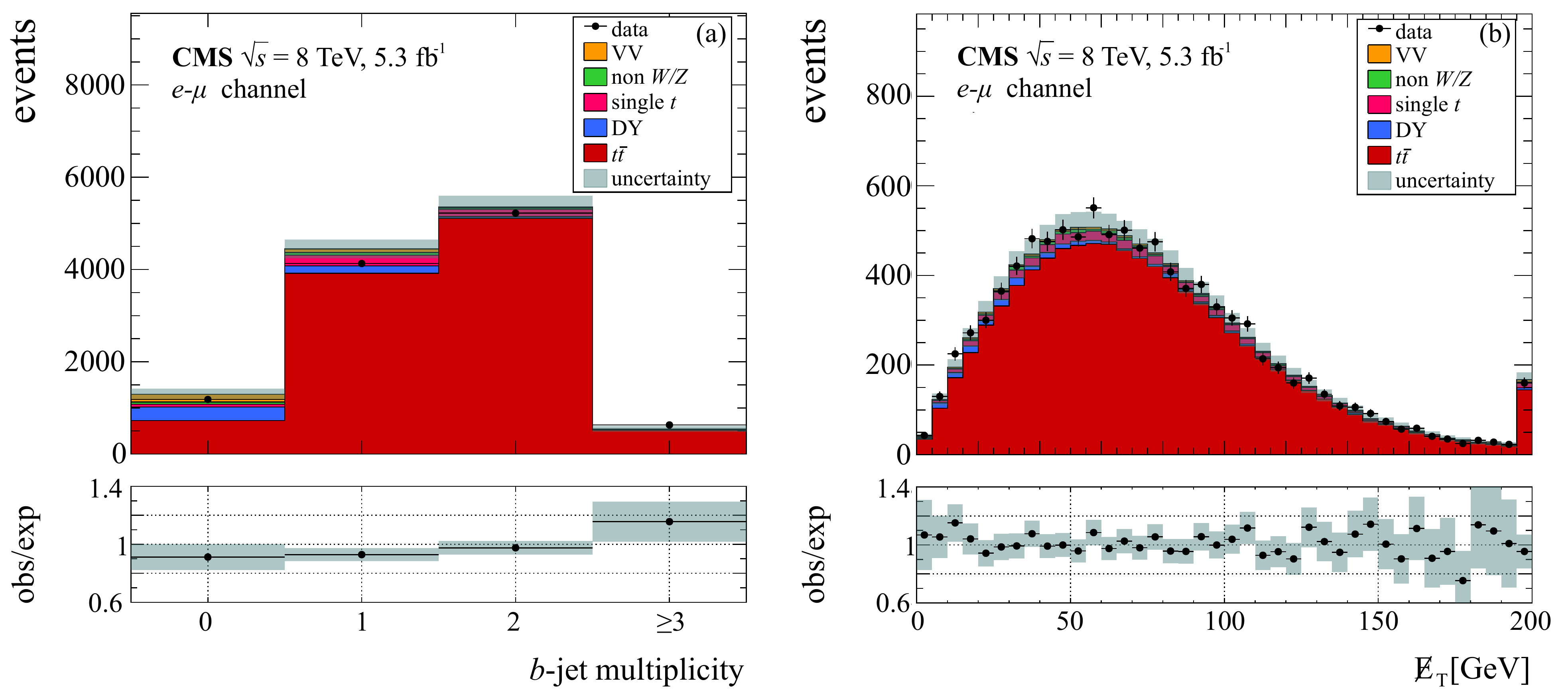} 
                \caption{Distributions of the \bq-tag multiplicity and missing transverse momentum of the CMS event sample used for the measurement of the inclusive \ttbar cross section at a centre-of-mass energy of \unit{8}{\TeV}.
                \textit{(Adapted from Ref.~\cite{Chatrchyan:2013faa}.)}                
                \label{fig:top:xsec:cms} 
} 
                \end{center}
\end{figure}

In the limit of large statistics, the most precise measurements of the inclusive \ttbar cross section 
can be obtained in the $e$-$\mu$ channel, as the backgrounds and associated uncertainties 
are minimal. The first published measurement of the top-pair cross section at a centre-of-mass energy 
of \unit{8}{\TeV} was performed by CMS~\cite{Chatrchyan:2013faa}. It makes use of the decay channel 
of top-quark pairs with two opposite-charge leptons, one electron and one muon, in the final state. 
In this channel, backgrounds
from non-top-quark events are minimal. Dominant contributions arise from 
Drell--Yan processes with two $\tau$ leptons in the final state that both decay into a lepton. 
These backgrounds are suppressed by requiring at least one of the two jets to be \bq-tagged.
Smaller background contributions come from single top-quark 
production (see \sect{\ref{sec:top:singletop}}) and from top-quark events in other decay 
channels where one of the jets is misidentified as a lepton.
In \fig{\ref{fig:top:xsec:cms}} the distributions of the \bq-tag multiplicity and the missing 
transverse momentum after the final event selection are shown. The final result of this measurement is 
$\sigma_{\ttbar} (\sqrt{s}=\unit{8}{\TeV})= 239.0 \pm 2.1 ~(\text{stat}) \pm 11.3~(\text{syst}) \pm 6.2~(\text{lumi}) \, \picobarn\, ,$ in good agreement with the NNLO prediction quoted in
\eqn{\eqref{eq:top:NNLOxsection}}. 

\begin{figure} [h] 
        \begin{center}
            \includegraphics[width=0.95\textwidth]{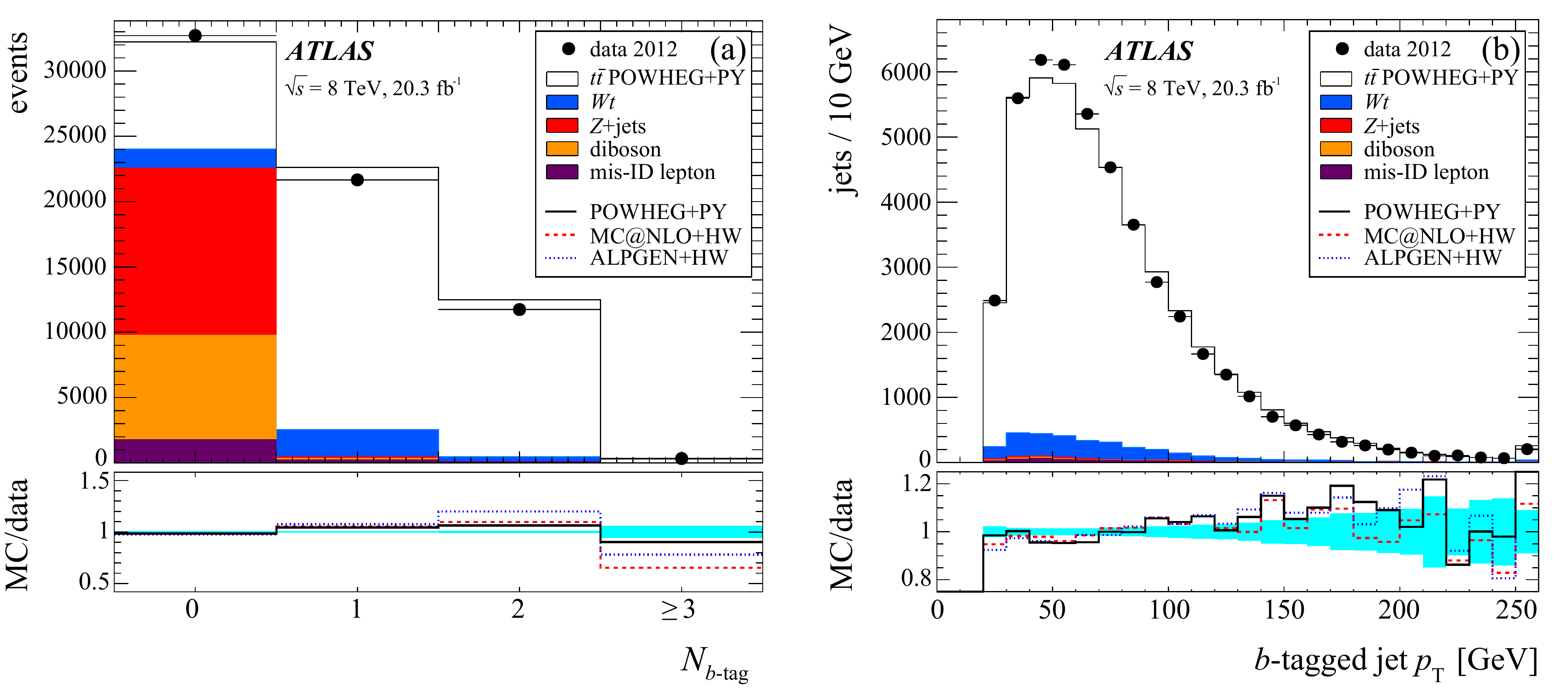} 
        \caption{Distributions of the \bq-tag multiplicity and \bq-jet transverse momentum of 
        the ATLAS event sample used for the measurement of the inclusive \ttbar cross section at 
        a centre-of-mass energy of \unit{8}{\TeV}.
        \textit{(Adapted from Ref.~\cite{Aad:2014kva}.)}
        \label{fig:top:xsec:atlas} 
        } 
        \end{center}
\end{figure}

Most recently, the ATLAS collaboration published a measurement yielding the most precise experimental 
result~\cite{Aad:2014kva}. In this analysis the numbers of events with exactly one and with exactly two 
\bq-tagged jets are counted and used to simultaneously determine $\sigma_{\ttbar}$ and 
the efficiency to reconstruct and \bq-tag a jet from a top-quark decay, thereby minimising 
the associated systematic uncertainties. 
In \fig{\ref{fig:top:xsec:atlas}} the distributions of the \bq-tag multiplicity and the transverse momentum 
of the \bq-tagged jets are displayed. The cross sections for centre-of-mass energies $\sqrt{s}=\unit{7}{\TeV}$ and \unit{8}{\TeV} are measured to be
$\sigma_{\ttbar} (\sqrt{s}=\unit{7}{\TeV})= 182.9 \pm 3.1 ~(\text{stat}) \pm 4.2~(\text{syst}) \pm 3.6~(\text{lumi}) \pm 3.3~(\text{beam}) \, \picobarn\, , $ and 
$\sigma_{\ttbar} (\sqrt{s}=\unit{8}{\TeV})= 242.4 \pm 1.7 ~(\text{stat}) \pm 5.5~(\text{syst}) \pm 7.5~(\text{lumi}) \pm 4.2~(\text{beam}) \, \picobarn\, ,$ respectively, where the latter uncertainty is due to the beam-energy uncertainty.


\subsection{Differential \ttbar Cross Sections}

Additional information about top-quark production and decay can be gained from 
measurements of differential distributions. These do not only probe QCD predictions and 
provide input to an improved choice of QCD model and scale parameters, but they also have the 
potential to constrain the parton distribution functions of gluons at large momentum fractions $x$.
Moreover, the differential distributions are potentially sensitive to new physics, e.g.\ to decays of 
massive \Zb-like bosons into top-quark pairs that would become visible at high \ttbar invariant 
masses (see also Chap.~11 of Ref.~\cite{thisbook}).

The kinematic properties of a top-quark pair are determined from the four-momenta of all final-state objects 
by means of reconstruction algorithms. For a general introduction to the different decay 
channels, see e.g.\ Ref.~\cite{PT:Top} and references therein.
In the single-lepton channels, kinematic-fitting algorithms are 
applied to obtain the kinematics of both top quarks. In the dilepton channels, due to the presence 
of two neutrinos, the kinematic reconstruction is under-constrained.
Ambiguities between several solutions are resolved by prioritisation, e.g.\ by the use of the expected 
neutrino energy distribution.

\begin{figure}[h]
        \begin{center}
                \includegraphics[width=0.95\textwidth]{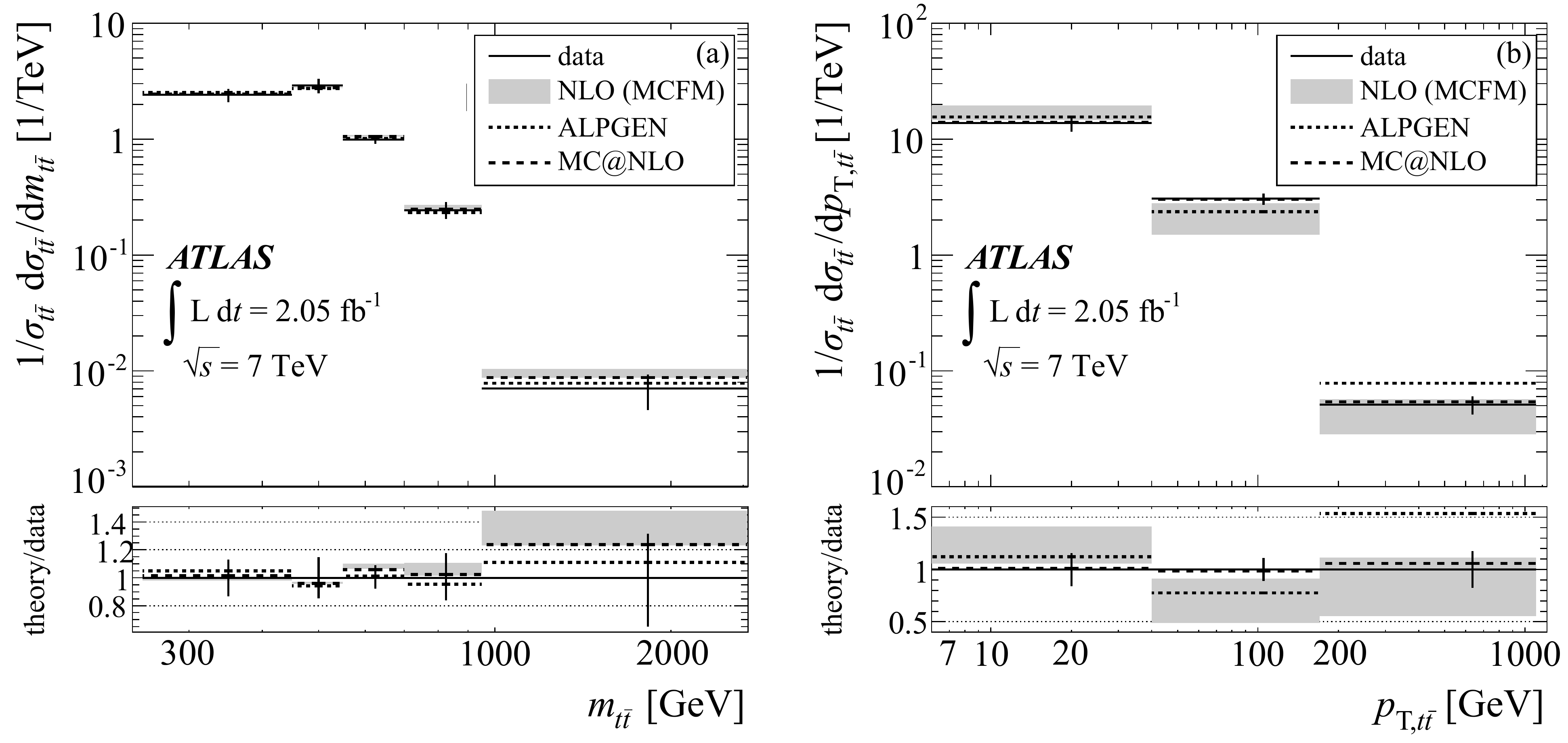} 
        \caption{Normalised differential \ttbar production cross sections, measured in \pp collisions at a centre-of-mass 
        energy of \unit{7}{\TeV} by the ATLAS collaboration, as a function of (a) the invariant mass of 
        the top-quark pair and (b) the transverse momentum of the top-quark pair system.
        \textit{(Adapted from Ref.~\cite{Aad:2012hg}.)}
        \label{fig:top:xsecdiff:atlas} }
        \end{center}
\end{figure}

\begin{figure}[h] 
        \begin{center}
                \includegraphics[width=0.95\textwidth]{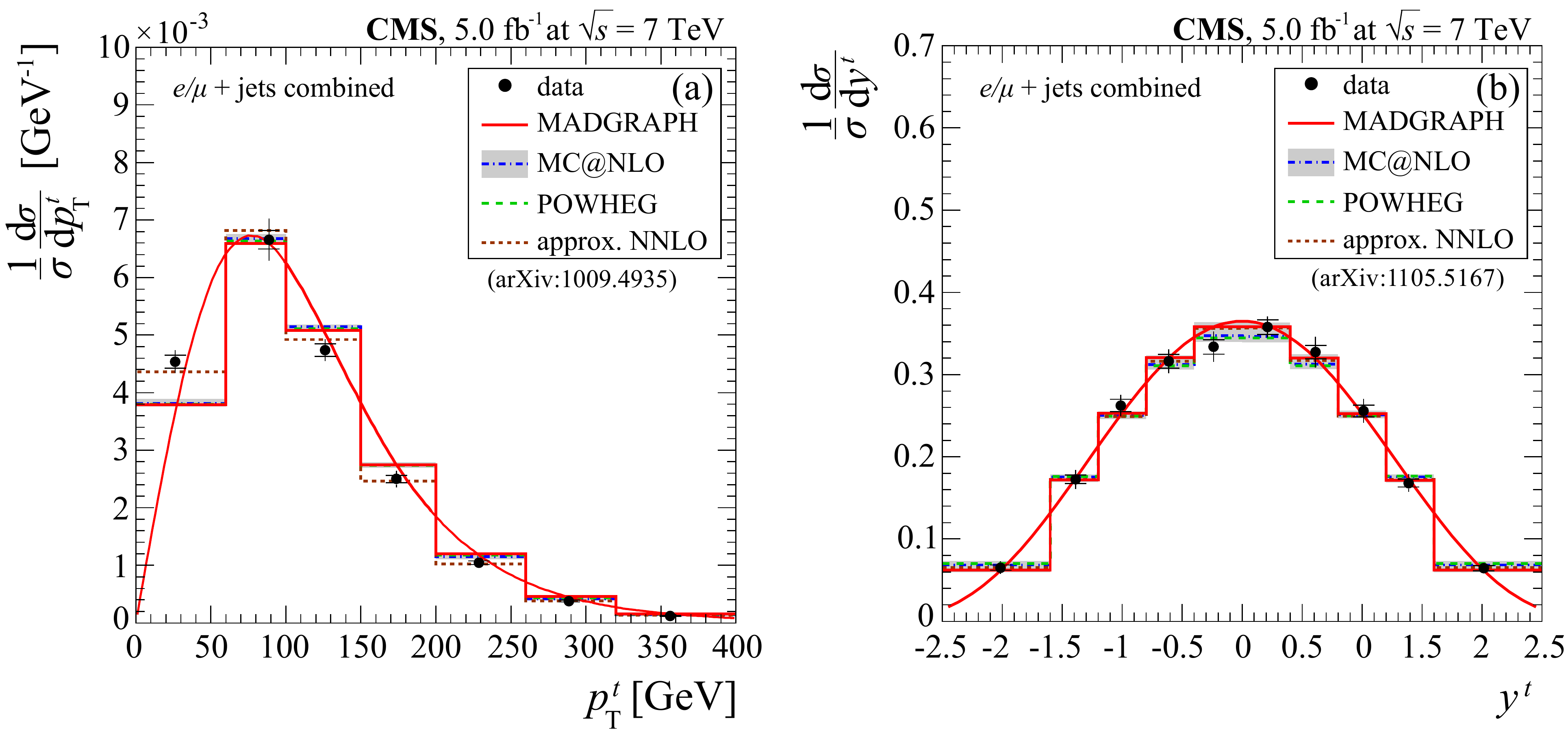} 
        \caption{Normalised differential \ttbar production cross sections, measured in \pp collisions at a 
        centre-of-mass energy of \unit{7}{\TeV} by the CMS collaboration, as a function of (a) the transverse 
        momentum and (b) the rapidity of each of the top quarks. 
        \textit{(Adapted from Ref.~\cite{Chatrchyan:2012saa}.)}
        \label{fig:top:xsecdiff:cms} }
        \end{center}
\end{figure}

A large number of distributions of the top quark and the top-quark pair system, as well as their 
decay products, has been measured at the LHC~\cite{Aad:2012hg,Chatrchyan:2012saa,Aad:2014zka}. 
In contrast to the situation at the \Tevatron, 
the large \ttbar production rate at the LHC leads to a substantial reduction of the statistical uncertainties in each bin.
The ATLAS and CMS collaborations report normalised differential cross sections, i.e.\ shape measurements, in 
which normalisation uncertainties are removed. In \fig{\ref{fig:top:xsecdiff:atlas}} the distributions of 
the invariant mass and of the transverse momentum of the top-quark pair system as measured by ATLAS 
are displayed~\cite{Aad:2012hg}. The data are very well described by the various calculations up 
to an energy scale of about \unit{1}{\TeV}.
The results from CMS agree with these findings~\cite{Chatrchyan:2012saa}. The transverse momentum and 
rapidity distribution of each of the top quarks were also measured and the results from CMS are presented in \fig{\ref{fig:top:xsecdiff:cms}}. Different theoretical predictions are confronted with the data, and they 
are generally found to give a good description of the data. 
However, most Monte Carlo simulations predict the transverse momentum distribution of the top quarks to be 
somewhat harder than what is seen in the data. This discrepancy between data and simulation
is presently under investigation. For the time being it constitutes an important source of uncertainty for many analyses.

%


\subsection{Top-Quark Pairs and Additional Jets}

\begin{figure}[h] 
  \begin{center}
    \includegraphics[width=0.95\textwidth]{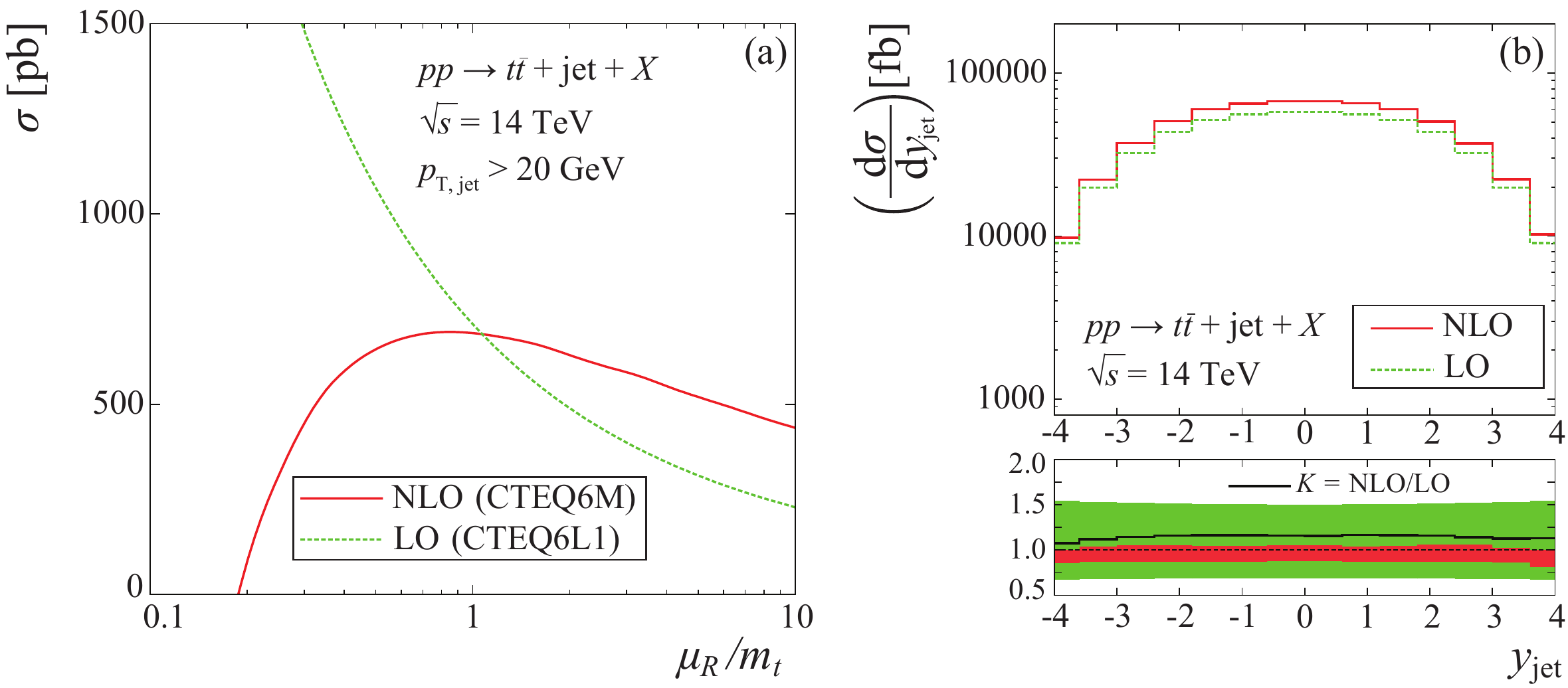}
  \caption{(a) Cross section
  for $\ttbar +\mbox{1-jet}+X$ production at $\sqrt{s}$ = 14 TeV 
  as a function of the
  renormalisation scale. 
  (b) Rapidity distribution of the additional jet at
  NLO. (\textit{Adapted from Refs.~\cite{Dittmaier:2007wz,Dittmaier:2008uj}.)}
    \label{fig:top:ttjets-nlo} }
  \end{center}
\end{figure}

%
At LHC energies, a large fraction of top-quark pairs is accompanied
by additional high-\pT jets. Demanding, for example, a minimal transverse
momentum of \unit{50}{\GeV} for such additional jets, about 30\% of all top-quark
pairs are produced together with at least one further jet~\cite{Dittmaier:2008uj}. 
From an experimental point of view, the jet
activity needs to be understood since the appearance of additional
jets affects the event reconstruction. Owing to the large rate,
\ttbar production with jets may also lead to sizeable
backgrounds for other SM studies or searches for new physics. 
As an example,
$\ttbar + \mbox{1-jet} + X$ production is the dominant background for
Higgs production via vector-boson fusion.  From a theoretical
perspective, the additional jet activity can be used for further tests
of the underlying production and decay mechanisms. 
Anomalous $\ttbar \glue$ couplings can be constrained, for example, through a detailed
analysis of the process $\pp \to \ttbar + \mbox{1-jet} + X$.
Assuming the validity of our theoretical understanding, \ttbar
production in association with a jet can also be used to measure the
top-quark mass~\cite{Alioli:2013mxa}.\index{top quark!mass}
Since the process 
$\pp\to \ttbar + \mbox{1-jet} + X$ is proportional to $\alpS^3$, NLO
contributions can easily give corrections of the order of 30\%. For a
precise understanding it is thus mandatory to take these corrections
into account. 
In Born approximation, the partonic processes $\glueglue\to
\ttbar \glue$, $\qbarq \to \ttbar \glue$, $\qq\glue\to \ttbar \qq$ and 
$\glue \qbar \to \ttbar \qbar$ contribute to \ttbar production in
association with a jet. The last three processes are related by
crossing. The leading-order partonic matrix elements can be
calculated e.g.\ with the help of \MADGRAPH\cite{Alwall:2011uj}.
The hadronic
cross sections are then calculable through a numerical Monte Carlo
integration, using again \eqn{\eqref{eq:top:Factorisation}}.  For
the evaluation of the NLO corrections, the one-loop corrections to the
aforementioned Born processes, together with real-emission processes,
need to be evaluated.  Since the two contributions are individually
infrared (IR) divergent---the divergences cancel only in the sum---a
method to organise this cancellation needs to be applied. In
Refs.~\cite{Dittmaier:2008uj,Dittmaier:2007wz}, the one-loop
amplitudes have been calculated using a traditional tensor reduction
for the one-loop integrals. The cancellation of the IR divergences is
achieved using the Catani--Seymour subtraction
method~\cite{Catani:1996vz,Catani:2002hc}. In
Ref.~\cite{Melnikov:2010iu}, an alternative calculation of the NLO
corrections, based on the unitarity method,
has been presented. In a subsequent study~\cite{Melnikov:2011qx}, also 
the on-shell decay of the top quark has
been taken into account.  NLO results for \ttbar production in
association with a photon are also available at
NLO QCD~\cite{Melnikov:2011ta}, since this process is closely related to \ttbar
production in association with jets.  In \fig{\ref{fig:top:ttjets-nlo}}(a),
the cross section for $\ttbar +\mbox{1-jet}+X$ production is shown as
a function of the renormalisation scale \murs which is set equal to
the factorisation scale. For the parton
distribution functions the CTEQ6 set~\cite{Pumplin:2002vw} has been
used.

The transverse momentum of the additional jet is required to be at least \unit{20}{\GeV}.
The leading-order result strongly depends on the
renormalisation scale~\murs, a fact that directly reflects the running of the coupling
constant $\alpS(\murs)$. The Born approximation can thus at best be
considered as a rough estimate of the cross section. The
NLO corrections as a function of the renormalisation 
scale, however, show a flat behaviour around $\murs=\Mt$. This may
be considered as an indication that $\Mt$ provides a natural scale for
this process. It can also be seen from \fig{\ref{fig:top:ttjets-nlo}}(a) that
the corrections are rather small for $\murs=\Mt$. Similar observations
can be made in \fig{\ref{fig:top:ttjets-nlo}}(b) where the rapidity
distribution of the additional jet, calculated at NLO, is shown.  
In Refs.~\cite{Dittmaier:2008uj,Melnikov:2011ta} a large variety of differential 
distributions have been investigated. In particular the transverse
momentum distribution of the top quark, the $t\bar t$ system and the 
additional jet have been calculated.
It turns out that for large transverse momentum \pT, significant QCD corrections
together with a large scale uncertainty are observed. In principle, this
is not surprising since at a large transverse momentum an additional
scale---different from $\Mt$---is introduced. It is conceivable that a phase-space-dependent
renormalisation scale could improve the behaviour of the perturbation
theory by effectively resumming large logarithmic corrections. 
In \tab{\ref{tab:top:ttbar+1jet-NLO-xsection}}, the dependence of
the cross section on the required minimal transverse momentum 
of the additional jet, $\pT^{\text{cut}}$, is shown. 
A strong dependence on $\pT^{\text{cut}}$ is found. For
$\pT^{\text{cut}} \to 0$ the cross section diverges
logarithmically (the divergence cancels a similar divergence in the cross section for
inclusive \ttbar production at NNLO, when $\ttbar +
\mbox{1-jet}+X$ production is combined with the two-loop corrections to
inclusive \ttbar production).
\begin{table}
  \begin{center}
    \begin{tabular}{c|l|l} \multicolumn{1}{c}{}
      &\multicolumn{2}{c}{$\sigma_{pp\to t\bar t + \mbox{1-jet}+X} [\picobarn]$} \\
      $\pT^{\mbox{\scriptsize \text{cut}}}$ [\GeV]&\multicolumn{1}{|c|}{LO}
      &\multicolumn{1}{|c}{NLO}\\ 
      \hline 
      \hline 
      20 & 710.8(8)$^{+358}_{-221}$ & 692(3)3$^{-40}_{-62}$ \\
      \hline 
      50 & 326.6(4)$^{+168}_{-103}$ & 376.2(6)$^{+17}_{-48}$ \\
      \hline 
      100& 146.7(2)$^{+77}_{-47}$   & 175.0(2)$^{+10}_{-24}$ \\
      \hline 
      200& 46.67(6)$^{+26}_{-15}$   & 52.81(8)$^{+0.8}_{-6.7}$ \\
\end{tabular}
\caption{Cross section $\sigma_{\pp\to \ttbar + \mbox{1-jet}+X}$
at the LHC for different values of $\pT^{\text{cut}}$ 
      for $\mu=\mufs=\murs = \Mt$~\cite{Dittmaier:2008uj}.
      The uncertainties correspond to changes in the scale, namely $\mu = \Mt/2$
      and $\mu = 2\Mt$. In parentheses the uncertainties due to Monte Carlo integrations are quoted.
    \label{tab:top:ttbar+1jet-NLO-xsection}}
  \end{center}
\end{table}
With the exception of $\pT^{\text{cut}}=\unit{20}{\GeV}$, where very
small and negative corrections are observed, the corrections are
positive and typically about 15--20\%.

Recently, NLO corrections for \ttbar production in association
with two additional jets were studied. For details, we refer to the
original work~\cite{Bredenstein:2009aj,Bredenstein:2008zb,Bredenstein:2010rs,%
Bevilacqua:2011aa,Bevilacqua:2012em,Hoeche:2014qda}. 

\begin{figure}[h] 
        \begin{center}
                \includegraphics[width=0.95\textwidth]{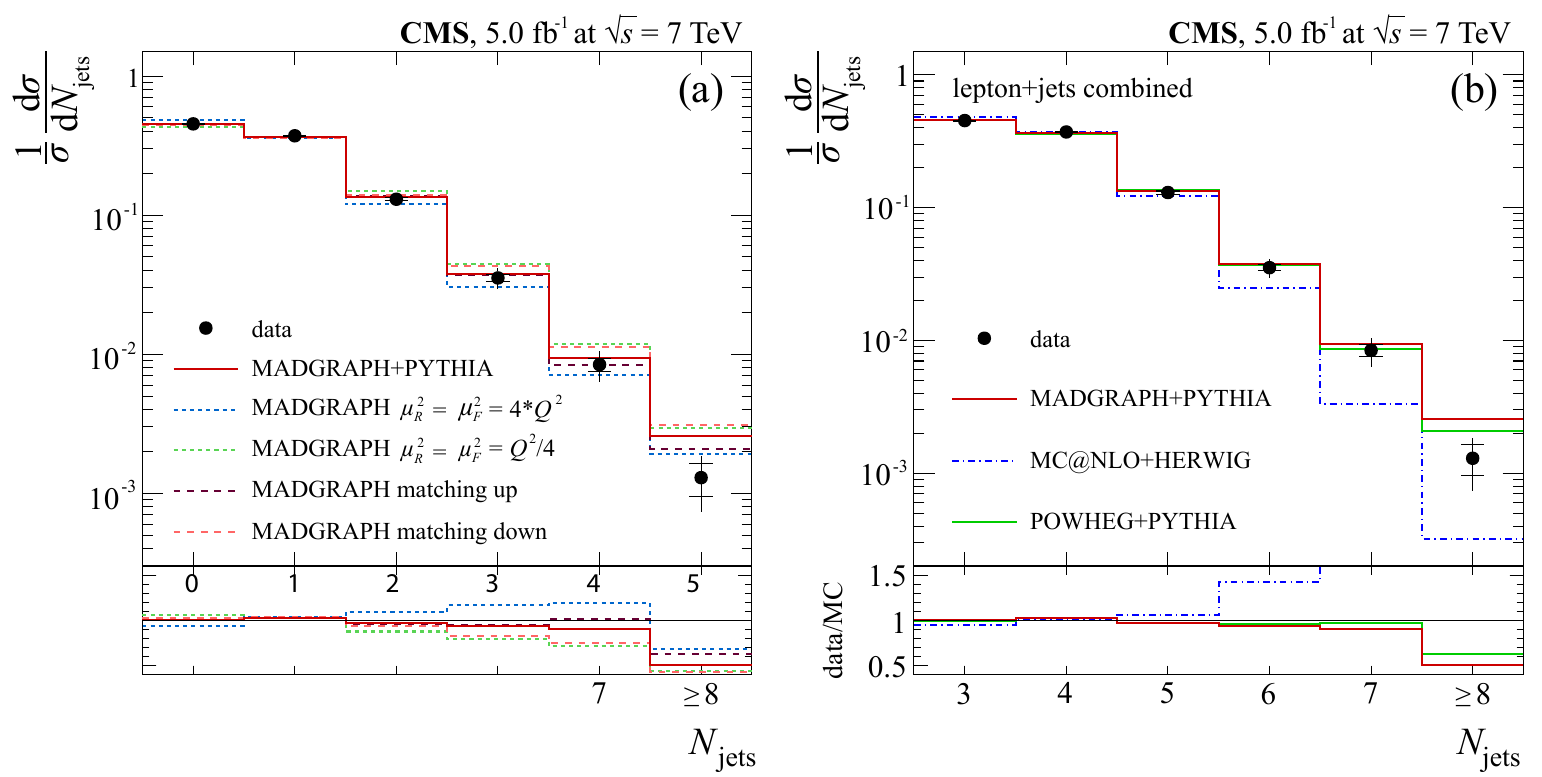} 
        \caption{Differential distributions of jet multiplicity in \ttbar events 
        with exactly one lepton and at least three jets in the final state. 
        (a) Data compared with the \MADGRAPH simulation for different choices 
        of renormalisation and factorisation scales and the matching scale between matrix 
        element and parton shower.
        (b) Data compared with several Monte Carlo simulations. 
        \textit{(Adapted from Ref.~\cite{Chatrchyan:2014gma}.)} 
        \label{fig:top:ttjets} }
        \end{center}
\end{figure}

The ATLAS and CMS collaborations studied the distributions of jet
multiplicities and additional jets due to QCD radiation in
detail~\cite{Chatrchyan:2014gma,ATLAS:2012al,Aad:2014iaa}. The multiplicity distributions
of jets for \ttbar events in the single-lepton channel as measured by the CMS collaboration is
shown in \fig{\ref{fig:top:ttjets}}.
The data are generally well described by the Monte Carlo predictions obtained using 
\MADGRAPH, \POWHEG and \MCATNLO. Towards very large multiplicities, 
the \MCATNLO generator interfaced 
with parton showers from \HERWIG predicts significantly less events than
\MADGRAPH or \POWHEG, which both use \PYTHIA to generate the parton showers. 

\begin{figure} [h] 
        \begin{center}
                \includegraphics[width=0.95\textwidth]{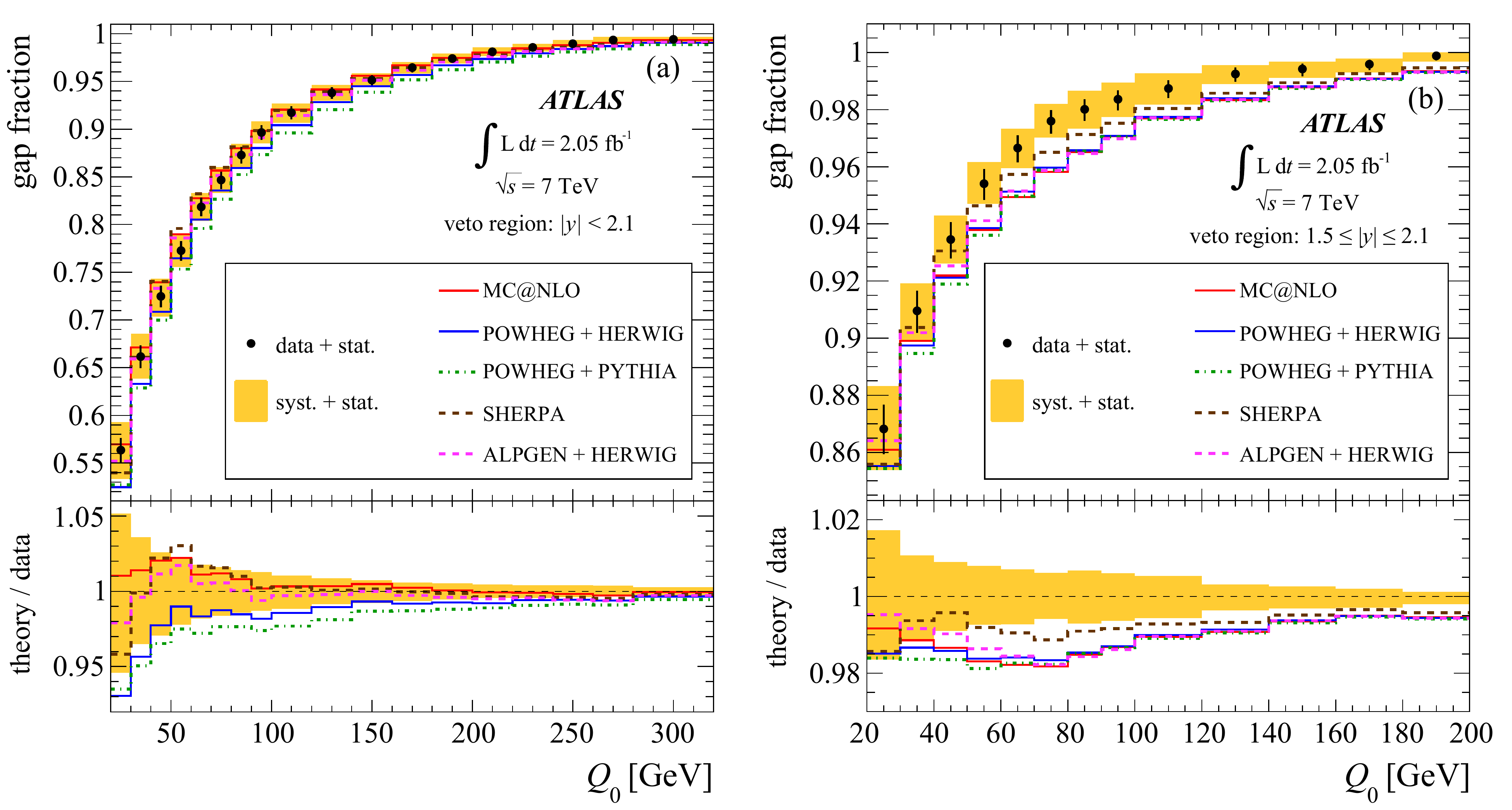} 
        \caption{The measured gap fraction as a function of the jet transverse 
        momentum threshold $Q_0$ above which there are no additional jets. 
        The data are compared with the 
        predictions from NLO and multi-leg LO MC generators, applying jet vetos in two
        different rapidity regions (a) $|y|<2.1$ and (b) $1.5<|y|<2.1$. 
        \textit{(Adapted from Ref.~\cite{ATLAS:2012al}.)}
        \label{fig:top:ttgap} }
        \end{center}
\end{figure}

An alternative way of investigating additional activity in \ttbar 
events is to study ``gap-fraction'' 
distributions~\cite{Chatrchyan:2014gma,ATLAS:2012al}. In these studies, events are vetoed 
if they contain an additional jet with transverse momentum above a given threshold in a central rapidity 
interval. The fraction of events surviving the jet veto, the gap fraction, is presented as a 
function of the threshold. The gap-fraction distributions for jets as measured by ATLAS are displayed in \fig{\ref{fig:top:ttgap}}. 
A qualitatively similar trend is observed as in the multiplicity distribution (\fig{\ref{fig:top:ttjets}})
in that the \MCATNLO generator predicts a larger fraction of events that have no jet activity 
beyond the jets originating directly from the top-quark decays. However, vetoing jets just in the forward region, at rapidities $|y|>1.5$, all simulations predict a smaller fraction of events with no additional jet than is seen in the data. 
These results can be used to improve the choice of models, scale parameters and tunes in Monte Carlo simulations for an optimal description of the data.

\begin{figure} [h] 
        \begin{center}
                \includegraphics[width=0.6\textwidth]{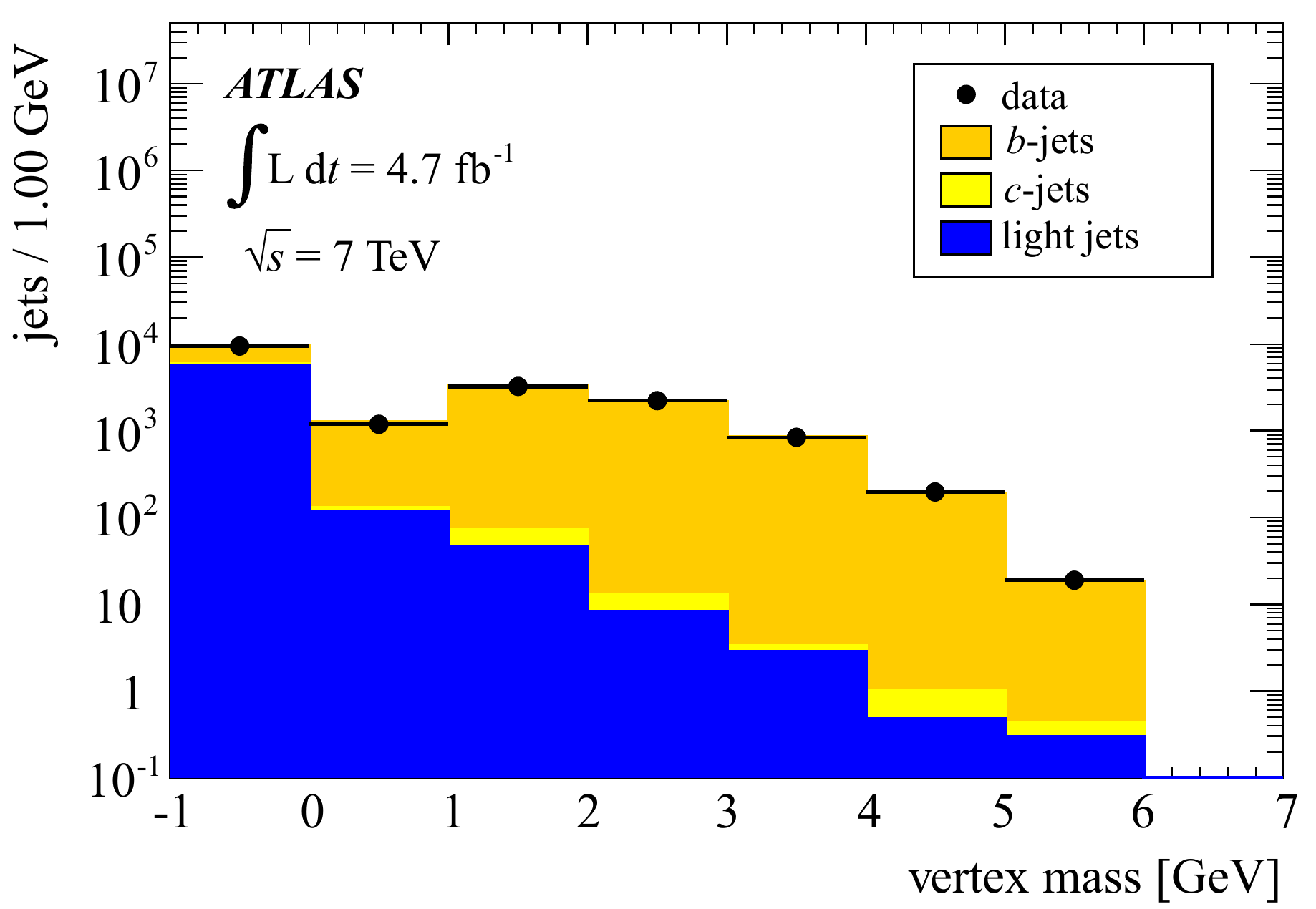} 
        \caption{Vertex-mass distribution for high-purity \bq-tagged jets in top-quark pair events with 
        two leptons in the final state. For this figure, no cut on the \bq-tagged jet multiplicity is 
        applied. 
        \textit{(Adapted from Ref.~\cite{Aad:2013tua}.)}
        \label{fig:top:ttHF} }
        \end{center}
\end{figure}

Building up on the insights about inclusive distributions of additional jets in top-quark pair events, 
both ATLAS and CMS have also been studying distributions and rates of events with additional jets 
originating from heavy quarks. These form an important background to events in which Higgs bosons 
are produced in association with top-quark pairs, with subsequent decay of the Higgs boson into 
$\bbbar$ or $\ccbar$ pairs, see Chap.~6 of Ref.~\cite{thisbook}.
The ATLAS experiment published a 
measurement of top-quark pairs together with heavy-flavour quarks, $\ttbar+\bq+X$ or $\ttbar+\cq+X$ 
at $\sqrt{s}=\unit{7}{\TeV}$~\cite{Aad:2013tua}. The separation between heavy-flavour and light-flavour contents of additional jets is achieved by a fit to the vertex-mass distribution of \bq-tagged jets, 
shown in \fig{\ref{fig:top:ttHF}}. From the fit, the relative contribution of heavy quarks 
is extracted to 
be $R_{\text{HF}}=[6.2\pm 1.1(\text{stat}) \pm 1.8 (\text{syst})]\%$, consistent within uncertainties with leading-order expectations, see Ref.~\cite{Aad:2013tua} and references therein.



\section{Top-Quark Mass}
\label{sec:top:mass}
\index{top quark!mass|(}

\begin{figure}[h] 
  \begin{center}
    \includegraphics[width=0.95\textwidth]{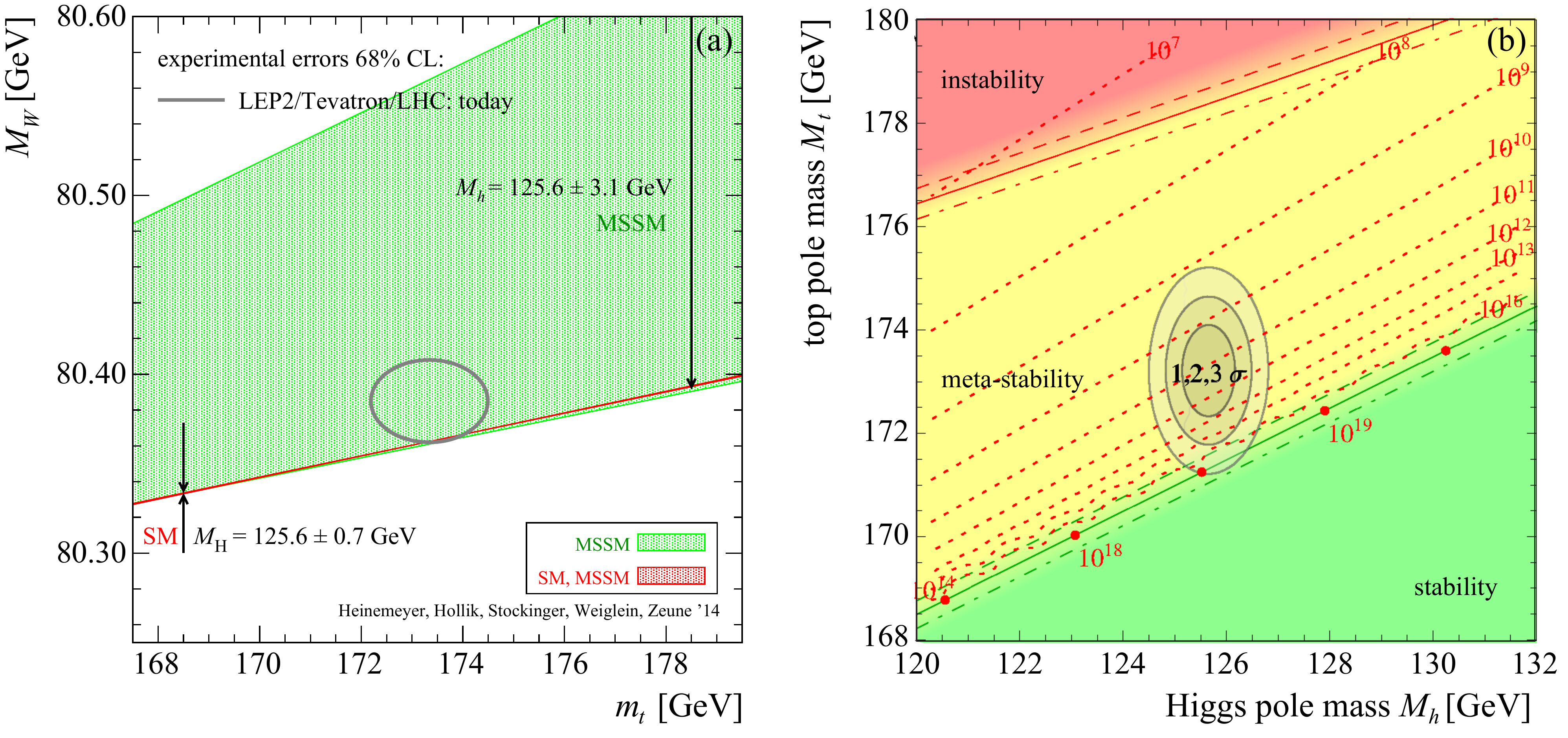}
  \caption{(a) Consistency of the Standard Model: The \Wb-boson
    mass as function of the top-quark mass for different Higgs-boson masses. 
    (b) Stability of the vacuum for different top-quark masses. 
    \text{(Adapted from Refs.~\cite{Heinemeyer:2013dia,Degrassi:2012ry,Degrassi:2014hoa}.)} 
  \label{fig:top:SMtests} }
  \end{center}
\end{figure}
In the SM, the couplings of the top quark to the gauge bosons 
are governed by the gauge structure (see also Chap.~4 of Ref.~\cite{thisbook}).
The only free parameters in top-quark
physics are thus the three corresponding 
CKM matrix elements
and the top-quark mass, $\Mt$.\index{CKM matrix} 
Instead of the top-quark mass, one may also use
the Yukawa coupling $\lambdat$ to the Higgs boson as a free parameter
since the two are related by $\Mt = v/\sqrt{2} \lambdat$, where $v$
is the vacuum expectation value of the Higgs field.
Once the CKM matrix elements and the
top-quark mass are known, the SM makes testable predictions for all
top-quark properties. Precise measurements of these properties can
thus be used to test the consistency of the SM.  A prominent example
is provided by the simultaneous measurements of the top-quark mass,
the \Wb-boson mass and the mass of the Higgs boson since these three
masses are related in the SM: The \Wb-boson mass can be calculated as
a function of the top-quark mass and the mass of the Higgs boson. A
comparison with the measured values thus allows the
mechanism of electroweak symmetry breaking predicted by the SM to be
indirectly tested. This is demonstrated in \fig{\ref{fig:top:SMtests}} where,
for comparison, also results within the minimal supersymmetric Standard 
Model (MSSM, see e.g. Chap.~10 of Ref.~\cite{thisbook})
are shown~\cite{Heinemeyer:2013dia}.\index{SUSY!MSSM} 

Recently, also the question of vacuum stability 
has attracted a lot of attention (see for example 
Refs.~\cite{Degrassi:2012ry,Degrassi:2014hoa}). 
Through quantum corrections, the top quark influences the effective Higgs
potential that is 
responsible for electroweak symmetry breaking. In
principle, it is conceivable that these quantum corrections modify the
effective potential such that it develops a second minimum or even
becomes unbounded from below. As a consequence, the electroweak vacuum
might become metastable or even unstable.  Calculating the corresponding
lifetime of the vacuum and comparing it with the age of the universe provides a
further consistency test of the SM. 

Experimentally, in most measurements, the mass of the top quark is determined
through the reconstruction of the top quark's decay products. The
top-quark mass can be estimated by comparing the measured values with
the value of the mass parameter used in the simulation. Measurements
employing this approach generally achieve the most precise results.

In contrast, from a theoretical point of view, a meaningful definition
of the top-quark mass requires to specify the renormalisation scheme
used to define the parameter in the theoretical predictions.  In this
respect the top-quark mass should be treated similar to a
coupling constant. In the theoretical description of hadronic
collisions, the so-called ``on-shell'' or ``pole mass'' scheme
and the ``minimal subtraction scheme''
(\MSbar) are commonly used.
Quantitatively, the pole mass and the mass measured from final state
reconstruction are expected to agree within
$\mathcal{O}(\unit{1}{\GeV})$~\cite{Buckley:2011ms,Moch:2014tta}.

In the pole mass scheme, the renormalised mass is defined as the
location of the pole of the renormalised quark propagator, including
higher-order corrections. In the (modified) minimal subtraction scheme,
the renormalised parameters are defined through a minimal subtraction of
the ultraviolet (UV) divergences. The renormalisation constants are
chosen such that they just cancel the UV divergences encountered in the
loop corrections (together with some irrelevant transcendental constants
in the modified minimal subtraction scheme).  The two definitions are
related in perturbation theory. At NLO accuracy, for example, the
relation reads
%
\begin{equation}
  \mtPole =  \mmu\left\{1 + {\alpS(\mu_R)\over \pi} \CF
    \left[1-{3\over 4}\ln\left({ \mmu^2\over\murs\squared}\right)\right] \right\}\, , 
    \label{eq:top:massms}
    \end{equation}
%
where \mtPole defines the pole mass and \mmu defines the mass in
the modified minimal subtraction scheme. $\CF=4/3$ denotes
the Casimir operator in the fundamental representation. The \MSbar
mass \mmu depends on the renormalisation scale, which coins the term
``running mass''.

The pole mass scheme is closely related to the intuitive understanding
of the mass of a free particle. While for many applications this
intuitive picture is a good approximation---for many purposes
the top quark behaves like a free quark---it should be clear that
this picture is doomed to fail if it comes to ultimate
precision. Indeed, it has been shown that the pole mass suffers from
the so-called ``renormalon ambiguity''
that leads
to an intrinsic uncertainty of the pole mass of the order
of \LambdaQCD~\cite{Bigi:1994em,Beneke:1994sw}.  Despite the fact that
the pole mass and the \MSbar mass definitions 
are related in perturbation theory, it is in
practice not straightforward to convert theoretical results from one scheme to the other.
So far, this translation
has only been calculated for the inclusive cross section for \ttbar
production and for a few differential distributions in \ttbar production.

Experimentally, the top-quark mass has been measured at the LHC 
using a large variety of methods and observables, and in different 
decay channels~\cite{
ATLAS:2012aj,
Chatrchyan:2012ea,
Chatrchyan:2013boa,
Chatrchyan:2013xza,
Chatrchyan:2012cz,
Chatrchyan:2011nb,
Aad:2014zea}. 
In addition, the difference of the masses of top quarks and antiquarks have been measured~\cite{Chatrchyan:2012uba,Aad:2013eva}.

\begin{figure}[h] 
        \begin{center}
            \includegraphics[width=0.95\textwidth]{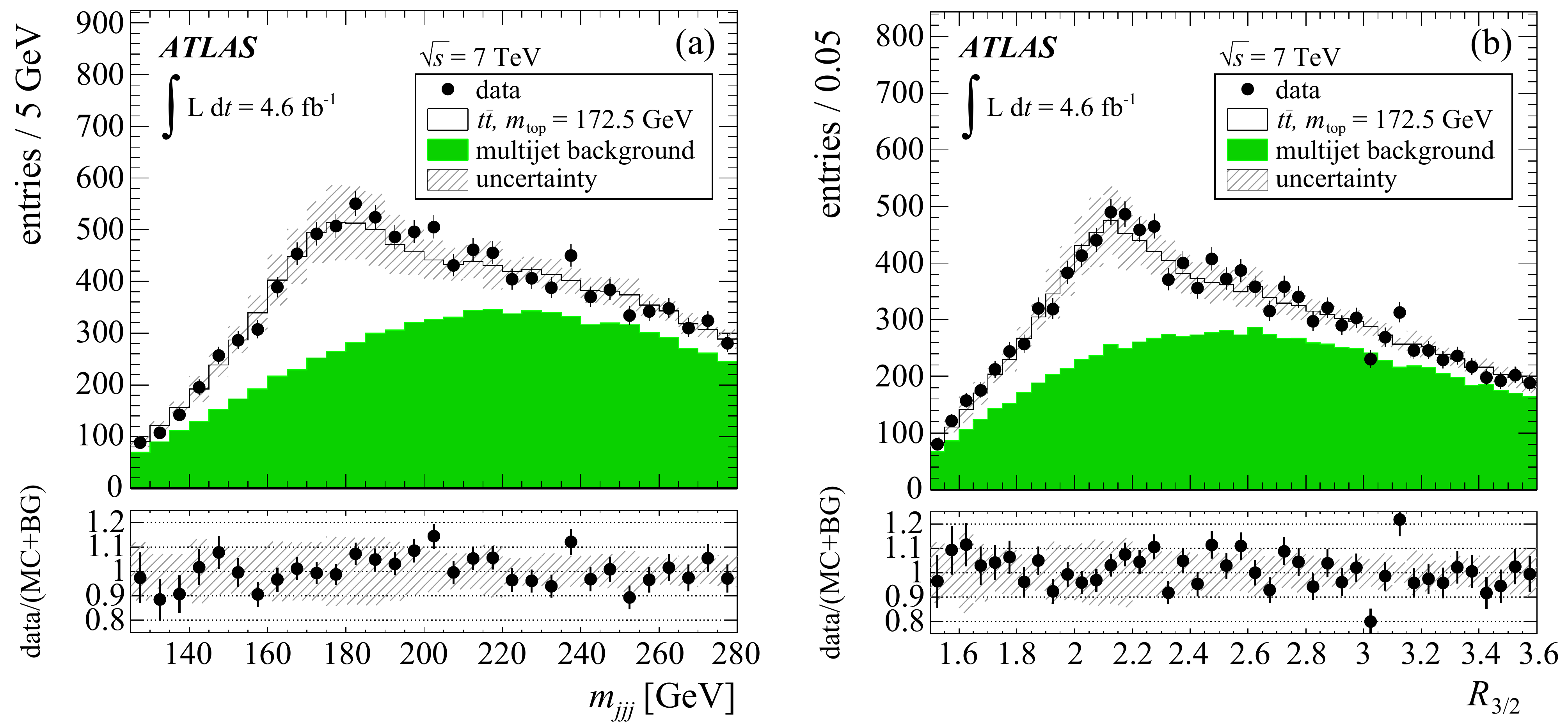} 
        \caption{Distributions of (a) the 3-jet invariant mass and (b) the ratio of the three-jet mass to 
        the dijet mass $R_{3/2}$, measured in data and compared to expectations after applying 
        all event selection criteria. The open histograms show the expected 
        distributions for \ttbar events for a top-quark mass value of \unit{172.5}{\GeV}. The shaded 
        histograms indicate the contribution from multi-jet backgrounds. 
        \textit{(Adapted from Ref.~\cite{Aad:2014zea}.)} 
        \label{fig:top:mass:atlas} }
        \end{center}
\end{figure}

A recent measurement using \ttbar events with fully hadronic final states is presented by the ATLAS collaboration~\cite{Aad:2014zea}. Events are selected if they contain at least six jets, and exactly two \bq-tagged jets are required to be among the four leading jets. 
The top-quark mass is extracted from a binned likelihood fit to the $R_{3/2}$ distribution, shown in \fig{\ref{fig:top:mass:atlas}}, where $R_{3/2}$ is the ratio of the reconstructed three-jet and two-jet masses. In this distribution systematic effects that are common to the 
masses of the reconstructed top quark and the associated \Wb boson cancel. 
The contribution from multi-jet backgrounds is determined from the data using the event 
yields in different regions of \bq-tag jet multiplicity and 6th-jet momentum. 
The measurement yields a value for the top-quark mass of
\begin{eqnarray*}
\Mt = 175.1 \pm 1.4~(\text{stat}) \pm 1.2~(\text{syst})\, \GeV \, .
\end{eqnarray*}
The systematic uncertainties are dominated by the residual uncertainties of the 
jet energy scale, in particular for \bq-quark jets, and by the uncertainties from hadronisation modelling.

\begin{figure}[h] 
        \begin{center}
                \includegraphics[width=0.95\textwidth]{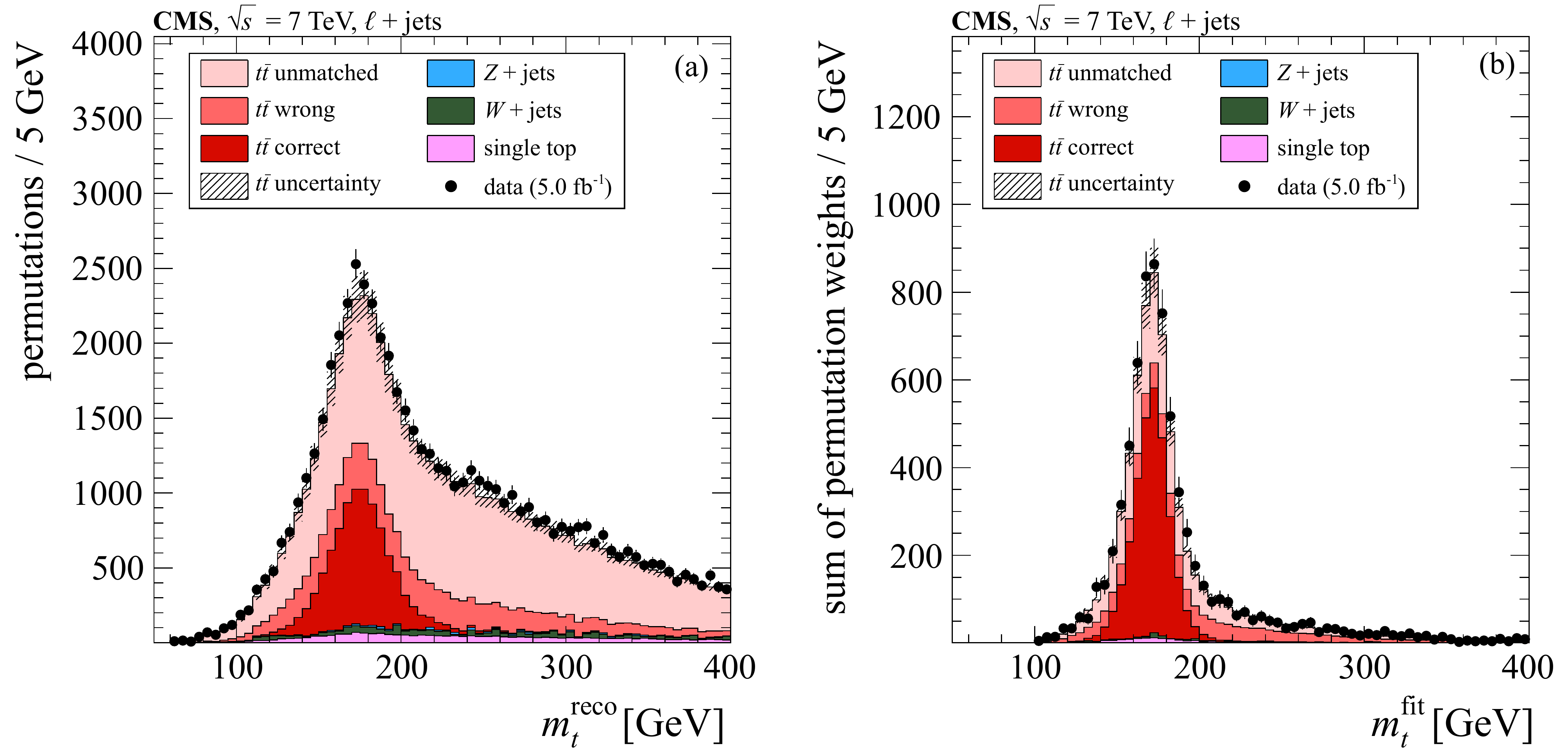} 
        \caption{Reconstructed top-quark mass distributions (a) before and (b) fitted top-quark mass distributions after kinematic reconstruction with  
        entries weighted by the reconstructed goodness-of-fit parameter $P_\text{gof}$. 
        \textit{(Adapted from Ref.~\cite{Chatrchyan:2012cz}.)} 
        \label{fig:top:mass:cms} }
        \end{center}
\end{figure}

The CMS collaboration presented a measurement of the top-quark mass using the full data set collected at a centre-of-mass energy of $\sqrt{s}=\unit{7}{\TeV}$ with one electron or muon and at least four jets in the final state~\cite{Chatrchyan:2012cz}. 
A kinematic fit to the four leading jets, the lepton and the missing transverse momentum is employed to constrain the selected events to the hypothesis of the production of two heavy particles of equal mass, each one decaying to a \Wb boson and a \bq quark. The reconstructed masses of the two \Wb bosons are constrained in the fit to \unit{80.4}{\GeV}. The reconstructed invariant-mass distribution is shown in \fig{\ref{fig:top:mass:cms}}.
Events can enter the distributions with different parton-jet assignments (permutations). For simulated \ttbar events, the parton-jet assignments can be classified as correct, wrong and unmatched permutations where, in the latter, at least one quark from the \ttbar decay is not matched to any of the four selected jets. The actual top-quark mass value is determined simultaneously with the 
jet energy scale using a joint likelihood fit. The joint likelihood is constructed based on the 
``ideogram method''
in which the likelihood for each event is evaluated from analytic expressions obtained from simulated events. Biases arising due to this method are determined using pseudo-experiments and corrections are applied accordingly.
The dominant uncertainty of the final result comes from the uncertainty of the difference in 
the jet energy responses for jets originating from light (\uq, \dq, \sq) or bottom quarks, as well as 
from statistical uncertainties in the determination of differences between different models 
for colour-reconnection
processes. The final result is
\begin{eqnarray*}
\Mt = 173.49 \pm 0.43~(\text{stat+JES}) \pm 0.98~(\text{syst})\, \GeV \, ,
\end{eqnarray*}
corresponding to an optimal jet energy scale correction of $0.994 \pm 0.003~(\text{stat}) \pm 0.008~(\text{syst})$ 
with respect to the CMS calibration. 

\begin{figure}[h] 
        \begin{center}
                \includegraphics[width=0.7\textwidth]{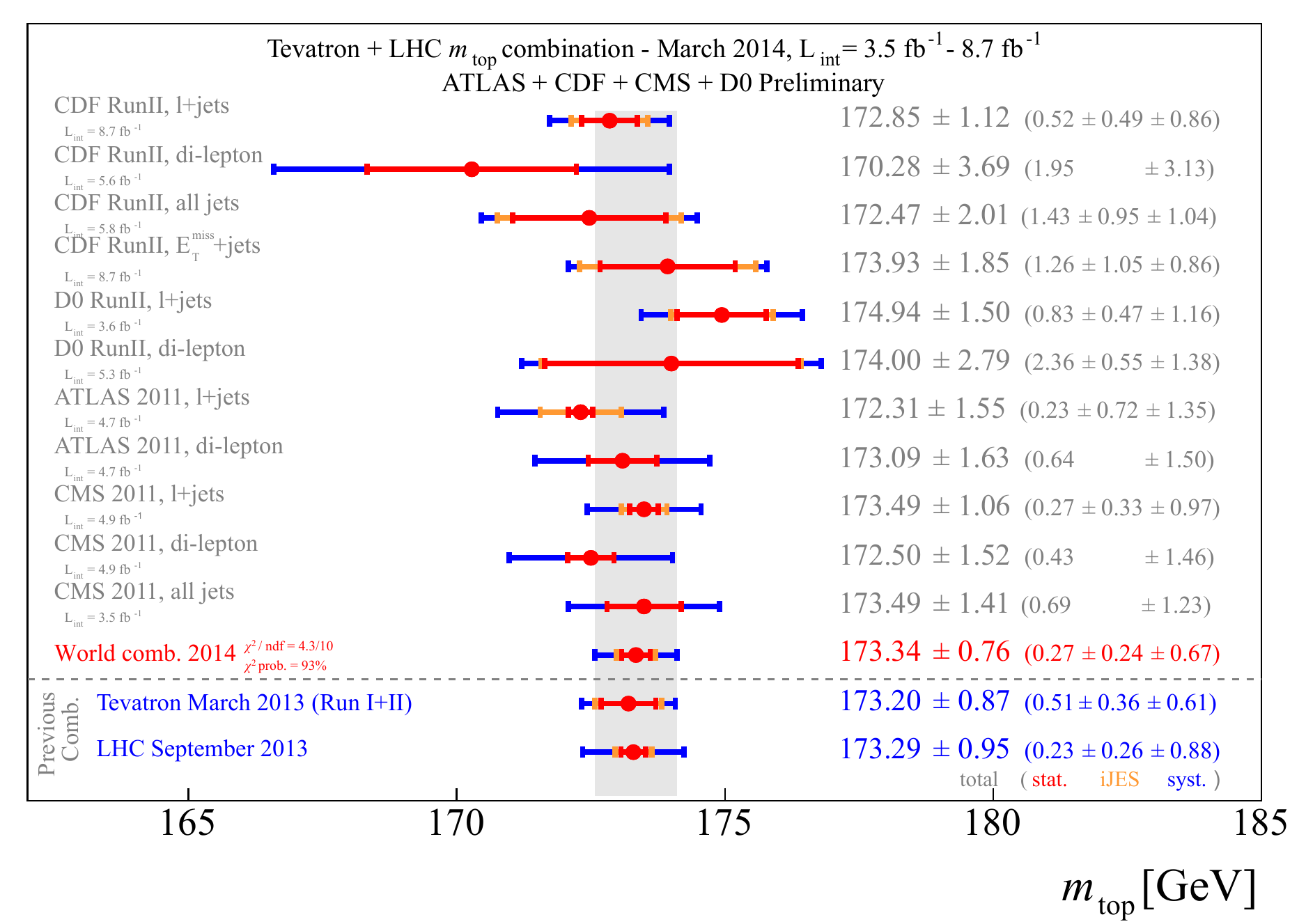} 
        \caption{Mass measurements included in the world combination 2014. 
        Also shown are the results from the combinations of LHC-only 
        data and \Tevatron-only data. 
        \textit{(Adapted from Ref.~\cite{ATLAS:2014wva}.)}
        \label{fig:top:masswa} }
        \end{center}
\end{figure}

In spring 2014, this and other precise results from the LHC and the \Tevatron, both preliminary and final, were 
combined to obtain a first world average of the top-quark mass~\cite{ATLAS:2014wva}. A summary is 
shown in \fig{\ref{fig:top:masswa}}.

\begin{figure}[h] 
        \begin{center}
                \includegraphics[width=0.7\textwidth]{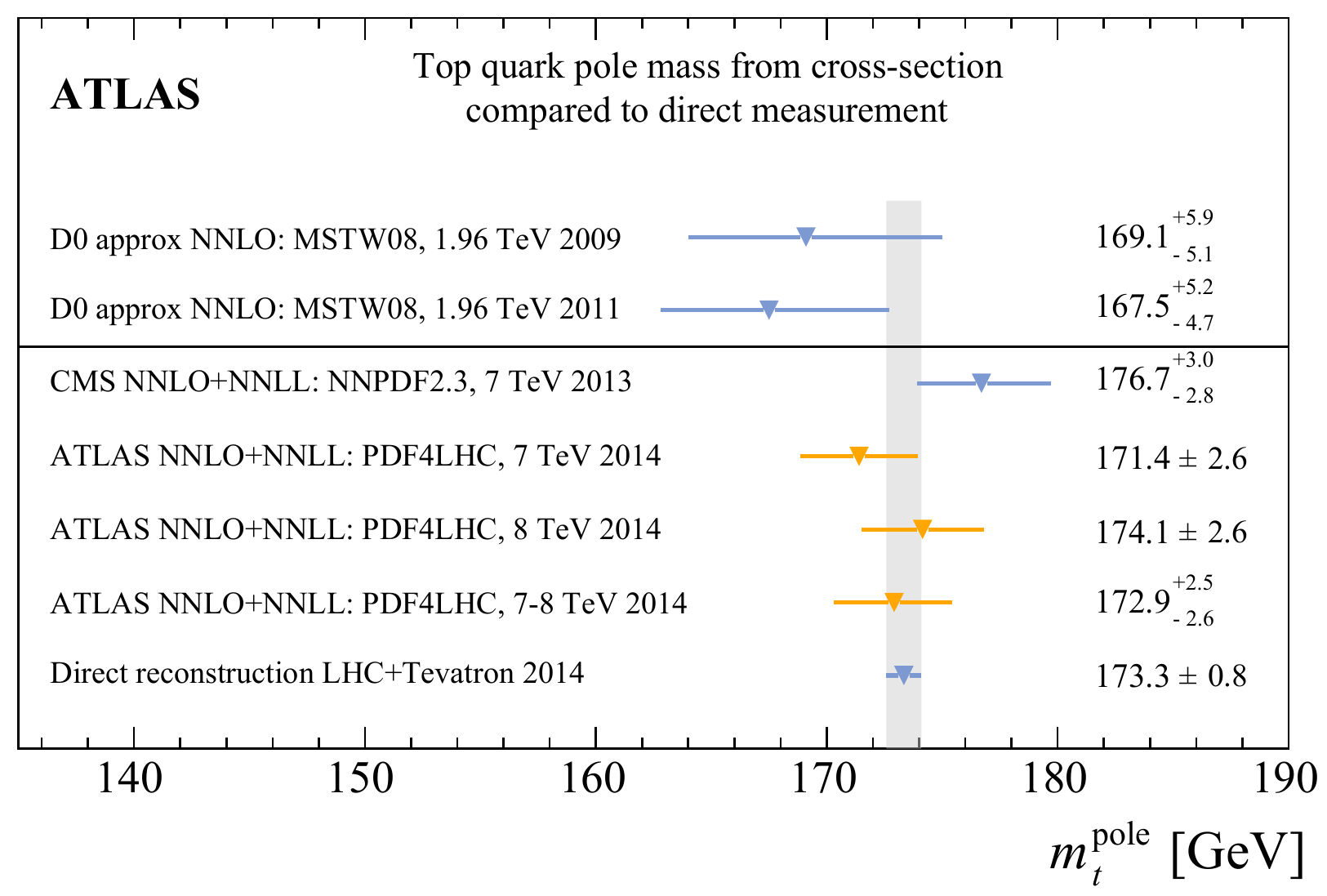} 
        \caption{Summary of determinations of the top-quark pole mass from cross-section 
        measurements. Also shown is the top-quark mass world average as 
        obtained from combination of direct experimental measurements~\cite{ATLAS:2014wva}. 
        \textit{(Adapted from Ref.~\cite{Aad:2014kva}.)} 
        \label{fig:top:massfromxsec} }
        \end{center}
\end{figure}

An alternative approach to the determination of the top-quark mass is to extract its value from 
the measured inclusive cross section. 
This approach has the advantage that the cross section and the 
pole mass are directly related, such that the extraction yields a theoretically well-defined quantity. Both ATLAS 
and CMS have used their cross-section measurements 
to extract the top-quark pole 
mass~\cite{Chatrchyan:2013haa,Aad:2014kva} as defined at NNLO accuracy~\cite{Czakon:2013goa}. 
The extractions are performed for different parton distribution functions and take into 
account the experimental dependence of the measured cross section on the assumed top-quark mass. A summary of the results is shown in \fig{\ref{fig:top:massfromxsec}}. 
Conversely, assuming equality between the pole mass and the directly measured mass within \unit{1}{\GeV}, 
the cross-section measurements can be used to determine the strong coupling 
constant $\alpS$~\cite{Chatrchyan:2013haa}. 
\index{top quark!mass|)}

\section{Tests of QCD Predictions}

Perturbative QCD (pQCD) is a fundamental building block for the understanding of physics processes at the LHC. In addition to cross-section predictions, several other aspects of pQCD can be probed in the top-quark sector. Examples are the small but finite charge asymmetry of \ttbar pairs and the correlation between the top-quark and antiquark spins in \ttbar production; both quantities are sensitive to the relative proportions of the different production processes of \ttbar pairs. Another example is the polarisation of top quarks in \ttbar events, which tests the \CCC and \CP structure of \ttbar production.

\subsection{Charge Asymmetry}

At leading order QCD, the reactions $\pp \to \ttbar$ and $\ppbar \to \ttbar$ are symmetric 
under the exchange of top quarks and antiquarks. The relevant hard scattering processes are 
quark-antiquark annihilation, $\qbarq \to \ttbar$, and gluon-fusion, $\glueglue \to \ttbar$. 
At NLO, two additional types of processes have to be considered which break this charge symmetry, 
namely quark-gluon scattering, $\qq\glue \to \ttbar \qq$, and radiative corrections to quark-antiquark 
annihilation~\cite{Kuhn:1998jr,Kuhn:1998kw}. In both cases, the charge asymmetry is induced by interference 
effects, e.g.\ interference between processes with initial-state and final-state radiation or interference 
between Born and box diagrams. All gluon-fusion processes are symmetric under the exchange of the 
top quark and antiquark. 

The charge asymmetry results in an asymmetry of the \ttbar event kinematics: Top quarks (antiquarks) are 
preferentially emitted in the direction of the incoming quark (antiquark).  
The observables with which the charge asymmetry can be measured are chosen depending on the colliding particles and the centre-of-mass energy. Quark-antiquark annihilation dominates the production of \ttbar pairs at the \Tevatron, and so the top quark and antiquark will preferentially be emitted in the direction of the incoming protons and antiprotons, respectively. The most common observable is thus a forward-backward asymmetry,
\def\Att{A^{t\bar t}}
\begin{eqnarray*}
        \Att = \frac{N(\Delta y > 0) - N(\Delta y < 0)}{N(\Delta y > 0) + N(\Delta y < 0)} \, , 
\end{eqnarray*}
%
where $\Delta y$ is the difference between the rapidities of the top quark and antiquark, 
i.e.\ $\Delta y = y_{\tq} - y_{\tbarq}$, and $N$ is the number of events. The predictions for this observable 
depend on several kinematic variables. The inclusive forward-backward asymmetry is predicted to be 
$\Att=0.088 \pm 0.006$~\cite{Bernreuther:2012sx}, 
while for invariant \ttbar masses larger than \unit{450}{\GeV} 
the prediction increases to $\Att(m_{t\bar t}>\unit{450}{\GeV})=0.129^{+0.008}_{-0.006}$~\cite{Bernreuther:2012sx}. 
The CDF and \Dzero experiments have measured the forward-backward asymmetry both inclusively and as 
a function of several kinematic quantities, e.g.\ $m_{t\bar t}$~\cite{Aaltonen:2012it,Abazov:2011rq}.
They found an excess compared to the NLO 
QCD predictions with significances of several standard deviations. Over the last few years these measurements gave rise to speculations about contributions to $t\bar{t}$ production due to physics beyond the Standard Model. Although refined theoretical studies and further measurements appear to have resolved the issue, measurements of the charge asymmetry are still in the focus of the LHC top-physics programme.

At the LHC with its symmetric \pp initial state, the charge asymmetry can not be measured 
as a forward-backward asymmetry. Instead, a central-decentral asymmetry is defined. As valence quarks 
carry a larger average momentum fraction and top (anti)quarks are produced preferentially in the direction 
of the incoming (anti)quark, 
the average top-quark rapidity is larger than that of top antiquarks. 
A useful observable is defined as
%
\begin{eqnarray*}
        \ACDC = \frac{N(\Delta |y|>0) - N(\Delta |y|<0)}{N(\Delta |y|>0) + N(\Delta |y|<0)} \, , 
\end{eqnarray*}
%
where $\Delta |y|$ is the difference between the absolute values of the top-quark and top-antiquark rapidities, 
i.e.\ $\Delta |y|=|y_{\tq}| - |y_{\tbarq}|$. The NLO QCD prediction including electroweak effects for the 
inclusive asymmetry is $\ACDC=0.0123\pm0.0005$~\cite{Bernreuther:2012sx}. Predictions are also available 
for different values of the invariant mass, the rapidity and the transverse momentum of the \ttbar pair, and 
they range between -0.6\% and 2.8\% (see discussion in Ref.~\cite{Aad:2013cea}). 
The asymmetry depends on the first two 
variables because they are correlated to the fraction of quark-antiquark annihilation in \ttbar production. 
It depends on the latter quantity because the amount  
of initial-state and finale-state radiation changes with 
increasing transverse momentum.

\begin{figure}[h] 
        \begin{center}
                        \includegraphics[width=0.95\textwidth]{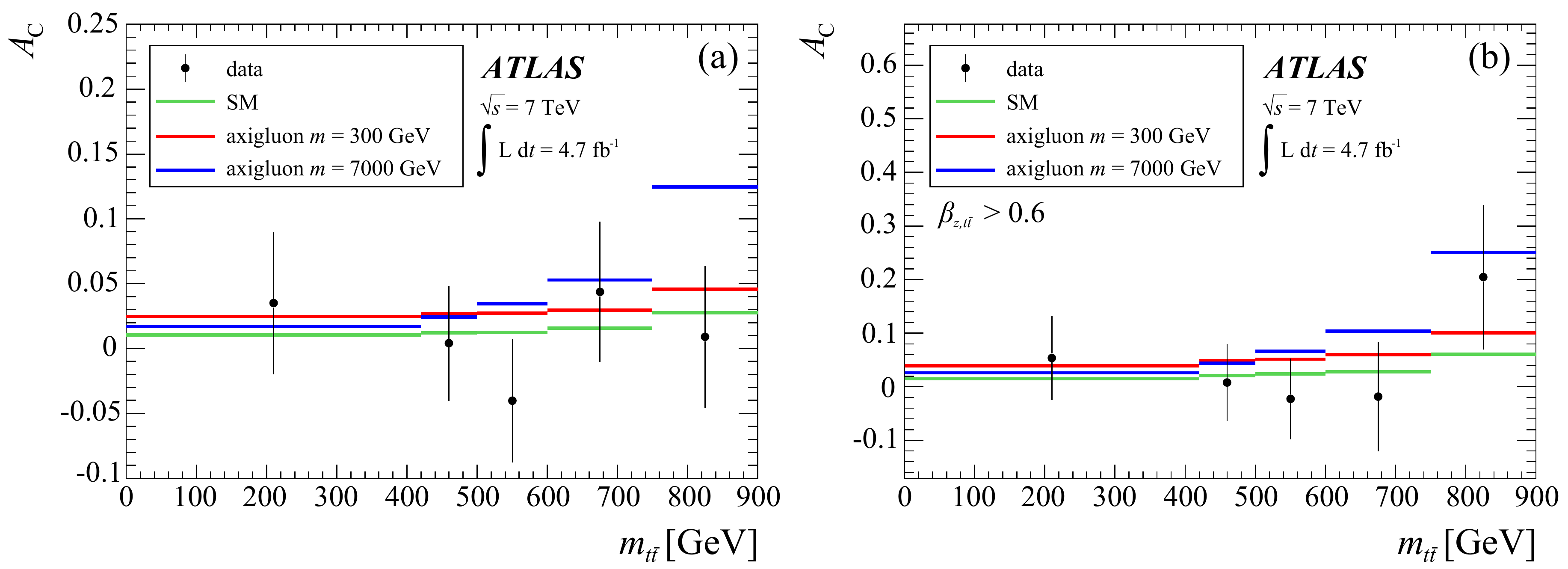} 
        \caption{Charge asymmetry measured by the ATLAS collaboration as a function of the 
        invariant \ttbar mass (a) without and (b) with the requirement of a large 
        longitudinal \ttbar velocity. 
        \textit{(Adapted from Ref.~\cite{Aad:2013cea}.)}         
        \label{fig:top:ACvsmtt}
        } 
        \end{center}
\end{figure}

After an initial measurement using only a subset of the available data recorded 
at $\sqrt{s}=\unit{7}{\TeV}$~\cite{ATLAS:2012an}, the ATLAS collaboration has studied the charge 
asymmetry based on the full data set, which corresponds to an integrated luminosity 
of \unit{4.7}{\invfb}~\cite{Aad:2013cea}. 
Events are selected that are consistent with the single-lepton decay mode of \ttbar production. For each event, 
the top-quark pair is reconstructed using a likelihood-based kinematic fit~\cite{Erdmann:2013rxa}, and the rapidities 
of the top quark and antiquark are reconstructed. The measured distribution of $\Delta |y|$ includes
background events and is distorted by detector and acceptance effects. A Bayesian unfolding 
technique~\cite{Choudalakis:2012hz} is applied on the background-subtracted spectrum to remove such effects.
The measured inclusive asymmetry is
\begin{eqnarray*}
\ACDC = 0.006 \pm 0.010 \, ,
\end{eqnarray*}
where the largest sources of uncertainty are the statistical uncertainty and the uncertainty due to lepton 
and jet reconstruction. In addition, the asymmetry for invariant \ttbar masses greater than \unit{600}{\GeV} 
is found to be $\ACDC(m_{\ttbar}>\unit{600}{\GeV})=0.018\pm0.022\, ,$ which is in good agreement with 
the predicted value of $0.0175^{+0.0005}_{-0.0004}$. The asymmetry
is also measured as a function of 
the transverse momentum, the absolute value of the rapidity and the invariant mass of the \ttbar pair. The 
latter measurement is repeated for a subset of the events featuring a high longitudinal \ttbar velocity, 
i.e.\ requiring $\beta_{z,\ttbar}>0.6$. 
The asymmetries as a function of the invariant \ttbar mass are shown in \fig{\ref{fig:top:ACvsmtt}} 
without (a) and with (b) the additional velocity requirement. All four differential measurements are in agreement 
with the SM predictions. 

\begin{figure}[h] 
        \begin{center}
                        \includegraphics[width=0.95\textwidth]{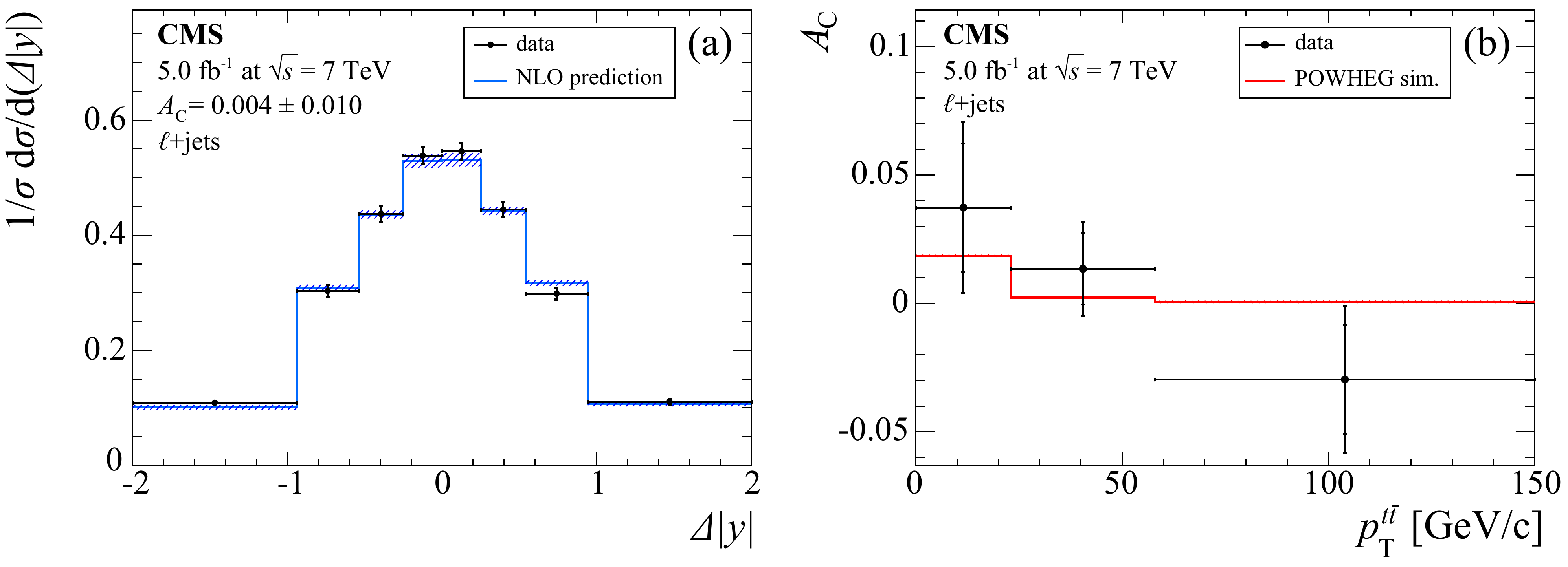} 
        \caption{(a) The background-subtracted and unfolded $\Delta |y|$ distribution obtained by 
        CMS for a single-lepton sample. 
        (b) The charge asymmetry as a function of the transverse momentum of the \ttbar pair.
        \textit{(Adapted from Ref.~\cite{Chatrchyan:2012cxa}.)} 
        \label{fig:top:deltaY}} 
        \end{center}
\end{figure}

The CMS collaboration has measured the charge asymmetry in a data set corresponding to an 
integrated luminosity of \unit{5.0}{\invfb}~\cite{Chatrchyan:2011hk,Chatrchyan:2012cxa} using events 
with exactly one charged lepton in the final state. After event reconstruction and calculation of the 
top-quark and top-antiquark rapidities, the $\Delta |y|$ distribution is determined.
Subsequently, the 
estimated background contributions are subtracted from the data and the spectra are corrected for 
detector and acceptance effects using a regularised unfolding procedure via 
matrix inversion~\cite{Blobel:2002pu}. The measured inclusive asymmetry is
\begin{eqnarray*}
        \ACDC = 0.004 \pm 0.010~(\text{stat}) \pm 0.011~(\text{syst}) \, , 
\end{eqnarray*}
where the major sources of systematic uncertainty are the residual model dependence of the unfolding procedure 
and the lepton reconstruction. As in the ATLAS measurement,
the asymmetry is also measured as a function of the transverse momentum, 
the rapidity and the invariant mass of the \ttbar system. Figure~\ref{fig:top:deltaY}(a) shows the 
background-subtracted and unfolded $\Delta |y|$ distribution for the inclusive case, and
\fig{\ref{fig:top:deltaY}}(b) shows the
charge asymmetry as a function of the transverse momentum of the \ttbar pair. 
All measurements are consistent with the SM predictions. 

In a further study using the full \unit{7}{\TeV} data set, the CMS collaboration measured the 
charge asymmetry in a sample of dileptonically decaying \ttbar pairs~\cite{Chatrchyan:2014yta}. 
In addition to the observable \ACDC, a ``lepton charge asymmetry'' is defined as
\begin{eqnarray*}
        \ALEPC = \frac{N(\Delta |\eta_{\ell}|>0) - N(\Delta |\eta_{\ell}|<0)}{N(\Delta |\eta_{\ell}|>0) + N(\Delta |\eta_{\ell}|<0)} \, , 
\end{eqnarray*}
where $\Delta |\eta_{\ell}| = |\eta_{\ell^{+}}| - |\eta_{\ell^{-}}|$ and where $\eta_{\ell^{\pm}}$ are the pseudo-rapidities 
of the positively and negatively charged leptons in each event. The SM prediction for this observable is 
$\ALEPC=0.0070\pm0.0003$~\cite{Bernreuther:2012sx}. The event reconstruction is performed using the 
``analytical matrix weighting technique'' (AMWT)~\cite{Chatrchyan:2011nb}. 
After subtracting all background contributions 
from the measured $\Delta |y|$ and $\Delta |\eta_{\ell}|$ distributions, the spectra are unfolded using singular 
value decomposition~\cite{Hocker:1995kb}. The measured
asymmetries are
\begin{eqnarray*}
        \ACDC & = & -0.010 \pm 0.017~(\text{stat}) \pm 0.008~(\text{syst}) ~\text{and} \\
        \ALEPC & = & \phantom{-}0.009 \pm 0.010~(\text{stat}) \pm 0.006~(\text{syst}) \, , 
\end{eqnarray*}
where the largest sources of systematic uncertainty are residual biases in the unfolding procedure 
and uncertainties in the \ttbar modelling and the jet reconstruction. The lepton charge asymmetry is also calculated as 
a function of the same three kinematic variables as for the single-lepton analysis. No deviations between the 
measurements and the predictions are found.

The charge asymmetry in \ttbar production is predicted to cause a
small effect on a variety of observables at the LHC.  None of the
measurements performed by the ATLAS and CMS collaborations are in
conflict with the SM predictions while the current experimental
precision is of the order of the size of the prediction itself.
Although a variety of models of physics beyond the Standard Model
can be excluded with the set of
measurements already performed, the analysis of the \unit{8}{\TeV}
data (which was not finished at the time of publication) will provide
further sensitivity to the predictions of perturbative QCD. It should
be noted, that the interpretation of the leptonic charge asymmetries
relies on a solid understanding of top-quark production and decay. In
particular, a non-standard top-quark polarisation could affect the
leptonic asymmetries. It is thus important to cross check the
polarisation through explicit measurements.


\subsection{Top-Quark Polarisation and Spin Correlation in Top-Quark Pairs}
\index{top quark!spin correlation|(}
\index{top quark!test of \CP violation|(}

Top-quark polarisation in \ttbar events and the correlation between the top-quark and antiquark spins in \ttbar production are probes of perturbative QCD and observables that are sensitive to anomalous production mechanisms. Measurements of such quantities are only feasible because of the extremely short lifetime of the top quark of $\tau_{\tq}\approx \unit{1.5 \cdot \power{10}{-25}}{\second}$. The lifetime is roughly one order of magnitude smaller than the time scale at which hadronisation takes place, 
$\tau_{\text{had}} \approx 1/\LambdaQCD \approx \unit{3 \cdot \power{10}{-24}}{\second}$, and in particular shorter than the time needed to decorrelate the spin configuration of the \ttbar pair, $\tau_{\text{decorr}} \approx \frac{\hbar \Mt}{\LambdaQCD^{2}} \approx \unit{3 \cdot \power{10}{-21}}{\second}$~\cite{Grossman:2008qh,Mahlon:2010gw}.
Top quarks will thus decay before they can form bound states. As mentioned before, the large top-quark width cuts off non-perturbative effects. The polarisation of the top-quark is thus not diluted by hadronisation effects and can be calculated reliably within perturbation theory. The parity-violating weak decay can then be used to analyse the top-quark polarisation through the angular distribution of the decay products. The correlation of the top-quark and antiquark spins is reflected in the angular correlation of the top-quark and antiquark decay products. This is a unique feature since all lighter quarks form hadrons for which---due to the hadronisation process---the initial spin information of the mother particle is diluted or even entirely lost.  In contrast, the spin information of the top quark is directly transferred to its decay products. 

In top-quark pairs, information about the polarisation of top quarks and the correlation between top-quark 
and antiquark spins can be obtained from the differential cross section
\begin{eqnarray}
        \label{eqn:top:diffangle} 
        \frac{1}{\sigma}\frac{\dif \sigma}{\dif \cos(\theta_{1}) \dif \cos(\theta_{2})} & = & \frac{1}{4} \left( 1 + \alpha_{1} P_{1} \cos(\theta_{1}) + \alpha_{2} P_{2} \cos(\theta_{2}) \right. \nonumber \\ 
 &  & \left. - \alpha_{1}\alpha_{2}A \cos(\theta_{1})\cos(\theta_{2}) \right) \, , 
\end{eqnarray}
where $\theta_{1}$($\theta_{2}$) are the angles between the momentum direction of a daughter particle of the top (anti)quark and a chosen reference axis. The coefficients $\alpha_{1}$($\alpha_{2}$) and $P_{1}$($P_{2}$) are the ``spin-analysing power''
of the daughter particle and the degree
of polarisation (with respect to the reference axis) of the top (anti)quark, respectively. 
The spin-analysing power quantifies the amount of spin information transferred to the daughter particle and depends on 
the particle type. It is approximately one for charged leptons and down-type quarks from the subsequent decay of 
the \Wb boson~\cite{Brandenburg:2002xr}. The coefficient $A$ is a measure of the spin correlation between top quark
and antiquark.

As the strong interaction conserves parity, the polarisation of top (anti)quarks in \ttbar production
within the production plane is expected to be zero. 
QCD absorptive parts, sometimes also called final-state interactions\footnote{These are due to imaginary parts of the loop integrals caused by the on-shell
production of intermediate states.}, 
introduce a tiny transverse polarisation at the one-loop
level~\cite{Bernreuther:1995cx,Dharmaratna:1989jr}.  Electroweak corrections lead to a small amount 
of net polarisation 
of $\alpha_{i}P_{i} = 0.003 \pm 0.001$~\cite{Bernreuther:2013aga}. It can be shown that polarisation can be induced by the imaginary part of 
a chromo-electric dipole moment which in turn can lead to \PARITY-odd and \CP-odd terms in the 
matrix elements~\cite{Bernreuther:2013aga}. 
Such effects can stem from processes beyond the SM.

The correlation coefficient $A$ can be expressed as an asymmetry variable in the number 
of events $N$ with parallel and antiparallel spin,

\begin{eqnarray*}
        A = \frac{N(\uparrow\uparrow) + N(\downarrow\downarrow) - N(\uparrow\downarrow) - N(\downarrow\uparrow)}{N(\uparrow\uparrow) + N(\downarrow\downarrow) + N(\uparrow\downarrow) + N(\downarrow\uparrow)} \, , 
\end{eqnarray*}
where $\uparrow$ and $\downarrow$ indicate the spin projections onto
the reference axis. The prediction of the correlation coefficient
depends on the particular choice of reference axis.  While for
measurements at the \Tevatron the ``off-diagonal
basis''
and the ``beam basis''
are suitable choices, measurements at
the LHC are most sensitive to the correlation coefficient in the
``helicity basis''.
In the beam basis, the direction of
the beam is used as reference axis for the top-quark as well as for
the antiquark. In the helicity basis, the direction of flight of the
top-quark/antiquark is used as respective reference axis. The strength of the correlation
predicted by the SM using the helicity basis is $A=0.031$~\cite{Bernreuther:2010ny} with an
uncertainty of approximately~1\%.

Close to the threshold, top-quark pairs produced via gluon fusion are in
a $^1{\rm S}_0$ 
state while top-quark pairs produced via
quark-antiquark annihilation are in a $^3{\rm S}_1$ state. As a
consequence, the spins of a top-quark pair produced in
quark-antiquark annihilation tend to be parallel while in gluon
fusion they tend to be antiparallel.  
A measurement
of the coefficient $A$ is thus a direct probe of the production mechanism.
Contributions from additional production mechanisms, e.g.\ yet unknown
intermediate vector bosons, can lead to altered predictions for $A$. Note
that the correlation coefficient depends both on the centre-of-mass
energy and on the initial-state particles, and so the measurements
conducted at the \Tevatron and the LHC are complementary.

\begin{figure}[h] 
        \begin{center}
                        \includegraphics[width=0.95\textwidth]{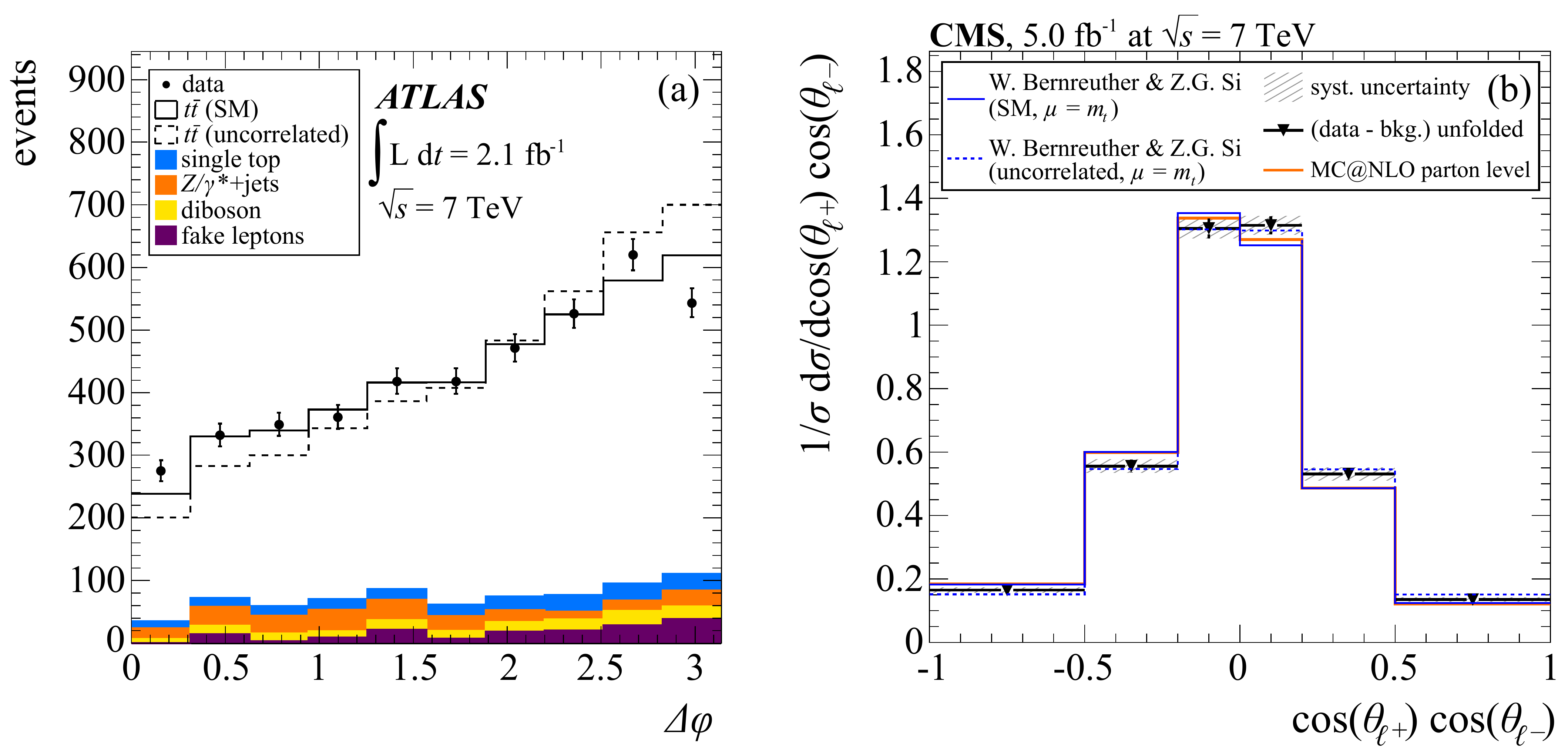}  
        \caption{(a) Observed and predicted $\Delta \phi$ spectrum in a sample of dileptonic \ttbar events recorded with the ATLAS experiment. 
        (b) Distribution of $\cos(\theta_{\ell^{+}})\cos(\theta_{\ell^{-}})$ in dileptonic \ttbar events as observed by 
        CMS together with predictions for two different correlation scenarios. 
        \textit{(Adapted from Refs.~\cite{ATLAS:2012ao,Chatrchyan:2013wua}.)} 
        \label{fig:top:deltaphi}} 
        \end{center}
\end{figure}

While an evidence for correlated spins in top-quark pairs was already reported by the \Dzero collaboration~\cite{Abazov:2011gi}, the hypothesis that the spins of the top quark and the antiquark are uncorrelated was fully disproved for the first time by a measurement of the ATLAS collaboration~\cite{ATLAS:2012ao}. The measurement is based on a data set corresponding to an integrated luminosity of \unit{2.1}{\invfb} recorded at $\sqrt{s}=\unit{7}{\TeV}$. Events consistent with the signature of \ttbar events decaying in the dilepton mode are selected, and the spin correlation is probed using the difference in azimuthal angle between the two charged leptons, $\Delta \phi$, calculated in the laboratory frame. The advantage of this observable compared to those in \eqn{\eqref{eqn:top:diffangle}} is that no kinematic reconstruction of the top-quark momenta is necessary while the sensitivity to the correlation strength is largely retained~\cite{Mahlon:2010gw}. Figure~\ref{fig:top:deltaphi}(a) shows the observed $\Delta\phi$ spectrum as well as the predictions for the assumption of SM correlations and the absence of correlations. 
The measured correlation coefficient is not compatible with zero,
\begin{eqnarray*}
        A = 0.40 \pm 0.04~(\text{stat}) ^{+0.08}_{-0.07}~(\text{syst}) \, , 
\end{eqnarray*}
with a significance of 5.1~standard deviations. The dominating sources
of systematic uncertainty are the estimate of events with
misidentified leptons and the jet reconstruction. In a second
publication based on the full \unit{7}{\TeV} data set~\cite{Aad:2014pwa}, the
spin correlation is measured in the dilepton channel using a variety
of alternative observables, and it it also measured in the
single-lepton channel.

The CMS collaboration has analysed a data set corresponding to an integrated luminosity of \unit{5.0}{\invfb} 
recorded at $\sqrt{s}=\unit{7}{\TeV}$~\cite{Chatrchyan:2013wua}. Events are selected that contain exactly 
two charged leptons with large transverse momenta. Three angular variables are calculated for each event---the 
$\Delta \phi$ variable defined earlier as well as the angles $\theta_{\ell^{+}}$ and $\theta_{\ell^{-}}$, which are defined as the angles of the positively and negatively charged lepton in the helicity frame, respectively.
From the distributions of these variables, two asymmetries are derived which provide discrimination 
between the two scenarios of SM correlations and no correlations:
\begin{eqnarray*}
        \ADELPHI & = & \frac{N(\Delta \phi > \pi/2) - N(\Delta \phi < \pi/2)}{N(\Delta \phi > \pi/2) + N(\Delta \phi < \pi/2)} \, , \\
        \ACOS & = & \frac{N(\cos(\theta_{\ell^{+}})\cos(\theta_{\ell^{-}})>0) - N(\cos(\theta_{\ell^{+}})\cos(\theta_{\ell^{-}})<0)}{N(\cos(\theta_{\ell^{+}})\cos(\theta_{\ell^{-}})>0) + N(\cos(\theta_{\ell^{+}})\cos(\theta_{\ell^{-}})<0)} \, . 
\end{eqnarray*}
The latter asymmetry is a measure for the correlation coefficient in the helicity basis, 
i.e.\ $A = -4 \cdot \ACOS$~\cite{Bernreuther:2013aga}. The predictions at NLO perturbation 
theory for the case of SM-like correlations (no correlations) are $\ADELPHI = 0.115^{+0.014}_{-0.016}$ 
($\ADELPHI = 0.210^{+0.013}_{-0.008}$) and $\ACOS=-0.078\pm0.006$ ($\ACOS=0$), see Refs.~\cite{Bernreuther:2010ny,Bernreuther:2013aga} and references in Ref.~\cite{Chatrchyan:2013wua}.
The angles $\theta_{\ell^{\pm}}$ require the explicit reconstruction of both the top quark and the antiquark, 
which is done using the AMWT technique.
In addition, the relation between the asymmetry variable and the correlation coefficient is valid 
only if no acceptance cuts and detector effects distort the measurement. The distributions of $\Delta \phi$ and $\cos(\theta_{\ell^{+}})\cos(\theta_{\ell^{-}})$ are thus unfolded using singular value decomposition. Figure~\ref{fig:top:deltaphi}(b) shows the 
unfolded $\cos(\theta_{\ell^{+}})\cos(\theta_{\ell^{-}})$ distribution and the predictions for the two correlation scenarios. The asymmetries are measured to be
\begin{eqnarray*}
        \ADELPHI & = & \phantom{-}0.133 \pm 0.010~(\text{stat}) \pm 0.007~(\text{syst}) \pm 0.012~(\text{top}~\pT) ~\text{and} \\
        \ACOS & = & -0.021 \pm 0.023~(\text{stat}) \pm 0.027~(\text{syst}) \pm 0.010~(\text{top}~\pT) \, , 
\end{eqnarray*}
where the statistical and systematic uncertainties as well as an uncertainty 
associated with the modelling of the \pT spectrum of the top quark are given. 
The largest systematic uncertainties come from the 
unfolding procedure
as well as the jet reconstruction, the background estimate and the modelling of \ttbar events. 

The measurements conducted by ATLAS and CMS show that the spins of the top quark and the antiquark in \ttbar events are 
indeed correlated and that the amount of correlation is as expected from perturbation theory at NLO. 
The production mechanisms of \ttbar events is thus consistent with that predicted by QCD, and no indications for 
additional production mechanisms are found.

The polarisation of top quarks in \ttbar events is studied by CMS using the same data set and 
event selection as for the measurement of the \ttbar spin correlation~\cite{Chatrchyan:2013wua}. The 
polarisation is estimated from the unfolded distribution of the angle $\theta_{\ell^{\pm}}$ and the resulting asymmetry,
\begin{eqnarray*}
        A_{P} = \frac{N(\cos(\theta_{\ell})>0) - N(\cos(\theta_{\ell})<0)}{N(\cos(\theta_{\ell})>0) + N(\cos(\theta_{\ell})<0)} \, . 
\end{eqnarray*}
The polarisation in the helicity basis is then $P=2 A_{P}$. The asymmetry is calculated using positively and negatively 
charged leptons under the assumption of \CP invariance and is measured to be
\begin{eqnarray*}
        A_{P} = 0.005 \pm 0.013~(\text{stat}) \pm 0.020~(\text{syst}) \pm 0.008~(\text{top}~\pT) \, , 
\end{eqnarray*}
where the uncertainties are again the statistical and systematic ones and the uncertainty due to the 
mismodelled 
\pT spectrum of the top quark. The two largest systematic uncertainties are 
uncertainties on the top-quark mass and on the jet reconstruction.

\begin{figure}[htb]
        \begin{center}
                        \includegraphics[width=0.8\textwidth]{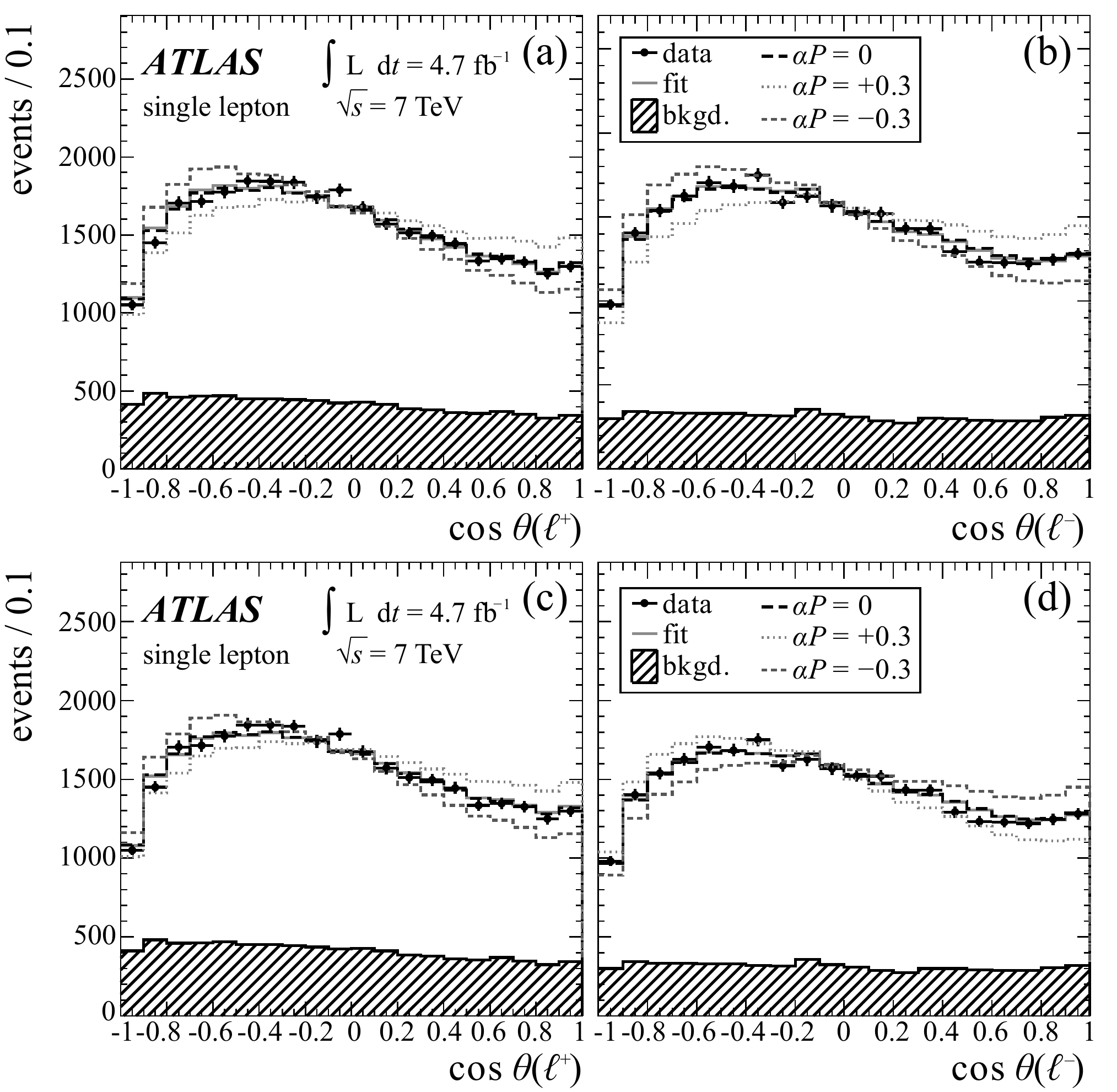} 
        \caption{ATLAS polarisation measurements in the single-lepton channel with three different polarisation assumptions. 
        (a) and (c) show the distributions for positively charged leptons, while (b) and (d) show the distributions for negatively charged leptons. (a),(b) assume \CP invariance, while for (c),(d) maximal \CP violation is assumed. 
        \textit{(Adapted from Ref.~\cite{Aad:2013ksa}.)} 
        \label{fig:top:polarization}} 
       \end{center}
 \end{figure}
 
 An ATLAS measurement of the top-quark polarisation~\cite{Aad:2013ksa} uses \ttbar events decaying 
in the single-lepton and dilepton mode. 
The data were collected at $\sqrt{s}=\unit{7}{\TeV}$ and correspond to an integrated luminosity 
of \unit{$4.66\pm0.08$}{\invfb}. The full \ttbar system is reconstructed using a 
kinematic likelihood~\cite{Erdmann:2013rxa} in 
the single-lepton channel and the neutrino-weighting method~\cite{Abbott:1997fv} in 
the dilepton channel. 
The angle $\theta_{\ell^{\pm}}$ is calculated in the helicity basis for each lepton, 
and the corresponding distributions are fitted with templates of partially polarised top quarks. The spin 
correlation is assumed to be that of the SM. The fits are done separately for each lepton type as well as for positively 
and negatively charged leptons. The latter is done in order to distinguish scenarios in which the polarisation comes from 
\CP-conserving processes and maximally \CP-violating processes. 
For such cases, the expressions $\alpha_{\ell} P$ are either the same for top quark and antiquark, or they differ by a sign. 
Figure~\ref{fig:top:polarization} shows the distributions of $\cos(\theta_{\ell^{\pm}})$ measured in single-lepton events for 
the two scenarios. The combined polarisation measured for the two scenarios is
\begin{eqnarray*}
        \alpha_{\ell} P & = & -0.035 \pm 0.014~(\text{stat}) \pm 0.037~(\text{syst}) ~\text{(\CP invariance)} \, , \\
        \alpha_{\ell} P & = & \phantom{-}0.020 \pm 0.016~(\text{stat}) ^{+0.013}_{-0.017}~(\text{syst}) ~\text{(\CP violation)} \, . 
\end{eqnarray*}
The major systematic uncertainty stems from the jet reconstruction. 

The ATLAS and CMS measurements are both consistent with the SM prediction that top quarks produced in \ttbar pairs via the strong interaction are not polarised.

\index{top quark!spin correlation|)}
\index{top quark!test of \CP violation|)}

\section{Tests of Electroweak Predictions}
\index{electroweak physics|(}

Studies of the electroweak couplings of the top quark comprise measurements of a number of different observables. 
While the polarisation of \Wb bosons from top-quark decays is a consequence of the \VminusA structure 
of the $\Wb\tq\bq$ vertex, the cross sections for single top-quark production
depend directly on the strength of the coupling to \Wb bosons 
(see \sect{\ref{sec:top:singletop}}). 
The coupling strength to \Zb bosons and photons can be probed by measurements of \ttbar production 
with additional such bosons.

\subsection{\Wb-Boson Polarisation}

The massive \Wb bosons produced in top-quark decays are real spin-1 particles and thus 
have three possible polarisation states. We will refer to them as longitudinally, left-handedly or right-handedly polarized \Wb bosons.
The net amount of polarisation is given by the fractions of the partial decay widths for 
differently polarised \Wb bosons, 
 $\Gamma_{0}$, $\Gamma_{L}$ and $\Gamma_{R}$, respectively. 
 These ``helicity fractions''
are
\begin{eqnarray*}
        F_{0/L/R} = \frac{\Gamma_{0/L/R}}{\Gamma_{0} + \Gamma_{L} + \Gamma_{R}} \, . 
\end{eqnarray*}
In perturbation theory at LO, and neglecting the mass of the bottom quark, these fractions depend solely 
on the masses of 
the top quark and the \Wb boson, i.e.\
\begin{eqnarray*}
        F_{0} & = & \frac{\Mt^{2}}{\Mt^{2} + 2 \MW^{2}} \approx 0.70 \, , \nonumber \\
        F_{L} & = & \frac{2 \MW^{2}}{\Mt^{2} + 2 \MW^{2}} \approx 0.30 \, , \nonumber \\
        F_{R} & = & 0 \, . 
\end{eqnarray*}
Calculations at NNLO that include electroweak corrections and assume a finite 
mass of the bottom quark yield $F_{0} = 0.687 \pm 0.005$, $F_{L} = 0.311 \pm 0.005$ and 
$F_{R} = 0.0017 \pm 0.0001$~\cite{Czarnecki:2010gb}. The helicity fractions can be altered if the structure of the 
$\Wb\tq\bq$ vertex differs from a pure \VminusA coupling. Such deviations are typically described by 
anomalous couplings
in effective field theory approaches~\cite{AguilarSaavedra:2006fy,Zhang:2010dr}.\index{anomalous couplings}\index{effective field theories}

Information about the polarisation of the \Wb boson can be obtained from angular distributions of the 
final-state particles. In the single-lepton and dilepton channels of \ttbar production, 
the angle $\theta^{*}$ is defined as the angle between the reverse 
momentum of the leptonically decaying top quark and the direction of the charged 
lepton, both evaluated in the rest frame of the corresponding \Wb boson~\cite{Kane:1991bg}. 
The differential decay width can then be written as
\begin{eqnarray}
        \label{eqn:top:diffdecay} \frac{1}{\Gamma}\frac{\dif \Gamma}{\dif \cos (\theta^{*})} & = & \sin (\theta^{*})^{2} F_{0} + \frac{3}{8} \left( 1 - \cos (\theta^{*}) \right)^{2} F_{L} \nonumber \\
 &  & + \frac{3}{8} \left( 1 + \cos (\theta^{*}) \right)^{2} F_{R} \, . 
\end{eqnarray}
An angle for the hadronically decaying top quark can be defined analogously. 
The ATLAS and CMS collaborations have both made use of the angular dependence described 
by \eqn{\eqref{eqn:top:diffdecay}} to estimate the helicity fractions.

The ATLAS collaboration has analysed a data set corresponding to an integrated luminosity of \unit{1.04}{\invfb} taken at 
a centre-of-mass energy of $\sqrt{s}=\unit{7}{\TeV}$~\cite{Aad:2012ky}. Two sets of selection criteria are defined so 
as to enrich samples with events stemming from \ttbar production with subsequent decay either 
in the single-lepton or dilepton decay 
modes. In both cases, the reconstruction of the top quarks is based on the final-state particles 
and assumptions on the detector performance. 
For both event types, an individual analysis strategy is followed. 
While the first strategy is based on a template fit of the $\cos(\theta^{*})$ distributions, the second makes use of the angular asymmetries derived from unfolded $\cos(\theta^{*})$ spectra. The individual results 
are all found to be in agreement and are combined using the BLUE method~\cite{Valassi:2003mu}. 
Although both analysis methods are based on the same data set, the combined result has a smaller overall uncertainty 
due to the different sensitivities to sources of systematic uncertainty. The largest systematic uncertainties are due to  
the signal and background modelling, to the jet reconstruction and to method-specific uncertainties. 
The combined helicity fractions are
\begin{eqnarray*}
        F_{0} & = & 0.67 \pm 0.03~(\text{stat}) \pm 0.06~(\text{syst}) \, , \nonumber \\
        F_{L} & = & 0.32 \pm 0.02~(\text{stat}) \pm 0.03~(\text{syst}) \, , \nonumber \\
        F_{R} & = & 0.01 \pm 0.01~(\text{sta.}) \pm 0.04~(\text{syst}) \, , \nonumber 
\end{eqnarray*}
with a correlation between $F_{0}$ and $F_{L}$ of $-0.96$.

The CMS collaboration has analysed the full \unit{7}{\TeV} data set corresponding to an integrated 
luminosity of \unit{5.0}{\invfb}~\cite{Chatrchyan:2013jna}. Events are selected that are compatible with the single-lepton 
decay mode, and top quarks are reconstructed from the final-state particles using a constrained fit. An estimate of 
the helicity fractions is obtained using a reweighting procedure and a subsequent fit to the $\cos(\theta^{*})$ distribution. 
The largest sources of systematic uncertainty are the background estimate and the jet reconstruction. 
The helicity fractions are estimated to be
\begin{eqnarray*}
        F_{0} & = & 0.682 \pm 0.030~(\text{stat}) \pm 0.033~(\text{syst}) \, , \nonumber \\
        F_{L} & = & 0.310 \pm 0.022~(\text{stat}) \pm 0.022~(\text{syst}) \, , \nonumber \\
        F_{R} & = & 0.008 \pm 0.012~(\text{stat}) \pm 0.014~(\text{syst}) \, , \nonumber 
\end{eqnarray*}
with a correlation between $F_{0}$ and $F_{L}$ of $-0.95$. A more recent measurement of the helicity fractions using a data sample collected at $\sqrt{s}=\unit{8}{\TeV}$ and enriched with single top-quark events is in good agreement with these values~\cite{Khachatryan:2014vma}.

The results obtained by the ATLAS and CMS collaborations are consistent with one another and with the NNLO predictions. 
Both measurements are more precise than those published by \Tevatron experiments. 

\begin{figure}[htb] 
        \begin{center}
                        \includegraphics[width=0.95\textwidth]{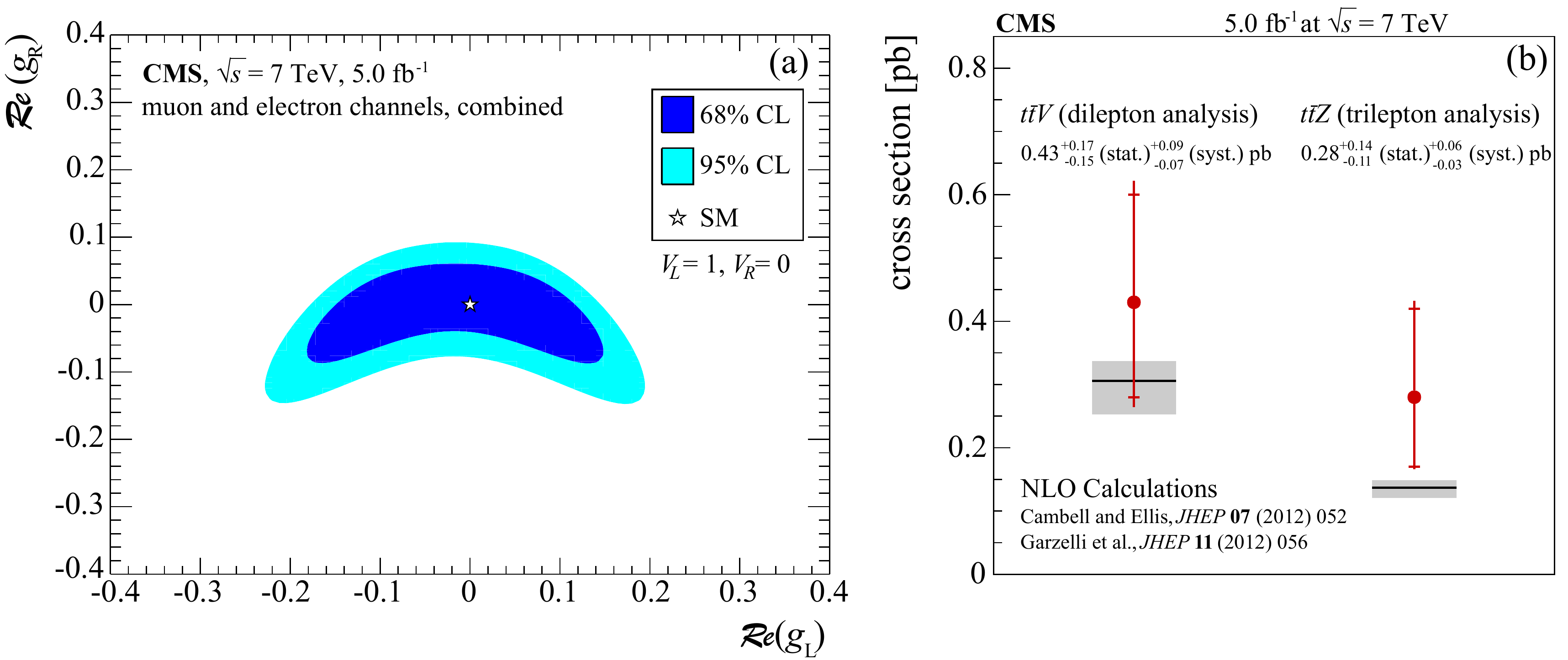}  
        \caption{(a) The 68\% and 95\% CL contours in the plane of the (real) tensor couplings $g_{R}$ and $g_{L}$ assuming 
        vector couplings of $V_{L}=1$ and $V_{R}=0$ as obtained by the CMS collaboration. 
        (b) Measured and predicted values of the cross sections for $\ttbar V$ 
        and $\ttbar \Zb$ production. 
        The inner and outer error bars indicate the statistical and total uncertainties, 
        respectively.
        \textit{(Adapted from Refs.~\cite{Chatrchyan:2013jna,Chatrchyan:2013qca}.)} 
        \label{fig:top:anomcoupl}} 
        \end{center}
\end{figure}

The measured helicity fractions are also interpreted in terms of anomalous couplings. Figure~\ref{fig:top:anomcoupl}(a) 
shows the 68\% and 95\% \CL contours in the plane of the (real) tensor couplings $g_{R}$ and $g_{L}$ assuming 
vector couplings of $V_{L}=1$ and $V_{R}=0$ as obtained by the more precise CMS result. 
A second region of solutions featuring large values of the real part of $g_{R}$
is excluded from the fit as it is not compatible with the measurement of the single-top $t$-channel cross section. 
The results are consistent with the absence of anomalous couplings, i.e.\ $g_{L}=g_{R}=0$, and are in very good 
agreement with the predicted \VminusA structure of the $\Wb\tq\bq$ vertex.


\subsection{Top-Quark Pairs and Additional Gauge Bosons}

While the production of a top-quark pair and an additional photon has only been observed at the 
Tevatron~\cite{Aaltonen:2011sp}, the CMS collaboration was the first to measure the production 
of top-quark pairs with additional \Zb bosons ($\ttbar\Zb$) and \Wb 
bosons ($\ttbar\Wb$)~\cite{Chatrchyan:2013qca, Khachatryan:2014ewa}. Such rare processes are 
expected in the SM. For a centre-of-mass energy of $\sqrt{s}=\unit{7}{\TeV}$, calculations 
at NLO yield predictions for the corresponding cross sections $\sigma_{\ttbar\Zb}$~\cite{Garzelli:2012bn} and $\sigma_{\ttbar\Wb}$~\cite{Campbell:2012dh} of
\begin{eqnarray*}
        \sigma_{\ttbar\Zb} & = & 0.137^{+0.012}_{-0.016} \, \picobarn \, , \\
        \sigma_{\ttbar\Wb} & = & 0.169^{+0.029}_{-0.051} \, \picobarn \, . 
\end{eqnarray*}

Here the analysis at $\sqrt{s}=7$ TeV is briefly described~\cite{Chatrchyan:2013qca}. The data set collected
corresponds to an integrated luminosity of \unit{5.0}{\invfb}. Events are selected according to the 
number of charged leptons in the final state. The process
\begin{eqnarray*}
        \pp \to \ttbar\Zb \to (\tq \to \bq \ell^{\pm}\nu)(\tq \to \bq jj)(\Zb \to \ell^{\pm}\ell^{\mp}) 
\end{eqnarray*}
is searched for by requiring two leptons with the same flavour but opposite electric charge and with 
an invariant mass compatible with the mass of the \Zb boson, and one additional charged lepton. After the 
event selection, nine events are observed while $3.2 \pm 0.8$ background events are expected. 
The resulting cross section is estimated to be
\begin{eqnarray*}
        \sigma_{\ttbar\Zb} = 0.28 ^{+0.14}_{-0.11}~(\text{stat}) ^{+0.06}_{-0.03}~(\text{syst}) \, \picobarn \, , 
\end{eqnarray*}
where the dominant source of systematic uncertainty is the background yield. 

On the other hand, the processes
\begin{eqnarray*}
        \pp & \to & \ttbar\Zb \to (\tq \to \bq \ell^{\pm}\nu)(t \to \bq jj)(\Zb \to \ell^{\pm}\ell^{\mp}) \, ,~\text{and} \\
        \pp & \to & \ttbar\Wb \to (\tq \to \bq \ell^{\pm}\nu)(t \to \bq jj)(\Wb \to \ell^{\pm}\nu) 
\end{eqnarray*}
are searched for by selecting events with exactly two leptons of the same electric charge. 
After the event selection, 16 events remain, while the background expectation is $9.2 \pm 2.6$. The combined 
cross section, $\sigma_{\ttbar V}$, where $V=\Wb,\Zb$, is  measured as
\begin{eqnarray*}
        \sigma_{\ttbar V} = 0.43 ^{+0.17}_{-0.15}~(\text{stat}) ^{+0.09}_{-0.07}~(\text{syst}) \, \picobarn \, . 
\end{eqnarray*}
Both measurements and their NLO predictions are shown in \fig{\ref{fig:top:anomcoupl}}(b).
The measured cross sections are in agreement with the predictions, which indicates that no deviation 
from the strength of the top-quark coupling to \Zb and \Wb bosons predicted by the SM is observed.
The results from the more recent CMS analysis at $\sqrt{s}=\unit{8}{\TeV}$~\cite{Khachatryan:2014ewa} 
are also consistent with this conclusion.

\index{electroweak physics|)}

\section{Single Top-Quark Production} \label{sec:top:singletop}

\begin{figure}[htbp]
  \begin{center}
    \includegraphics[width=0.95\textwidth]{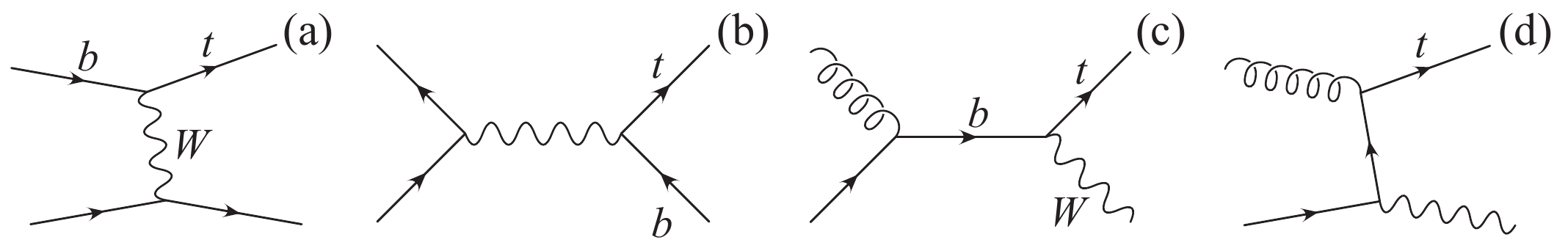}
    \caption{Leading-order Feynman diagrams of processes contributing to single top-quark production: (a) $t$-channel, (b) $s$-channel, (c, d) $tW$-channel.
    \label{fig:top:FeynmanDiagramsSingleTopProduction}}
  \end{center}
\end{figure}

Single top-quark production, in contrast to top-quark pair-production which proceeds through the strong interactions, takes place by virtue of charged-current interactions. The production rate for single top quarks is suppressed with respect to top-quark pair production by a factor of 2--3 due to the different couplings strengths $\alpW$ and $\alpS$ of weak and strong interactions, which is only partially compensated by the larger partonic fluxes due to the lower production threshold. Feynman diagrams of the LO processes contributing to the scattering amplitudes are shown
in \fig{\ref{fig:top:FeynmanDiagramsSingleTopProduction}}.

\begin{figure}[htbp]
  \begin{center}
    \includegraphics[width=0.2\textwidth]{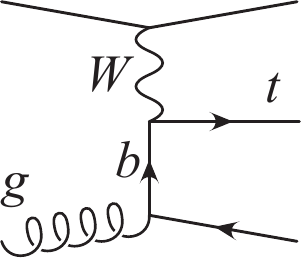}
    \caption{Leading-order Feynman diagram for $t$-channel single top-quark production in the
      four-flavour scheme.
    \label{fig:top:FeynmanDiagramsSingleTopProductionFourFlavor}}
  \end{center}
\end{figure}

Depending on whether the \Wb boson is space-like (\fig{\ref{fig:top:FeynmanDiagramsSingleTopProduction}}(a)), time-like (\fig{\ref{fig:top:FeynmanDiagramsSingleTopProduction}}(b)) or real (\fig{\ref{fig:top:FeynmanDiagramsSingleTopProduction}}(c,d)), one distinguishes between the $t$-channel, the $s$-channel and the $\tq\Wb$-channel. In the latter case, a single top quark is produced in association with a \Wb boson in the final state.  The dominant contribution to single top-quark production at the \Tevatron and the LHC is the $t$-channel. As can be seen from \fig{\ref{fig:top:FeynmanDiagramsSingleTopProduction}}(a), this channel
assumes the existence of a \bq quark inside the proton and thus
requires in the theoretical description 
the so-called five-flavour scheme,
in which $\uq,\dq,\sq,\cq,\bq$ are
treated as active flavours inside the proton. In the four-flavour
scheme,
$t$-channel production occurs formally at higher orders of the
QCD coupling, as is illustrated in
\fig{\ref{fig:top:FeynmanDiagramsSingleTopProductionFourFlavor}}.
In general, calculations in these two schemes should give similar
results. Differences occur if logarithmic corrections of the form
$\ln(\Mb\squared/Q\squared)$ become large, where $Q$ denotes a generic energy
scale. The five-flavour scheme may partially resum these logarithms
through the evolution of the \bq-quark parton distribution function,
while in the four-flavour scheme these logarithms are kept at fixed
order only. In Refs.~\cite{Campbell:2009ss,Campbell:2009gj} it has
been checked through an explicit calculation that the two schemes
lead indeed to consistent results for the cross section.

At the LHC, the second important production channel for single top quarks is the $\tq\Wb$ channel.  Due to phase-space suppression and the small gluon luminosity, this channel gives only a tiny contribution at the \Tevatron. In contrast, $s$-channel production, which is roughly responsible for one third of the cross section at the \Tevatron, leads only to a contribution of a few percent at the
LHC. Single top-quark production at the LHC is thus to some extent complementary to that at the \Tevatron. A further major difference arises from the fact that the initial state at the \Tevatron is a \CP eigenstate. Since \CP-violating effects are negligible in single top-quark production within the SM, the numbers of produced single top quarks and antiquarks are identical at the \Tevatron.  At the LHC, however, the initial state is not a \CP eigenstate, and more top quarks than antiquarks are produced because in \pp collisions the flux of up-type quarks is larger than the flux for down-type quarks.

The NLO QCD corrections for inclusive single top-quark production have been presented in Refs.~\cite{Bordes:1994ki,Stelzer:1997ns,Stelzer:1998ni} for the $t$-channel, in Ref.~\cite{Smith:1996ij} for the $s$-channel and in Refs.~\cite{Giele:1995kr,Zhu:2002uj} for the $\tq\Wb$ channel.
\def\LO{{\mbox{\scriptsize LO}}}
\def\NLO{{\mbox{\scriptsize NLO}}}
\begin{table}[htbp]\renewcommand{\arraystretch}{1.4}
  \begin{center}
    \begin{tabular}[htbp]{l|c|c|c|c|c|c}
      &\multicolumn{4}{c|}{LHC \unit{8}{\TeV}} &\multicolumn{2}{c}{\Tevatron} \\ 
      & $\sigma_\tq^\LO$ & $\sigma_{\tbarq}^\LO$ & 
      $\sigma_\tq^\NLO$ & $\sigma_{\tbarq}^\NLO$ & $\sigma_{\tq,\tbarq}^\LO$ &
      $\sigma_{\tq,\tbarq}^\NLO$ \\ 
      \hline 
      \hline 
      $t$ &53.8 &29.1 &55.2\,$^{+1.6}_{-0.9}$ $^{+0.35}_{-0.32}$
 &
         30.1\,$^{+0.9}_{-0.5}$ $^{+0.29}_{-0.32}$
        &$1.03$
        & 0.998\,$^{+0.025}_{-0.022}$ $^{+0.029}_{-0.032}$\\ \hline
      $s$ &2.22 &1.24 &
      3.30\,$^{-0.06}_{+0.08}$ $^{+0.07}_{-0.05}$
        &1.90\,$^{-0.04}_{+0.05}$ $^{+0.04}_{-0.03}$ &0.28 &0.442\,$^{-0.023}_{+
0.025}$ $^{+0.015}_{-0.011}$\\ \hline
      $\tq\Wb$ &8.86 &8.85 &9.12\,$^{+0.21}_{-0.38}$ $^{+0.29}_{-0.36}$ 
      &9.11\,$^{+0.21}_{-0.38}$ $^{+0.29}_{-0.36}$&0.069 & 0.070\,$^{-0.002}_{-0
.001}$ 
      $^{+0.008}_{-0.009}$
    \end{tabular}
    \caption{Cross sections for single top-quark production at LO and
      NLO QCD using $\Mt = \unit{173.3}{\GeV}$ and the MSTW2008lo/nlo PDF
      set. See text for references. \newline
      The notation is  $\sigma(\mu=\Mt)\,\,^{\sigma(\mu=2\Mt) -
  \sigma(\mu=\Mt)}_{\sigma(\mu={\Mt\over2}) - \sigma(\mu=\Mt)}
\mbox{ $^{+\mbox{ \scriptsize PDF err. up}}_{-\mbox{ \scriptsize PDF err. down}}$}$.
    \label{tab:top:SingleTopCrossSection}}
  \end{center}
\end{table}

These cross sections are shown in
\tab{\ref{tab:top:SingleTopCrossSection}}, assuming $\Mt =
\unit{173.3}{\GeV}$ and using the MSTW2008lo/nlo68cl pdf
set~\cite{Martin:2009iq}. It turns out that the NLO corrections 
are only a few percent for the
$t$-channel and slightly larger for the $\tq\Wb$ channel. In contrast, the
NLO corrections to the $s$-channel contribution are about
30\%. However, as has been pointed out before, this channel gives only
a small contribution at the LHC. Since at NLO, no colour transfer
between the two quark lines is allowed (the corresponding box
contributions vanish after interference with the Born amplitude), it
is conceivable that the small corrections are accidental and that the
small scale uncertainty observed at NLO underestimates the possible
size of the NNLO corrections. Very recently, the vertex corrections
for the $t$-channel have been calculated at NNLO
QCD~\cite{Brucherseifer:2014ama}. The corrections are found to be
comparable in size to the NLO corrections.

Fully differential results at NLO accuracy for single top-quark
production have been presented in
Refs.~\cite{Harris:2002md,Sullivan:2004ie,Sullivan:2005ar}.  In
Refs.~\cite{Campbell:2004ch,Cao:2004ky,Cao:2005pq,Campbell:2005bb},
the analysis has been extended by including the semileptonic decay of
the top quark in the narrow-width approximation. In addition, the
systematic combination of the NLO corrections with the parton shower
has been investigated in \MCATNLO~\cite{Frixione:2005vw,Frixione:2008yi}
as well as in the \POWHEG framework~\cite{Alioli:2009je,Re:2010bp}.
Beyond fixed-order perturbation theory, the impact of logarithmic
corrections due to soft-gluon emission has been
studied~\cite{Mrenna:1997wp,Kidonakis:2006bu,Kidonakis:2007ej,Kidonakis:2010ux,%
Kidonakis:2010dk,Kidonakis:2011wy}.

As will be seen below, the experimental study of single top-quark
production is challenging---despite the sizeable cross 
sections---because of the complicated event signature and the large
backgrounds. 
In contrast to \ttbar production,
singly-produced top quarks are highly polarised---a consequence of
their \VminusA coupling to the \Wb boson. The measurement of the top-quark
polarisation thus provides a further test of the \VminusA  structure of the $\tq\Wb\bq$ vertex.
Assuming that the production mechanism is well understood,
$t$-channel single top-quark production can also be used to constrain
the \bq parton distribution in the
proton. Furthermore, it can be used for a direct measurement of the
CKM matrix element \Vtb which is otherwise only indirectly
accessible, via the assumptions of CKM matrix unitarity and of the existence of only three quark families.

\subsection{$t$-Channel Production}

Measurements of the $t$-channel production cross section
have been performed using events with exactly one isolated lepton (electron or
muon) originating from the decay of the \Wb boson and two or three jets
in the final state. One of the jets has to be identified as a
\bq jet. Additional requirements on kinematic observables, such as the missing
transverse momentum or the transverse mass of the \Wb boson, are imposed
in order to further remove background. Detailed and precise results are
available from ATLAS and CMS~\cite{Aad:2012ux,Chatrchyan:2011vp,Chatrchyan:2012ep,Khachatryan:2014iya,Aad:2014fwa}.

\begin{figure}[h] 
        \begin{center}
                \includegraphics[width=0.95\textwidth]{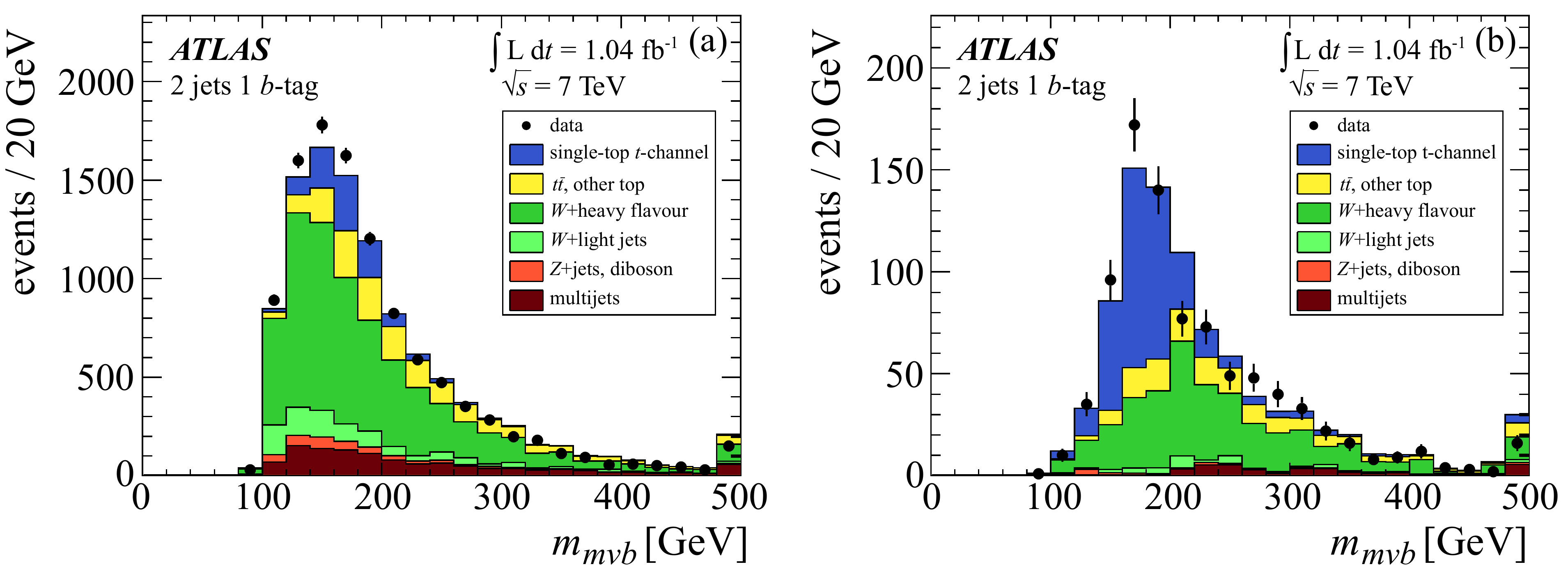} 
        \caption{Distribution of the invariant mass of the \bq-tagged jet, the charged lepton 
        and the neutrino in the \bq-tagged sample for 
        two-jet events (a) for the inclusive sample and (b) after cuts. The last histogram 
        bin includes overflow. 
        \textit{(Adapted from Ref.~\cite{Aad:2012ux}.)}
        \label{fig:top:t-chan:atlas} }
        \end{center}
\end{figure}

An early analysis of $t$-channel single-top production was performed by the ATLAS collaboration based on 
an integrated luminosity of about \unit{1}{\invfb}~\cite{Aad:2012ux}. 
Events with two or three jets are selected. To separate $t$-channel single top-quark signal events from backgrounds, 
several kinematic variables are combined into one discriminant by employing a neural network that also 
exploits correlations between the variables. The most discriminating variable for the two-jet sample is the invariant mass of 
the system formed by the \bq-tagged jet, the charged lepton and the 
neutrino, $m_{\ell\nu\bq}$, see \fig{\ref{fig:top:t-chan:atlas}}(a).
In the three-jet category, the invariant mass of the two leading jets and the absolute value of the 
 difference in pseudo-rapidity of the leading and the lowest-\pT jet are among 
 the most discriminating variables. Multi-jet event yields are determined with data-driven techniques, while contributions from $\Wb+$jets events are derived from simulation and normalised to data in control regions
using a cut-based analysis.
 All other backgrounds and the $t$-channel signal expectation are normalised to theoretical cross sections.
To extract the signal content of the selected sample, a maximum-likelihood fit is performed to the 
output distributions of 
the neural net in the two-jet and three-jet data sets. From a simultaneous 
measurement in the two channels, a cross section of 
\begin{eqnarray*}
\sigma_{t+\bar{t}} = 83\pm 4~(\text{stat}) ^{+20}_{-19}~(\text{syst})\,\picobarn
\end{eqnarray*}
is measured. This result is confirmed in a cut-based analysis which is illustrated by the distribution in \fig{\ref{fig:top:t-chan:atlas}}(b).

\begin{figure}[h] 
        \begin{center}
                \includegraphics[width=1.0\textwidth]{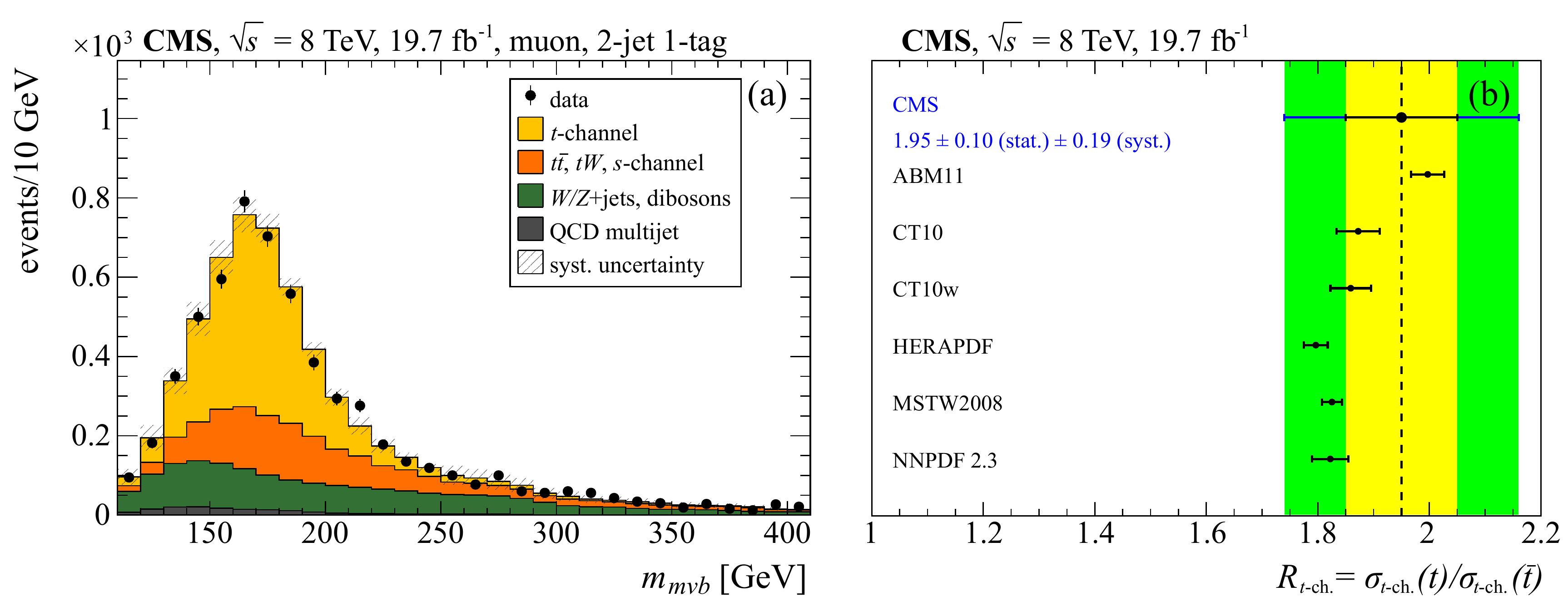} 
        \caption{(a) Distribution of the invariant mass $m_{\ell\nu\bq}$ for the muon decay channel in the region 
        with $|\eta_j′| > 2.5$ as measured by CMS. 
        (b) Comparison of the measured $t$-channel top quark/antiquark production ratio with the predictions obtained using different 
        PDF sets. 
        \textit{(Adapted from Refs.~\cite{Khachatryan:2014iya,Khachatryan:2014iya}.)}
        \label{fig:top:t-chan:cms} }
        \end{center}
\end{figure}

More recently, the CMS collaboration performed an analysis based on the full statistics available 
for $\sqrt{s}=\unit{8}{\TeV}$~\cite{Khachatryan:2014iya}. The event sample is selected by the application of simple criteria:
The events must contain exactly one muon or electron with large transverse momentum.
They are categorised 
according to the numbers of jets and \bq-tagged jets, and the category enriched with $t$-channel signal is the 
one with two jets and one tag.
One of the jets, denoted by $j'$,
is expected to not originate from \bq quarks, and its pseudo-rapidity distribution is typical of the $t$-channel 
processes where a light parton recoils against a much more massive particle like the top quark. 
Signal events populate forward regions in the $|\eta_{j'}|$ spectrum, and this feature is used to distinguish
the signal from background. Background events from \ttbar and from $\Wb+$jets processes are determined by the use of 
control categories.
In all categories the invariant mass $m_{\ell\nu\bq}$ 
is used to define a signal region and a side-band region that contain events inside 
and outside the reconstructed top-quark mass window of $130 < m_{\ell\nu\bq} < \unit{220}{\GeV}$, respectively. 
To determine the contribution from signal events, a binned maximum-likelihood fit is performed to the $|\eta_{j'}|$ 
distribution of the events in the signal region of the category with two jets and one tag.
In \fig{\ref{fig:top:t-chan:cms}} (a) the $m_{\ell\nu\bq}$ distribution is shown for events 
with forward jets.
The figure illustrates that large-purity samples of $t$-channel single-top quark events can be isolated at the LHC using simple selection criteria. 
The measured cross section for this process is 
\begin{eqnarray*}
\sigma_{t+\bar{t}} = 83.6 \pm 2.3~(\text{stat}) \pm 7.4~(\text{syst})\,\picobarn \, .
\end{eqnarray*}
The largest contributions to the systematic uncertainty come from the choice of 
the renormalisation and factorisation scales in the simulation of the signal samples and from uncertainties on the jet energy scale and resolutions.
From another fit, the cross sections for \tq quarks and \tbarq quarks and the corresponding ratio, $R_{\tq} = \sigma_{\tq}/\sigma_{\tbarq}$, are obtained:
\begin{eqnarray*}
\sigma_{\tq} & = & 53.8 \pm 4.4~(\text{stat}) \pm 8 (\text{syst})\,\picobarn \, , \\
\sigma_{\tbarq} & = & 27.6 \pm 1.3~(\text{stat.}) \pm 3.7~(\text{syst})\,\picobarn \, , \\
R_{\tq} & = & 1.95 \pm 0.10~(\text{stat}) \pm 0.19~(\text{syst}) \, .
\end{eqnarray*}
%
In \fig{\ref{fig:top:t-chan:cms}}(b) the measured ratio is compared with predictions using different PDF sets. 

The largeness of the LHC data samples and of the $t$-channel single top cross section give access 
to detailed studies of differential distributions~\cite{Aad:2014fwa} and properties, such as the top-quark 
polarisation, \Wb helicity distributions, and mass measurements. At the time of preparation of this volume, 
the publication of the latter measurements is still in progress.


\subsection{Single Top-Quark Production in Association with a \Wb Boson}

\begin{figure}[h] 
        \begin{center}
                \includegraphics[width=0.9\textwidth]{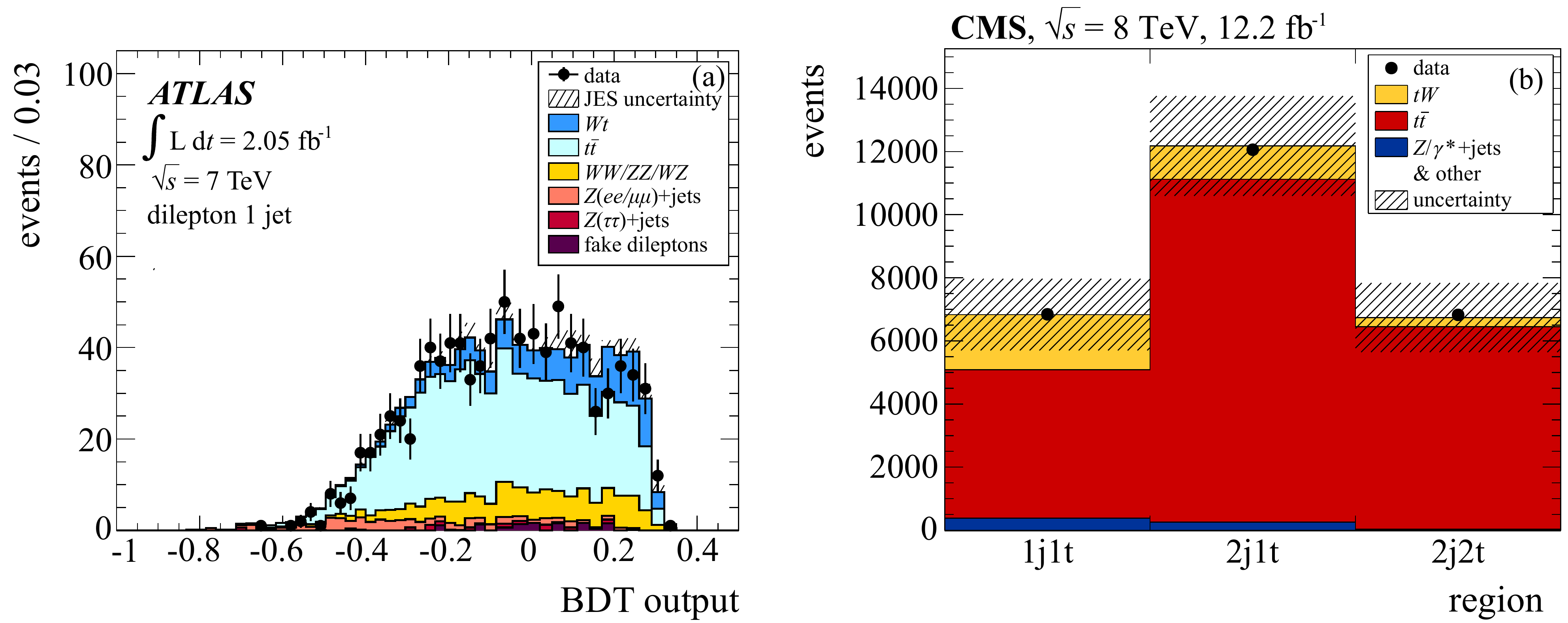} 
        \caption{(a) BDT output for selected events in the 1-jet category 
        of the ATLAS analysis.
        The $\tq\Wb$ signal is normalised to the theory prediction. 
        (b) Event yields in the signal region with one jet and one \bq-tagged jet (1j1t) and the 
        control regions (2j1t and 2j2t) for the CMS analysis. The hatched band represents uncertainties 
        of the predicted event yields.
        \textit{(Adapted from Refs.~\cite{Aad:2012xca,Chatrchyan:2014tua}.)} 
        \label{fig:top:tW-chan} 
        }
        \end{center}
\end{figure}

At the LHC, the production of single top quarks in association with \Wb bosons becomes 
experimentally accessible for the first time. First evidence was reported by ATLAS using about 
the first half of the \unit{7}{\TeV} data recorded in 
2011~\cite{Aad:2012xca} and was confirmed by CMS~\cite{Chatrchyan:2012zca}.
The ATLAS analysis makes use of dileptonic final states with events featuring 
two isolated leptons (electron or muon) with significant transverse missing momentum and at least one 
jet. A boosted decision tree (BDT) is used to discriminate single top-quark $\tq\Wb$ events 
from background events, which mostly arise from top-quark pair production.\index{boosted decision tree}
The result is extracted from a template fit to the BDT output discriminant distribution, 
which is shown in \fig{\ref{fig:top:tW-chan}}(a). 
It is incompatible with the background-only hypothesis at the $3.3\,\sigma$ level. The 
expected sensitivity assuming the Standard Model production rate is $3.4\,\sigma$. The measured cross section is
\begin{eqnarray*}
\sigma_{\tq\Wb} = 16.8 \pm 2.9~(\text{stat}) \pm 4.9~(\text{syst}) \, \picobarn \, .
\end{eqnarray*}
The uncertainty of the jet energy scale and of the modelling of the production process are dominant sources of systematic uncertainty.

Most recently, the CMS collaboration reported an observation of the process based 
on \unit{8}{\TeV} data~\cite{Chatrchyan:2014tua}. Similarly to previous analyses, a multivariate 
analysis technique makes use of kinematic and topological properties to separate the 
signal from the dominant \ttbar background. 
An excess consistent with the signal hypothesis is observed, with an observed (expected) 
significance of 6.1 (5.4) standard deviations above a background-only hypothesis. 
In \fig{\ref{fig:top:tW-chan}}b the distribution of $\tq\Wb$ events over different event categories is shown.
The measured production cross section is \unit{23.4$\pm$5.4}{\pico\barn}, in agreement with the Standard Model prediction. 

\subsection{Determination of \Vtb}

Single top-quark production measurements also give access to the
determination of the modulus of the CKM matrix element $\Vtb$ since in
the SM $t$-channel and $s$-channel production depend on
$|\Vtb|\squared$. 
The extraction is independent of assumptions about the number of quark generations and about the 
unitarity of the CKM matrix. The only assumptions required are that $|\Vtb| \gg |\Vts|, |\Vtd|$ and 
that the $\Wb\tq\bq$ interaction is a SM-like left-handed weak coupling. 
Using the single-top cross section measurements, $|\Vtb|\squared$ can
be extracted from a comparison
of the measured and the predicted cross section. 
Assuming unitarity of the CKM matrix, 
$|\Vtb| \le 1$, a limit can be set. The most precise results are summarised in \tab{\ref{tab:top:vtb}}.
\begin{table}
\begin{center}
    \begin{tabular}{l|c|c} 
        Measurement & $|\Vtb|\squared$ & $|\Vtb|$ limit at 95\% \CL\\
    \hline 
    \hline 
    $t$-ch., ATLAS~\cite{Aad:2014fwa} 
    & $1.02\pm0.07$
        & $> 0.88\phantom{0}$ \\
    \hline 
    $t$-ch., CMS~\cite{Khachatryan:2014iya} 
    & $0.998 \pm 0.038 (\text{exp}) \pm 0.016 (\text{theo}) $
        & $> 0.92$\phantom{0} \\
    \hline 
    $\tq\Wb$-ch., ATLAS~\cite{Aad:2012xca} 
    & $1.03^{+0.16}_{-0.19}$ 
    & n/a \\
    \hline 
    $\tq\Wb$-ch., CMS~\cite{Chatrchyan:2014tua} 
    & $ 1.03 \pm 0.12 (\text{exp}) \pm 0.04 (\text{theo})$ 
    & $ >0.78\phantom{0}$ \\
    \hline 
    $R_{\Bmeson}$, CMS~\cite{Khachatryan:2014nda} 
    & $ 1.014 \pm 0.003 (\text{stat}) \pm 0.032 (\text{syst})$ 
    & $ >0.975$ \\
\end{tabular}
\caption{Most precise values and limits for $|\Vtb|\squared$ as extracted from single 
top-quark cross-section measurements and from $R_{\Bmeson}$.
    \label{tab:top:vtb}}
\end{center}
\end{table}

\begin{figure}[h] 
        \begin{center}
                \includegraphics[width=0.4\textwidth]{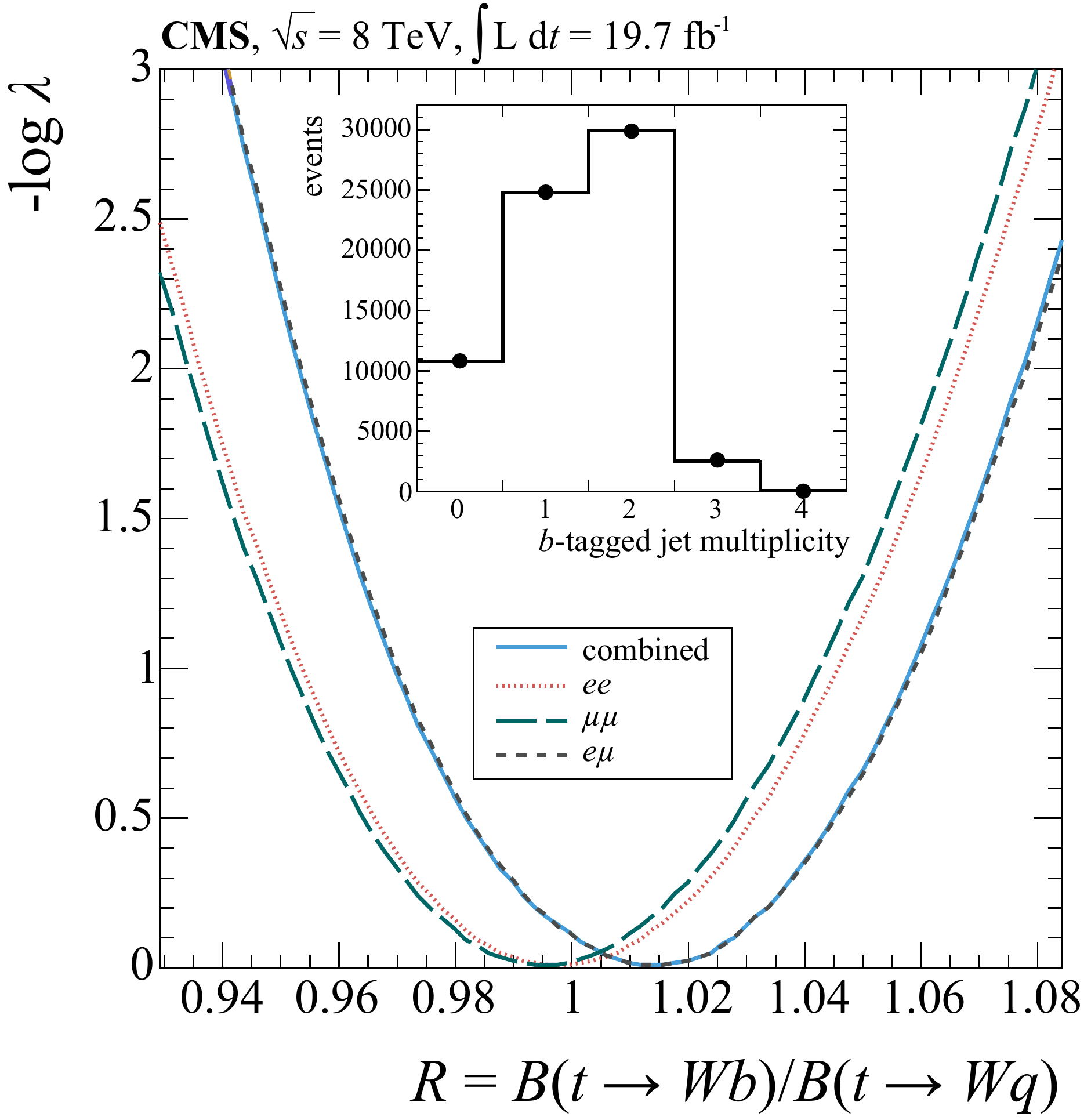} 
        \caption{Variations of the likelihood ratio used to extract $R_{\Bmeson}$ from the data. The variations 
        observed in the combined fit and in the exclusive $ee$, $\mu\mu$, and $e\mu$ channels 
        are shown separately. The inset shows the inclusive \bq-tagged jet multiplicity distribution and the fit distribution.
        \textit{(Adapted from Ref.~\cite{Khachatryan:2014nda}.)} 
        \label{fig:top:rb} 
        }
        \end{center}
\end{figure}

A significantly more precise determination of $\Vtb$ can be obtained from the measurement of the 
ratio $R_{\Bmeson} = B(\tq \to \Wb\bq)/B(\tq \to \Wb\qq)$. Measurements of $R_{\Bmeson}$ have also 
been performed at the \Tevatron~\cite{Abazov:2011zk,Aaltonen:2013doa,Aaltonen:2014yua}. 
CMS presented a measurement of $R_{\Bmeson}$ using a binned-likelihood function of the observed \bq-tagging multiplicity distributions in events with two, three, or four observed jets in the different dilepton channels~\cite{Khachatryan:2014nda}. In \fig{\ref{fig:top:rb}} the variation of the profile likelihood ratio is shown. 
The fit yields a value $R_{\Bmeson} = 1.014 \pm 0.003~(\text{stat}) \pm 0.032~(\text{syst})$.
Assuming the CKM matrix to be unitary, a lower limit for $\Vtb$ of 0.975 is set at the 95\% confidence level.

\index{CKM matrix}

\section{Conclusions} 

LHC data have brought a large boost to the field of top-quark physics, and since the 
LHC start in 2009, tremendous progress has been made. 

On the theoretical side, a major breakthrough has been made with the
calculation of the NNLO QCD corrections for inclusive production of top-quark pairs. 
Very recently, the theoretical calculations have been
extended to include also predictions for differential distributions.
The NNLO QCD corrections have been further improved by including weak
corrections, the resummation of logarithmically enhanced contributions
due to soft-gluon emission, and bound-state effects. In addition, 
finite-width effects have been studied in ground-breaking calculations by
studying the one-loop QCD corrections including the decay of the top
quark. For the experimental analysis, NLO predictions matched and
merged to take into account parton-shower corrections and predictions
for larger jet multiplicities are available. For single-top-quark
production, NLO QCD corrections are known including also the decay of
the top quark. For $t$-channel production, the theoretical predictions
have been extended recently to include also partial results at NNLO
QCD. In addition, conceptual differences using four or five flavours in
the initial state have been studied in detail.  For many observables
the theoretical uncertainties are at the level of ten percent, and in some
cases even the level of a few percent is reached. Beyond the
steadily improving quality of theoretical predictions, a variety of new
observables that allow precise tests of the underlying theory have been
proposed. New methods to measure the top-quark mass are currently
under development.

On the experimental side, during LHC Run~1, a rich and diverse program of top-quark measurements and studies has been performed, yielding a comprehensive spectrum of detailed and precise results.

Many of the measurements previously carried out at the \Tevatron have been repeated at the LHC, and in many areas the size of the LHC top-quark data samples has facilitated in-depth studies at unprecedented levels of precision. Detailed and precise measurements of differential top-quark cross sections have been performed at the LHC, providing new insights in the top-quark production process and in QCD. Precision measurements of top-quark properties have as yet confirmed the SM expectations. Some of these properties, such as the polarisation of top quarks, the correlation of the top-quark spins in pair production, or the production of top quarks with additional gauge bosons were measured at the LHC for the first time. Stringent limits were set on anomalous couplings, e.g.\ flavour-changing neutral currents in the top-quark sector.

To date, most LHC top-quark measurements are no longer limited by statistical, but by systematic uncertainties.
The latter are related to both the detector and the modelling of top-quark production and decay. Prime examples for the former are the jet energy scale and \bq-tagging uncertainties, while examples for the latter include the modelling of initial-state and final-state radiation as well as scale and hadronisation uncertainties. Substantial effort is being invested to reduce these uncertainties, through auxiliary measurements, through tuning of Monte Carlo generators and through further improvements of the theoretical calculations. The wealth of the LHC Run~1 results is the avant-garde of many more exciting improvements in the future, both in precision measurements as well as in searches for rare processes associated with top quarks.

\index{top quark|)}

\bibliography{lhc_top_arxiv}{}

\providecommand{\href}[2]{#2}\begingroup\raggedright\begin{thebibliography}{100}%
\makeatletter
\providecommand{\hrefCMSnoop }[0]{\@secondoftwo}%
\makeatother
\providecommand{\doi}{\texttt{doi:}\begingroup \urlstyle{tt}\Url}

\bibitem{thisbook}
T.~Sch{\"o}rner-Sadenius, ed., ``{The Large Hadron Collider --- Harvest of
  Run~1}''.
\newblock {Springer}, {Heidelberg, Germany},
2015.
\newblock

\bibitem{Kobayashi:1973fv}
\hrefCMSnoop {}{M.~Kobayashi and T.~Maskawa, ``{CP Violation in the
  Renormalizable Theory of Weak Interaction}'',} \textit{ Prog. Theor. Phys.}
  \textbf{ 49} (1973)
652.

\bibitem{Abe:1995hr}
\hrefCMSnoop {}{{CDF Collaboration}, ``{Observation of top quark production in
  $\bar{p}p$ collisions}'',} \textit{ Phys. Rev. Lett.} \textbf{ 74} (1995)
  2626,
\href{http://www.arXiv.org/abs/hep-ex/9503002}{\texttt{arXiv:hep-ex/9503002}}.

\bibitem{Abachi:1995iq}
\hrefCMSnoop {}{{\Dzero Collaboration}, ``{Observation of the top quark}'',}
  \textit{ Phys. Rev. Lett.} \textbf{ 74} (1995) 2632,
\href{http://www.arXiv.org/abs/hep-ex/9503003}{\texttt{arXiv:hep-ex/9503003}}.

\bibitem{Aaltonen:2013wca}
\hrefCMSnoop {}{{CDF and \Dzero Collaborations}, ``{Combination of measurements
  of the top-quark pair production cross section from the Tevatron
  Collider}'',} \textit{ Phys. Rev. D} \textbf{ 89} (2014) 072001,
  \href{http://dx.doi.org/10.1103/PhysRevD.89.072001}{\doi{10.1103/PhysRevD.89.072001}},
\href{http://www.arXiv.org/abs/1309.7570}{\texttt{arXiv:1309.7570}}.

\bibitem{Tevatron:2014cka}
\hrefCMSnoop {}{{Tevatron Electroweak Working Group}, ``{Combination of CDF and
  \Dzero results on the mass of the quark using up to 9.7 fb$^{-1}$ at the
  Tevatron}'',}
\href{http://www.arXiv.org/abs/1407.2682}{\texttt{arXiv:1407.2682}}.

\bibitem{Nason:1987xz}
\hrefCMSnoop {}{P.~Nason, S.~Dawson, and R.~K. Ellis, ``{The Total
  Cross-Section for the Production of Heavy Quarks in Hadronic Collisions}'',}
  \textit{ Nucl. Phys. B} \textbf{ 303} (1988)
607.

\bibitem{Beenakker:1988bq}
\hrefCMSnoop {}{W.~Beenakker {et~al.}, ``{QCD Corrections to Heavy Quark
  Production in p anti-p Collisions}'',} \textit{ Phys. Rev. D} \textbf{ 40}
  (1989)
54.

\bibitem{Catani:1996dj}
\hrefCMSnoop {}{S.~Catani {et~al.}, ``{The Top cross-section in hadronic
  collisions}'',} \textit{ Phys. Lett. B} \textbf{ 378} (1996) 329,
\href{http://www.arXiv.org/abs/hep-ph/9602208}{\texttt{arXiv:hep-ph/9602208}}.

\bibitem{Bonciani:1998vc}
\hrefCMSnoop {}{R.~Bonciani {et~al.}, ``{NLL resummation of the heavy quark
  hadroproduction cross-section}'',} \textit{ Nucl. Phys. B} \textbf{ 529}
  (1998) 424,
\href{http://www.arXiv.org/abs/hep-ph/9801375}{\texttt{arXiv:hep-ph/9801375}}.

\bibitem{Kidonakis:2001nj}
\hrefCMSnoop {}{N.~Kidonakis {et~al.}, ``{Sudakov resummation and finite order
  expansions of heavy quark hadroproduction cross-sections}'',} \textit{ Phys.
  Rev. D} \textbf{ 64} (2001) 114001,
\href{http://www.arXiv.org/abs/hep-ph/0105041}{\texttt{arXiv:hep-ph/0105041}}.

\bibitem{Kidonakis:2003qe}
\hrefCMSnoop {}{N.~Kidonakis and R.~Vogt, ``{Next-to-next-to-leading order soft
  gluon corrections in top quark hadroproduction}'',} \textit{ Phys. Rev. D}
  \textbf{ 68} (2003) 114014,
\href{http://www.arXiv.org/abs/hep-ph/0308222}{\texttt{arXiv:hep-ph/0308222}}.

\bibitem{Moch:2008qy}
\hrefCMSnoop {}{S.~Moch and P.~Uwer, ``{Theoretical status and prospects for
  top-quark pair production at hadron colliders}'',} \textit{ Phys. Rev. D}
  \textbf{ 78} (2008) 034003,
\href{http://www.arXiv.org/abs/0804.1476}{\texttt{arXiv:0804.1476}}.

\bibitem{Czakon:2008cx}
\hrefCMSnoop {}{M.~Czakon and A.~Mitov, ``{On the Soft-Gluon Resummation in Top
  Quark Pair Production at Hadron Colliders}'',} \textit{ Phys. Lett. B}
  \textbf{ 680} (2009) 154,
\href{http://www.arXiv.org/abs/0812.0353}{\texttt{arXiv:0812.0353}}.

\bibitem{Cacciari:2011hy}
\hrefCMSnoop {}{M.~Cacciari {et~al.}, ``{Top-pair production at hadron
  colliders with next-to-next-to-leading logarithmic soft-gluon
  resummation}'',} \textit{ Phys. Lett. B} \textbf{ 710} (2012) 612,
\href{http://www.arXiv.org/abs/1111.5869}{\texttt{arXiv:1111.5869}}.

\bibitem{Beneke:2011mq}
\hrefCMSnoop {}{M.~Beneke {et~al.}, ``{Hadronic top-quark pair production with
  NNLL threshold resummation}'',} \textit{ Nucl. Phys. B} \textbf{ 855} (2012)
  695,
\href{http://www.arXiv.org/abs/1109.1536}{\texttt{arXiv:1109.1536}}.

\bibitem{Hagiwara:2008df}
\hrefCMSnoop {}{K.~Hagiwara, Y.~Sumino, and H.~Yokoya, ``{Bound-state Effects
  on Top Quark Production at Hadron Colliders}'',} \textit{ Phys. Lett. B}
  \textbf{ 666} (2008) 71,
\href{http://www.arXiv.org/abs/0804.1014}{\texttt{arXiv:0804.1014}}.

\bibitem{Kiyo:2008bv}
\hrefCMSnoop {}{Y.~Kiyo {et~al.}, ``{Top-quark pair production near threshold
  at LHC}'',} \textit{ Eur. Phys. J. C} \textbf{ 60} (2009) 375,
\href{http://www.arXiv.org/abs/0812.0919}{\texttt{arXiv:0812.0919}}.

\bibitem{Beenakker:1993yr}
\hrefCMSnoop {}{W.~Beenakker {et~al.}, ``{Electroweak one loop contributions to
  top pair production in hadron colliders}'',} \textit{ Nucl. Phys. B} \textbf{
  411} (1994)
343.

\bibitem{Kuhn:2005it}
\hrefCMSnoop {}{J.~H. Kuhn, A.~Scharf, and P.~Uwer, ``{Electroweak corrections
  to top-quark pair production in quark-antiquark annihilation}'',} \textit{
  Eur. Phys. J. C} \textbf{ 45} (2006) 139,
\href{http://www.arXiv.org/abs/hep-ph/0508092}{\texttt{arXiv:hep-ph/0508092}}.

\bibitem{Bernreuther:2005is}
\hrefCMSnoop {}{W.~Bernreuther, M.~Fuecker, and Z.~Si, ``{Mixed QCD and weak
  corrections to top quark pair production at hadron colliders}'',} \textit{
  Phys. Lett. B} \textbf{ 633} (2006) 54,
\href{http://www.arXiv.org/abs/hep-ph/0508091}{\texttt{arXiv:hep-ph/0508091}}.

\bibitem{Kuhn:2006vh}
\hrefCMSnoop {}{J.~H. Kuhn, A.~Scharf, and P.~Uwer, ``{Electroweak effects in
  top-quark pair production at hadron colliders}'',} \textit{ Eur. Phys. J. C}
  \textbf{ 51} (2007) 37,
\href{http://www.arXiv.org/abs/hep-ph/0610335}{\texttt{arXiv:hep-ph/0610335}}.

\bibitem{Bernreuther:2006vg}
\hrefCMSnoop {}{W.~Bernreuther, M.~Fuecker, and Z.-G. Si, ``{Weak interaction
  corrections to hadronic top quark pair production}'',} \textit{ Phys. Rev. D}
  \textbf{ 74} (2006) 113005,
\href{http://www.arXiv.org/abs/hep-ph/0610334}{\texttt{arXiv:hep-ph/0610334}}.

\bibitem{Hollik:2007sw}
\hrefCMSnoop {}{W.~Hollik and M.~Kollar, ``{NLO QED contributions to top-pair
  production at hadron collider}'',} \textit{ Phys. Rev. D} \textbf{ 77} (2008)
  014008,
\href{http://www.arXiv.org/abs/0708.1697}{\texttt{arXiv:0708.1697}}.

\bibitem{Bernreuther:2008md}
\hrefCMSnoop {}{W.~Bernreuther, M.~Fuecker, and Z.-G. Si, ``{Weak interaction
  corrections to hadronic top quark pair production: Contributions from
  quark-gluon and b anti-b induced reactions}'',} \textit{ Phys. Rev. D}
  \textbf{ 78} (2008) 017503,
\href{http://www.arXiv.org/abs/0804.1237}{\texttt{arXiv:0804.1237}}.

\bibitem{Czakon:2013goa}
\hrefCMSnoop {}{M.~Czakon, P.~Fiedler, and A.~Mitov, ``{Total Top-Quark
  Pair-Production Cross Section at Hadron Colliders Through
  $O(\alpha\frac{4}{S})$}'',} \textit{ Phys. Rev. Lett.} \textbf{ 110} (2013),
  no.~25, 252004,
\href{http://www.arXiv.org/abs/1303.6254}{\texttt{arXiv:1303.6254}}.

\bibitem{Martin:2009iq}
\hrefCMSnoop {}{A.~Martin {et~al.}, ``{Parton distributions for the LHC}'',}
  \textit{ Eur. Phys. J. C} \textbf{ 63} (2009) 189,
\href{http://www.arXiv.org/abs/0901.0002}{\texttt{arXiv:0901.0002}}.

\bibitem{Campbell:2012uf}
\hrefCMSnoop {}{J.~M. Campbell and R.~K. Ellis, ``{Top-quark processes at NLO
  in production and decay}'',} \textit{ J.Phys.} \textbf{ G42} (2015) 015005,
  \href{http://dx.doi.org/10.1088/0954-3899/42/1/015005}{\doi{10.1088/0954-3899/42/1/015005}},
\href{http://www.arXiv.org/abs/1204.1513}{\texttt{arXiv:1204.1513}}.

\bibitem{Frixione:2002ik}
\hrefCMSnoop {}{S.~Frixione and B.~R. Webber, ``{Matching NLO QCD computations
  and parton shower simulations}'',} \textit{ JHEP} \textbf{ 06} (2002) 029,
\href{http://www.arXiv.org/abs/hep-ph/0204244}{\texttt{arXiv:hep-ph/0204244}}.

\bibitem{Nason:2004rx}
\hrefCMSnoop {}{P.~Nason, ``{A New method for combining NLO QCD with shower
  Monte Carlo algorithms}'',} \textit{ JHEP} \textbf{ 11} (2004) 040,
\href{http://www.arXiv.org/abs/hep-ph/0409146}{\texttt{arXiv:hep-ph/0409146}}.

\bibitem{Frixione:2003ei}
\hrefCMSnoop {}{S.~Frixione, P.~Nason, and B.~R. Webber, ``{Matching NLO QCD
  and parton showers in heavy flavor production}'',} \textit{ JHEP} \textbf{
  08} (2003) 007,
\href{http://www.arXiv.org/abs/hep-ph/0305252}{\texttt{arXiv:hep-ph/0305252}}.

\bibitem{Frixione:2007nw}
\hrefCMSnoop {}{S.~Frixione, P.~Nason, and G.~Ridolfi, ``{A Positive-weight
  next-to-leading-order Monte Carlo for heavy flavour hadroproduction}'',}
  \textit{ JHEP} \textbf{ 09} (2007) 126,
\href{http://www.arXiv.org/abs/0707.3088}{\texttt{arXiv:0707.3088}}.

\bibitem{Melnikov:1995fx}
\hrefCMSnoop {}{K.~Melnikov and O.~I. Yakovlev, ``{Final state interaction in
  the production of heavy unstable particles}'',} \textit{ Nucl. Phys. B}
  \textbf{ 471} (1996) 90,
\href{http://www.arXiv.org/abs/hep-ph/9501358}{\texttt{arXiv:hep-ph/9501358}}.

\bibitem{Denner:2010jp}
\hrefCMSnoop {}{A.~Denner {et~al.}, ``{NLO QCD corrections to WWbb production
  at hadron colliders}'',} \textit{ Phys. Rev. Lett.} \textbf{ 106} (2011)
  052001,
\href{http://www.arXiv.org/abs/1012.3975}{\texttt{arXiv:1012.3975}}.

\bibitem{Denner:2012yc}
\hrefCMSnoop {}{A.~Denner {et~al.}, ``{NLO QCD corrections to off-shell
  top-antitop production with leptonic decays at hadron colliders}'',} \textit{
  JHEP} \textbf{ 10} (2012) 110,
\href{http://www.arXiv.org/abs/1207.5018}{\texttt{arXiv:1207.5018}}.

\bibitem{Bevilacqua:2010qb}
\hrefCMSnoop {}{G.~Bevilacqua {et~al.}, ``{Complete off-shell effects in top
  quark pair hadroproduction with leptonic decay at next-to-leading order}'',}
  \textit{ JHEP} \textbf{ 02} (2011) 083,
\href{http://www.arXiv.org/abs/1012.4230}{\texttt{arXiv:1012.4230}}.

\bibitem{Aad:2010ey}
\hrefCMSnoop {}{{ATLAS Collaboration}, ``{Measurement of the top quark-pair
  production cross section with ATLAS in $pp$ collisions at $\sqrt{s}=7$
  TeV}'',} \textit{ Eur. Phys. J. C} \textbf{ 71} (2011) 1577,
\href{http://www.arXiv.org/abs/1012.1792}{\texttt{arXiv:1012.1792}}.

\bibitem{Khachatryan:2010ez}
\hrefCMSnoop {}{{CMS Collaboration}, ``{First Measurement of the Cross Section
  for Top-Quark Pair Production in Proton-Proton Collisions at $\sqrt{s}=7$
  TeV}'',} \textit{ Phys. Lett. B} \textbf{ 695} (2011) 424,
\href{http://www.arXiv.org/abs/1010.5994}{\texttt{arXiv:1010.5994}}.

\bibitem{Aad:2012qf}
\hrefCMSnoop {}{{ATLAS Collaboration}, ``{Measurement of the top quark pair
  production cross-section with ATLAS in the single lepton channel}'',}
  \textit{ Phys. Lett. B} \textbf{ 711} (2012) 244,
\href{http://www.arXiv.org/abs/1201.1889}{\texttt{arXiv:1201.1889}}.

\bibitem{Aad:2011yb}
\hrefCMSnoop {}{{ATLAS Collaboration}, ``{Measurement of the top quark pair
  production cross section in $pp$ collisions at $\sqrt{s}=7$ TeV in dilepton
  final states with ATLAS}'',} \textit{ Phys. Lett. B} \textbf{ 707} (2012)
  459,
\href{http://www.arXiv.org/abs/1108.3699}{\texttt{arXiv:1108.3699}}.

\bibitem{Chatrchyan:2011nb}
\hrefCMSnoop {}{{CMS Collaboration}, ``{Measurement of the $t\bar{t}$
  production cross section and the top quark mass in the dilepton channel in
  $pp$ collisions at $\sqrt{s}=7$ TeV}'',} \textit{ JHEP} \textbf{ 07} (2011)
  049,
\href{http://www.arXiv.org/abs/1105.5661}{\texttt{arXiv:1105.5661}}.

\bibitem{Chatrchyan:2011ew}
\hrefCMSnoop {}{{CMS Collaboration}, ``{Measurement of the Top-antitop
  Production Cross Section in $pp$ Collisions at $\sqrt{s}=7$ TeV using the
  Kinematic Properties of Events with Leptons and Jets}'',} \textit{ Eur. Phys.
  J. C} \textbf{ 71} (2011) 1721,
\href{http://www.arXiv.org/abs/1106.0902}{\texttt{arXiv:1106.0902}}.

\bibitem{Chatrchyan:2011yy}
\hrefCMSnoop {}{{CMS Collaboration}, ``{Measurement of the $t \bar{t}$
  Production Cross Section in $pp$ Collisions at 7 TeV in Lepton + Jets Events
  Using $b$-quark Jet Identification}'',} \textit{ Phys. Rev. D} \textbf{ 84}
  (2011) 092004,
\href{http://www.arXiv.org/abs/1108.3773}{\texttt{arXiv:1108.3773}}.

\bibitem{Aad:2012vip}
\hrefCMSnoop {}{{ATLAS Collaboration}, ``{Measurement of the ttbar production
  cross section in the tau+jets channel using the ATLAS detector}'',} \textit{
  Eur. Phys. J. C} \textbf{ 73} (2013) 2328,
\href{http://www.arXiv.org/abs/1211.7205}{\texttt{arXiv:1211.7205}}.

\bibitem{Aad:2012mza}
\hrefCMSnoop {}{{ATLAS Collaboration}, ``{Measurement of the top quark pair
  cross section with ATLAS in $pp$ collisions at $\sqrt(s) = 7$~TeV using final
  states with an electron or a muon and a hadronically decaying $\tau$
  lepton}'',} \textit{ Phys. Lett. B} \textbf{ 717} (2012) 89,
\href{http://www.arXiv.org/abs/1205.2067}{\texttt{arXiv:1205.2067}}.

\bibitem{ATLAS:2012aa}
\hrefCMSnoop {}{{ATLAS Collaboration}, ``{Measurement of the cross section for
  top-quark pair production in $pp$ collisions at $\sqrt{s}=7$ TeV with the
  ATLAS detector using final states with two high-pt leptons}'',} \textit{
  JHEP} \textbf{ 05} (2012) 059,
\href{http://www.arXiv.org/abs/1202.4892}{\texttt{arXiv:1202.4892}}.

\bibitem{Chatrchyan:2012ria}
\hrefCMSnoop {}{{CMS Collaboration}, ``{Measurement of the $t\bar{t}$
  production cross section in $pp$ collisions at $\sqrt{s}=7$ TeV with lepton +
  jets final states}'',} \textit{ Phys. Lett. B} \textbf{ 720} (2013) 83,
\href{http://www.arXiv.org/abs/1212.6682}{\texttt{arXiv:1212.6682}}.

\bibitem{Chatrchyan:2012bra}
\hrefCMSnoop {}{{CMS Collaboration}, ``{Measurement of the $t\bar{t}$
  production cross section in the dilepton channel in $pp$ collisions at
  $\sqrt{s}=7$ TeV}'',} \textit{ JHEP} \textbf{ 11} (2012) 067,
\href{http://www.arXiv.org/abs/1208.2671}{\texttt{arXiv:1208.2671}}.

\bibitem{Chatrchyan:2013ual}
\hrefCMSnoop {}{{CMS Collaboration}, ``{Measurement of the $t\bar{t}$
  production cross section in the all-jet final state in $pp$ collisions at
  $\sqrt{s}$ = 7 TeV}'',} \textit{ JHEP} \textbf{ 05} (2013) 065,
\href{http://www.arXiv.org/abs/1302.0508}{\texttt{arXiv:1302.0508}}.

\bibitem{Chatrchyan:2013kff}
\hrefCMSnoop {}{{CMS Collaboration}, ``{Measurement of the top-antitop
  production cross section in the tau+jets channel in $pp$ collisions at
  $\sqrt(s) = 7$~TeV}'',} \textit{ Eur. Phys. J. C} \textbf{ 73} (2013) 2386,
\href{http://www.arXiv.org/abs/1301.5755}{\texttt{arXiv:1301.5755}}.

\bibitem{Chatrchyan:2012vs}
\hrefCMSnoop {}{{CMS Collaboration}, ``{Measurement of the top quark pair
  production cross section in $pp$ collisions at $\sqrt{s} = 7$ TeV in dilepton
  final states containing a $\tau$}'',} \textit{ Phys. Rev. D} \textbf{ 85}
  (2012) 112007,
\href{http://www.arXiv.org/abs/1203.6810}{\texttt{arXiv:1203.6810}}.

\bibitem{Khachatryan:2014loa}
\hrefCMSnoop {}{{CMS} Collaboration, ``{Measurement of the $t \bar t$
  production cross section in $pp$ collisions at $\sqrt s = 8$ TeV in dilepton
  final states containing one $\tau$ lepton}'',} \textit{ Phys. Lett. B}
  \textbf{ 739} (2014) 23,
  \href{http://dx.doi.org/10.1016/j.physletb.2014.10.032}{\doi{10.1016/j.physletb.2014.10.032}},
\href{http://www.arXiv.org/abs/1407.6643}{\texttt{arXiv:1407.6643}}.

\bibitem{Aad:2014jra}
\hrefCMSnoop {}{{ATLAS Collaboration}, ``{Simultaneous measurements of the
  $t\bar{t}$, $W^+W^-$, and $Z/\gamma^{*}\rightarrow\tau\tau$ production
  cross-sections in $pp$ collisions at $\sqrt{s} = 7$ TeV with the ATLAS
  detector}'',}
\href{http://www.arXiv.org/abs/1407.0573}{\texttt{arXiv:1407.0573}}.

\bibitem{Aad:2014kva}
\hrefCMSnoop {}{{ATLAS} Collaboration, ``{Measurement of the $t\bar{t}$
  production cross-section using $e\mu$ events with $b$-tagged jets in $pp$
  collisions at $\sqrt{s}=7$ and 8 TeV with the ATLAS detector}'',} \textit{
  Eur. Phys. J. C} \textbf{ 74} (2014) 3109,
  \href{http://dx.doi.org/10.1140/epjc/s10052-014-3109-7}{\doi{10.1140/epjc/s10052-014-3109-7}},
\href{http://www.arXiv.org/abs/1406.5375}{\texttt{arXiv:1406.5375}}.

\bibitem{Chatrchyan:2013faa}
\hrefCMSnoop {}{{CMS Collaboration}, ``{Measurement of the $t \bar{t}$
  production cross section in the dilepton channel in $pp$ collisions at
  $\sqrt{s}$ = 8 TeV}'',} \textit{ JHEP} \textbf{ 02} (2014) 024,
\href{http://www.arXiv.org/abs/1312.7582}{\texttt{arXiv:1312.7582}}.

\bibitem{PT:Top}
\hrefCMSnoop {}{P.~Uwer and W.~Wagner, ``{Top Quarks: The Peak of the Mass
  Hierarchy? \emph{in} ``Physics at the Terascale'', eds. Brock, I. and
  Sch{\"o}rner-Sadenius, T.}''}, pp.~187--209.
\newblock Wiley-VCH, Weinheim (Germany), 2011.

\bibitem{Aad:2012hg}
\hrefCMSnoop {}{{ATLAS Collaboration}, ``{Measurements of top quark pair
  relative differential cross-sections with ATLAS in $pp$ collisions at
  $\sqrt{s}=7$ TeV}'',} \textit{ Eur. Phys. J. C} \textbf{ 73} (2013) 2261,
\href{http://www.arXiv.org/abs/1207.5644}{\texttt{arXiv:1207.5644}}.

\bibitem{Chatrchyan:2012saa}
\hrefCMSnoop {}{{CMS Collaboration}, ``{Measurement of differential top-quark
  pair production cross sections in $pp$ colisions at $\sqrt{s}=7$ TeV}'',}
  \textit{ Eur. Phys. J. C} \textbf{ 73} (2013) 2339,
\href{http://www.arXiv.org/abs/1211.2220}{\texttt{arXiv:1211.2220}}.

\bibitem{Aad:2014zka}
\hrefCMSnoop {}{{ATLAS} Collaboration, ``{Measurements of normalized
  differential cross-sections for $t\bar{t}$ production in $pp$ collisions at
  $\sqrt{s}$=7 TeV using the ATLAS detector}'',} \textit{ Phys. Rev. D}
  \textbf{ 90} (2014) 072004,
  \href{http://dx.doi.org/10.1103/PhysRevD.90.072004}{\doi{10.1103/PhysRevD.90.072004}},
\href{http://www.arXiv.org/abs/1407.0371}{\texttt{arXiv:1407.0371}}.

\bibitem{Dittmaier:2008uj}
\hrefCMSnoop {}{S.~Dittmaier, P.~Uwer, and S.~Weinzierl, ``{Hadronic top-quark
  pair production in association with a hard jet at next-to-leading order QCD:
  Phenomenological studies for the Tevatron and the LHC}'',} \textit{ Eur.
  Phys. J. C} \textbf{ 59} (2009) 625,
\href{http://www.arXiv.org/abs/0810.0452}{\texttt{arXiv:0810.0452}}.

\bibitem{Alioli:2013mxa}
\hrefCMSnoop {}{S.~Alioli {et~al.}, ``{A new observable to measure the
  top-quark mass at hadron colliders}'',} \textit{ Eur. Phys. J. C} \textbf{
  73} (2013) 2438,
\href{http://www.arXiv.org/abs/1303.6415}{\texttt{arXiv:1303.6415}}.

\bibitem{Alwall:2011uj}
\hrefCMSnoop {}{J.~Alwall {et~al.}, ``MadGraph 5 : Going Beyond'',} \textit{
  JHEP} \textbf{ 06} (2011) 128,
  \href{http://dx.doi.org/10.1007/JHEP06(2011)128}{\doi{10.1007/JHEP06(2011)128}},
\href{http://www.arXiv.org/abs/1106.0522}{\texttt{arXiv:1106.0522}}.

\bibitem{Dittmaier:2007wz}
\hrefCMSnoop {}{S.~Dittmaier, P.~Uwer, and S.~Weinzierl, ``{NLO QCD corrections
  to t anti-t + jet production at hadron colliders}'',} \textit{ Phys. Rev.
  Lett.} \textbf{ 98} (2007) 262002,
\href{http://www.arXiv.org/abs/hep-ph/0703120}{\texttt{arXiv:hep-ph/0703120}}.

\bibitem{Catani:1996vz}
\hrefCMSnoop {}{S.~Catani and M.~Seymour, ``{A General algorithm for
  calculating jet cross-sections in NLO QCD}'',} \textit{ Nucl. Phys. B}
  \textbf{ 485} (1997) 291,
\href{http://www.arXiv.org/abs/hep-ph/9605323}{\texttt{arXiv:hep-ph/9605323}}.

\bibitem{Catani:2002hc}
\hrefCMSnoop {}{S.~Catani {et~al.}, ``{The Dipole formalism for next-to-leading
  order QCD calculations with massive partons}'',} \textit{ Nucl. Phys. B}
  \textbf{ 627} (2002) 189,
\href{http://www.arXiv.org/abs/hep-ph/0201036}{\texttt{arXiv:hep-ph/0201036}}.

\bibitem{Melnikov:2010iu}
\hrefCMSnoop {}{K.~Melnikov and M.~Schulze, ``{NLO QCD corrections to top quark
  pair production in association with one hard jet at hadron colliders}'',}
  \textit{ Nucl. Phys. B} \textbf{ 840} (2010) 129,
\href{http://www.arXiv.org/abs/1004.3284}{\texttt{arXiv:1004.3284}}.

\bibitem{Melnikov:2011qx}
\hrefCMSnoop {}{K.~Melnikov, A.~Scharf, and M.~Schulze, ``{Top quark pair
  production in association with a jet: QCD corrections and jet radiation in
  top quark decays}'',} \textit{ Phys. Rev. D} \textbf{ 85} (2012) 054002,
\href{http://www.arXiv.org/abs/1111.4991}{\texttt{arXiv:1111.4991}}.

\bibitem{Melnikov:2011ta}
\hrefCMSnoop {}{K.~Melnikov, M.~Schulze, and A.~Scharf, ``{QCD corrections to
  top quark pair production in association with a photon at hadron
  colliders}'',} \textit{ Phys. Rev. D} \textbf{ 83} (2011) 074013,
\href{http://www.arXiv.org/abs/1102.1967}{\texttt{arXiv:1102.1967}}.

\bibitem{Pumplin:2002vw}
\hrefCMSnoop {}{J.~Pumplin {et~al.}, ``{New generation of parton distributions
  with uncertainties from global QCD analysis}'',} \textit{ JHEP} \textbf{ 07}
  (2002) 012,
\href{http://www.arXiv.org/abs/hep-ph/0201195}{\texttt{arXiv:hep-ph/0201195}}.

\bibitem{Bredenstein:2009aj}
\hrefCMSnoop {}{A.~Bredenstein {et~al.}, ``{NLO QCD corrections to $pp \to
  \ttbar \bbar + X$ at the LHC}'',} \textit{ Phys. Rev. Lett.} \textbf{ 103}
  (2009) 012002,
\href{http://www.arXiv.org/abs/0905.0110}{\texttt{arXiv:0905.0110}}.

\bibitem{Bredenstein:2008zb}
\hrefCMSnoop {}{A.~Bredenstein {et~al.}, ``{NLO QCD corrections to t anti-t b
  anti-b production at the LHC: 1. Quark-antiquark annihilation}'',} \textit{
  JHEP} \textbf{ 08} (2008) 108,
\href{http://www.arXiv.org/abs/0807.1248}{\texttt{arXiv:0807.1248}}.

\bibitem{Bredenstein:2010rs}
\hrefCMSnoop {}{A.~Bredenstein {et~al.}, ``{NLO QCD Corrections to Top Anti-Top
  Bottom Anti-Bottom Production at the LHC: 2. full hadronic results}'',}
  \textit{ JHEP} \textbf{ 03} (2010) 021,
\href{http://www.arXiv.org/abs/1001.4006}{\texttt{arXiv:1001.4006}}.

\bibitem{Bevilacqua:2011aa}
\hrefCMSnoop {}{G.~Bevilacqua {et~al.}, ``{Hadronic top-quark pair production
  in association with two jets at Next-to-Leading Order QCD}'',} \textit{ Phys.
  Rev. D} \textbf{ 84} (2011) 114017,
\href{http://www.arXiv.org/abs/1108.2851}{\texttt{arXiv:1108.2851}}.

\bibitem{Bevilacqua:2012em}
\hrefCMSnoop {}{G.~Bevilacqua and M.~Worek, ``{Constraining BSM Physics at the
  LHC: Four top final states with NLO accuracy in perturbative QCD}'',}
  \textit{ JHEP} \textbf{ 07} (2012) 111,
\href{http://www.arXiv.org/abs/1206.3064}{\texttt{arXiv:1206.3064}}.

\bibitem{Hoeche:2014qda}
\hrefCMSnoop {}{S.~Hoeche {et~al.}, ``{Next-to-leading order QCD predictions
  for top-quark pair production with up to two jets merged with a parton
  shower}'',}
\href{http://www.arXiv.org/abs/1402.6293}{\texttt{arXiv:1402.6293}}.

\bibitem{Chatrchyan:2014gma}
\hrefCMSnoop {}{{CMS} Collaboration, ``{Measurement of jet multiplicity
  distributions in $\mathrm {t}\overline{\mathrm {t}}$ production in pp
  collisions at $\sqrt{s} = 7\,\text {TeV} $}'',} \textit{ Eur. Phys. J. C}
  \textbf{ 74} (2014) 3014,
  \href{http://dx.doi.org/10.1140/epjc/s10052-014-3014-0}{\doi{10.1140/epjc/s10052-014-3014-0}},
\href{http://www.arXiv.org/abs/1404.3171}{\texttt{arXiv:1404.3171}}.

\bibitem{ATLAS:2012al}
\hrefCMSnoop {}{{ATLAS Collaboration}, ``{Measurement of $t \bar{t}$ production
  with a veto on additional central jet activity in $pp$ collisions at
  $\sqrt(s) = 7$~TeV using the ATLAS detector}'',} \textit{ Eur. Phys. J. C}
  \textbf{ 72} (2012) 2043,
\href{http://www.arXiv.org/abs/1203.5015}{\texttt{arXiv:1203.5015}}.

\bibitem{Aad:2014iaa}
\hrefCMSnoop {}{{ATLAS} Collaboration, ``{Measurement of the $ t\overline{t} $
  production cross-section as a function of jet multiplicity and jet transverse
  momentum in 7 TeV proton-proton collisions with the ATLAS detector}'',}
  \textit{ JHEP} \textbf{ 1501} (2015) 020,
  \href{http://dx.doi.org/10.1007/JHEP01(2015)020}{\doi{10.1007/JHEP01(2015)020}},
\href{http://www.arXiv.org/abs/1407.0891}{\texttt{arXiv:1407.0891}}.

\bibitem{Aad:2013tua}
\hrefCMSnoop {}{{ATLAS} Collaboration, ``{Study of heavy-flavor quarks produced
  in association with top-quark pairs at $\sqrt{s}=7$~TeV using the ATLAS
  detector}'',} \textit{ Phys. Rev. D} \textbf{ 89} (2014) 072012,
  \href{http://dx.doi.org/10.1103/PhysRevD.89.072012}{\doi{10.1103/PhysRevD.89.072012}},
\href{http://www.arXiv.org/abs/1304.6386}{\texttt{arXiv:1304.6386}}.

\bibitem{Heinemeyer:2013dia}
\hrefCMSnoop {}{S.~Heinemeyer, W.~Hollik, G.~Weiglein, and L.~Zeune,
  ``{Implications of LHC search results on the W boson mass prediction in the
  MSSM}'',} \textit{ JHEP} \textbf{ 12} (2013) 084,
  \href{http://dx.doi.org/10.1007/JHEP12(2013)084}{\doi{10.1007/JHEP12(2013)084}},
\href{http://www.arXiv.org/abs/1311.1663}{\texttt{arXiv:1311.1663}}.

\bibitem{Degrassi:2012ry}
G.~Degrassi\hrefCMSnoop {}{ {et~al.}, ``{Higgs mass and vacuum stability in the
  Standard Model at NNLO}'',} \textit{ JHEP} \textbf{ 1208} (2012) 098,
  \href{http://dx.doi.org/10.1007/JHEP08(2012)098}{\doi{10.1007/JHEP08(2012)098}},
\href{http://www.arXiv.org/abs/1205.6497}{\texttt{arXiv:1205.6497}}.

\bibitem{Degrassi:2014hoa}
\hrefCMSnoop {}{G.~Degrassi, ``{The role of the top quark in the stability of
  the SM Higgs potential}'',} \textit{ Nuovo Cim. C} \textbf{ 37} (2014) 47,
  \href{http://dx.doi.org/10.1393/ncc/i2014-11735-1}{\doi{10.1393/ncc/i2014-11735-1}},
\href{http://www.arXiv.org/abs/1405.6852}{\texttt{arXiv:1405.6852}}.

\bibitem{Buckley:2011ms}
\hrefCMSnoop {}{A.~Buckley {et~al.}, ``{General-purpose event generators for
  LHC physics}'',} \textit{ Phys. Rept.} \textbf{ 504} (2011) 145,
  \href{http://dx.doi.org/10.1016/j.physrep.2011.03.005}{\doi{10.1016/j.physrep.2011.03.005}},
\href{http://www.arXiv.org/abs/1101.2599}{\texttt{arXiv:1101.2599}}.

\bibitem{Moch:2014tta}
\hrefCMSnoop {}{S.~Moch {et~al.}, ``{High precision fundamental constants at
  the TeV scale}'',}
\href{http://www.arXiv.org/abs/1405.4781}{\texttt{arXiv:1405.4781}}.

\bibitem{Bigi:1994em}
\hrefCMSnoop {}{I.~I.~Y. Bigi {et~al.}, ``{The Pole mass of the heavy quark.
  Perturbation theory and beyond}'',} \textit{ Phys. Rev. D} \textbf{ 50}
  (1994) 2234,
\href{http://www.arXiv.org/abs/hep-ph/9402360}{\texttt{arXiv:hep-ph/9402360}}.

\bibitem{Beneke:1994sw}
\hrefCMSnoop {}{M.~Beneke and V.~M. Braun, ``{Heavy quark effective theory
  beyond perturbation theory: Renormalons, the pole mass and the residual mass
  term}'',} \textit{ Nucl. Phys. B} \textbf{ 426} (1994) 301,
\href{http://www.arXiv.org/abs/hep-ph/9402364}{\texttt{arXiv:hep-ph/9402364}}.

\bibitem{ATLAS:2012aj}
\hrefCMSnoop {}{{ATLAS Collaboration}, ``{Measurement of the top quark mass
  with the template method in the $t \bar{t}$ -> lepton + jets channel using
  ATLAS data}'',} \textit{ Eur. Phys. J. C} \textbf{ 72} (2012) 2046,
\href{http://www.arXiv.org/abs/1203.5755}{\texttt{arXiv:1203.5755}}.

\bibitem{Chatrchyan:2012ea}
\hrefCMSnoop {}{{CMS Collaboration}, ``{Measurement of the top-quark mass in
  $t\bar{t}$ events with dilepton final states in $pp$ collisions at
  $\sqrt{s}=7$ TeV}'',} \textit{ Eur. Phys. J. C} \textbf{ 72} (2012) 2202,
\href{http://www.arXiv.org/abs/1209.2393}{\texttt{arXiv:1209.2393}}.

\bibitem{Chatrchyan:2013boa}
\hrefCMSnoop {}{{CMS Collaboration}, ``{Measurement of masses in the $t
  \bar{t}$ system by kinematic endpoints in $pp$ collisions at $\sqrt{s}$ = 7
  TeV}'',} \textit{ Eur. Phys. J. C} \textbf{ 73} (2013) 2494,
\href{http://www.arXiv.org/abs/1304.5783}{\texttt{arXiv:1304.5783}}.

\bibitem{Chatrchyan:2013xza}
\hrefCMSnoop {}{{CMS} Collaboration, ``{Measurement of the top-quark mass in
  all-jets $t\bar{t}$ events in pp collisions at $\sqrt{s}$=7 TeV}'',} \textit{
  Eur. Phys. J. C} \textbf{ 74} (2014) 2758,
  \href{http://dx.doi.org/10.1140/epjc/s10052-014-2758-x}{\doi{10.1140/epjc/s10052-014-2758-x}},
\href{http://www.arXiv.org/abs/1307.4617}{\texttt{arXiv:1307.4617}}.

\bibitem{Chatrchyan:2012cz}
\hrefCMSnoop {}{{CMS Collaboration}, ``{Measurement of the top-quark mass in
  $t\bar{t}$ events with lepton+jets final states in $pp$ collisions at
  $\sqrt{s}=7$ TeV}'',} \textit{ JHEP} \textbf{ 12} (2012) 105,
\href{http://www.arXiv.org/abs/1209.2319}{\texttt{arXiv:1209.2319}}.

\bibitem{Aad:2014zea}
\hrefCMSnoop {}{{ATLAS Collaboration}, ``{Measurement of the top-quark mass in
  the fully hadronic decay channel from ATLAS data at $\sqrt{s}=7$ TeV}'',}
\href{http://www.arXiv.org/abs/1409.0832}{\texttt{arXiv:1409.0832}}.

\bibitem{Chatrchyan:2012uba}
\hrefCMSnoop {}{{CMS Collaboration}, ``{Measurement of the mass difference
  between top and antitop quarks}'',} \textit{ JHEP} \textbf{ 06} (2012) 109,
\href{http://www.arXiv.org/abs/1204.2807}{\texttt{arXiv:1204.2807}}.

\bibitem{Aad:2013eva}
\hrefCMSnoop {}{{ATLAS Collaboration}, ``{Measurement of the mass difference
  between top and anti-top quarks in $pp$ collisions at $\sqrt(s) = 7$ TeV
  using the ATLAS detector}'',} \textit{ Phys. Lett. B} \textbf{ 728} (2014)
  363,
\href{http://www.arXiv.org/abs/1310.6527}{\texttt{arXiv:1310.6527}}.

\bibitem{ATLAS:2014wva}
\hrefCMSnoop {}{{ATLAS Collaboration, CDF Collaboration, CMS Collaboration,
  \Dzero} Collaboration, ``{First combination of \Tevatron and LHC measurements
  of the top-quark mass}'',}
\href{http://www.arXiv.org/abs/1403.4427}{\texttt{arXiv:1403.4427}}.

\bibitem{Chatrchyan:2013haa}
\hrefCMSnoop {}{{CMS Collaboration}, ``{Determination of the top-quark pole
  mass and strong coupling constant from the $\ttbar$ production cross section
  in $pp$ collisions at $\sqrt{s}$ = 7 TeV}'',} \textit{ Phys. Lett. B}
  \textbf{ 728} (2014) 496,
  \href{http://dx.doi.org/10.1016/j.physletb.2014.08.040,
  10.1016/j.physletb.2013.12.009}{\doi{10.1016/j.physletb.2014.08.040,
  10.1016/j.physletb.2013.12.009}},
\href{http://www.arXiv.org/abs/1307.1907}{\texttt{arXiv:1307.1907}}.

\bibitem{Kuhn:1998jr}
\hrefCMSnoop {}{J.~H. Kuhn and G.~Rodrigo, ``{Charge asymmetry in
  hadroproduction of heavy quarks}'',} \textit{ Phys. Rev. Lett.} \textbf{ 81}
  (1998) 49,
\href{http://www.arXiv.org/abs/hep-ph/9802268}{\texttt{arXiv:hep-ph/9802268}}.

\bibitem{Kuhn:1998kw}
\hrefCMSnoop {}{J.~H. Kuhn and G.~Rodrigo, ``{Charge asymmetry of heavy quarks
  at hadron colliders}'',} \textit{ Phys. Rev. D} \textbf{ 59} (1999) 054017,
\href{http://www.arXiv.org/abs/hep-ph/9807420}{\texttt{arXiv:hep-ph/9807420}}.

\bibitem{Bernreuther:2012sx}
\hrefCMSnoop {}{W.~Bernreuther and Z.-G. Si, ``{Top quark and leptonic charge
  asymmetries for the Tevatron and LHC}'',} \textit{ Phys. Rev. D} \textbf{ 86}
  (2012) 034026,
\href{http://www.arXiv.org/abs/1205.6580}{\texttt{arXiv:1205.6580}}.

\bibitem{Aaltonen:2012it}
\hrefCMSnoop {}{{CDF Collaboration}, ``{Measurement of the top quark
  forward-backward production asymmetry and its dependence on event kinematic
  properties}'',} \textit{ Phys. Rev. D} \textbf{ 87} (2013) 092002,
\href{http://www.arXiv.org/abs/1211.1003}{\texttt{arXiv:1211.1003}}.

\bibitem{Abazov:2011rq}
\hrefCMSnoop {}{{\Dzero Collaboration}, ``{Forward-backward asymmetry in top
  quark-antiquark production}'',} \textit{ Phys. Rev. D} \textbf{ 84} (2011)
  112005,
\href{http://www.arXiv.org/abs/1107.4995}{\texttt{arXiv:1107.4995}}.

\bibitem{Aad:2013cea}
\hrefCMSnoop {}{{ATLAS Collaboration}, ``{Measurement of the top quark pair
  production charge asymmetry in proton-proton collisions at $\sqrt{s}$ = 7 TeV
  using the ATLAS detector}'',} \textit{ JHEP} \textbf{ 02} (2014) 107,
\href{http://www.arXiv.org/abs/1311.6724}{\texttt{arXiv:1311.6724}}.

\bibitem{ATLAS:2012an}
\hrefCMSnoop {}{{ATLAS Collaboration}, ``{Measurement of the charge asymmetry
  in top quark pair production in $pp$ collisions at $\sqrt{s}=7$ TeV using the
  ATLAS detector}'',} \textit{ Eur. Phys. J. C} \textbf{ 72} (2012) 2039,
\href{http://www.arXiv.org/abs/1203.4211}{\texttt{arXiv:1203.4211}}.

\bibitem{Erdmann:2013rxa}
\hrefCMSnoop {}{J.~Erdmann {et~al.}, ``{A likelihood-based reconstruction
  algorithm for top-quark pairs and the KLFitter framework}'',} \textit{ Nucl.
  Instrum. Meth. A} \textbf{ 748} (2014) 18,
  \href{http://dx.doi.org/10.1016/j.nima.2014.02.029}{\doi{10.1016/j.nima.2014.02.029}},
\href{http://www.arXiv.org/abs/1312.5595}{\texttt{arXiv:1312.5595}}.

\bibitem{Choudalakis:2012hz}
\hrefCMSnoop {}{G.~Choudalakis, ``{Fully Bayesian Unfolding}'',}
\href{http://www.arXiv.org/abs/1201.4612}{\texttt{arXiv:1201.4612}}.

\bibitem{Chatrchyan:2011hk}
\hrefCMSnoop {}{{CMS Collaboration}, ``{Measurement of the charge asymmetry in
  top-quark pair production in proton-proton collisions at $\sqrt{s}=7$
  TeV}'',} \textit{ Phys. Lett. B} \textbf{ 709} (2012) 28,
\href{http://www.arXiv.org/abs/1112.5100}{\texttt{arXiv:1112.5100}}.

\bibitem{Chatrchyan:2012cxa}
\hrefCMSnoop {}{{CMS Collaboration}, ``{Inclusive and differential measurements
  of the $t \bar{t}$ charge asymmetry in proton-proton collisions at 7 TeV}'',}
  \textit{ Phys. Lett. B} \textbf{ 717} (2012) 129,
\href{http://www.arXiv.org/abs/1207.0065}{\texttt{arXiv:1207.0065}}.

\bibitem{Blobel:2002pu}
\hrefCMSnoop {}{V.~Blobel, ``{An Unfolding method for high-energy physics
  experiments}'',}
\href{http://www.arXiv.org/abs/hep-ex/0208022}{\texttt{arXiv:hep-ex/0208022}}.

\bibitem{Chatrchyan:2014yta}
\hrefCMSnoop {}{{CMS Collaboration}, ``{Measurements of the $t\bar{t}$ charge
  asymmetry using the dilepton decay channel in $pp$ collisions at $\sqrt{s} =$
  7 TeV}'',} \textit{ JHEP} \textbf{ 04} (2014) 191,
  \href{http://dx.doi.org/10.1007/JHEP04(2014)191}{\doi{10.1007/JHEP04(2014)191}},
\href{http://www.arXiv.org/abs/1402.3803}{\texttt{arXiv:1402.3803}}.

\bibitem{Hocker:1995kb}
\hrefCMSnoop {}{A.~Hocker and V.~Kartvelishvili, ``{SVD approach to data
  unfolding}'',} \textit{ Nucl. Instrum. Meth. A} \textbf{ 372} (1996) 469,
\href{http://www.arXiv.org/abs/hep-ph/9509307}{\texttt{arXiv:hep-ph/9509307}}.

\bibitem{Grossman:2008qh}
\hrefCMSnoop {}{Y.~Grossman and I.~Nachshon, ``{Hadronization, spin, and
  lifetimes}'',} \textit{ JHEP} \textbf{ 07} (2008) 016,
\href{http://www.arXiv.org/abs/0803.1787}{\texttt{arXiv:0803.1787}}.

\bibitem{Mahlon:2010gw}
\hrefCMSnoop {}{G.~Mahlon and S.~J. Parke, ``{Spin Correlation Effects in Top
  Quark Pair Production at the LHC}'',} \textit{ Phys. Rev. D} \textbf{ 81}
  (2010) 074024,
\href{http://www.arXiv.org/abs/1001.3422}{\texttt{arXiv:1001.3422}}.

\bibitem{Brandenburg:2002xr}
\hrefCMSnoop {}{A.~Brandenburg, Z.~Si, and P.~Uwer, ``{QCD corrected spin
  analyzing power of jets in decays of polarized top quarks}'',} \textit{ Phys.
  Lett. B} \textbf{ 539} (2002) 235,
\href{http://www.arXiv.org/abs/hep-ph/0205023}{\texttt{arXiv:hep-ph/0205023}}.

\bibitem{Bernreuther:1995cx}
\hrefCMSnoop {}{W.~Bernreuther, A.~Brandenburg, and P.~Uwer, ``{Transverse
  polarization of top quark pairs at the Tevatron and the large hadron
  collider}'',} \textit{ Phys. Lett. B} \textbf{ 368} (1996) 153,
  \href{http://dx.doi.org/10.1016/0370-2693(95)01475-6}{\doi{10.1016/0370-2693(95)01475-6}},
\href{http://www.arXiv.org/abs/hep-ph/9510300}{\texttt{arXiv:hep-ph/9510300}}.

\bibitem{Dharmaratna:1989jr}
\hrefCMSnoop {}{W.~G. Dharmaratna and G.~R. Goldstein, ``{Gluon Fusion as a
  Source for Massive Quark Polarization}'',} \textit{ Phys. Rev. D} \textbf{
  41} (1990) 1731,
\href{http://dx.doi.org/10.1103/PhysRevD.41.1731}{\doi{10.1103/PhysRevD.41.1731}}.

\bibitem{Bernreuther:2013aga}
\hrefCMSnoop {}{W.~Bernreuther and Z.-G. Si, ``{Top quark spin correlations and
  polarization at the LHC: standard model predictions and effects of anomalous
  top chromo moments}'',} \textit{ Phys. Lett. B} \textbf{ 725} (2013) 115,
\href{http://www.arXiv.org/abs/1305.2066}{\texttt{arXiv:1305.2066}}.

\bibitem{Bernreuther:2010ny}
\hrefCMSnoop {}{W.~Bernreuther and Z.-G. Si, ``{Distributions and correlations
  for top quark pair production and decay at the Tevatron and LHC.}'',}
  \textit{ Nucl. Phys. B} \textbf{ 837} (2010) 90,
\href{http://www.arXiv.org/abs/1003.3926}{\texttt{arXiv:1003.3926}}.

\bibitem{Abazov:2011gi}
\hrefCMSnoop {}{{\Dzero Collaboration}, ``{Evidence for spin correlation in
  $t\bar{t}$ production}'',} \textit{ Phys. Rev. Lett.} \textbf{ 108} (2012)
  032004,
\href{http://www.arXiv.org/abs/1110.4194}{\texttt{arXiv:1110.4194}}.

\bibitem{ATLAS:2012ao}
\hrefCMSnoop {}{{ATLAS Collaboration}, ``{Observation of spin correlation in $t
  \bar{t}$ events from $pp$ collisions at $\sqrt(s) = 7~\TeV$ using the ATLAS
  detector}'',} \textit{ Phys. Rev. Lett.} \textbf{ 108} (2012) 212001,
\href{http://www.arXiv.org/abs/1203.4081}{\texttt{arXiv:1203.4081}}.

\bibitem{Aad:2014pwa}
\hrefCMSnoop {}{{ATLAS Collaboration}, ``{Measurements of spin correlation in
  top-antitop quark events from proton-proton collisions at $\sqrt{s}=7$ TeV
  using the ATLAS detector}'',}
\href{http://www.arXiv.org/abs/1407.4314}{\texttt{arXiv:1407.4314}}.

\bibitem{Chatrchyan:2013wua}
\hrefCMSnoop {}{{CMS} Collaboration, ``{Measurements of $t\bar{t}$ spin
  correlations and top-quark polarization using dilepton final states in pp
  collisions at $\sqrt{s}$ = 7 TeV}'',} \textit{ Phys. Rev. Lett.} \textbf{
  112} (2014) 182001,
  \href{http://dx.doi.org/10.1103/PhysRevLett.112.182001}{\doi{10.1103/PhysRevLett.112.182001}},
\href{http://www.arXiv.org/abs/1311.3924}{\texttt{arXiv:1311.3924}}.

\bibitem{Aad:2013ksa}
\hrefCMSnoop {}{{ATLAS} Collaboration, ``{Measurement of Top Quark Polarization
  in Top-Antitop Events from Proton-Proton Collisions at $\sqrt{s}$ = 7??TeV
  Using the ATLAS Detector}'',} \textit{ Phys. Rev. Lett.} \textbf{ 111} (2013)
  232002,
  \href{http://dx.doi.org/10.1103/PhysRevLett.111.232002}{\doi{10.1103/PhysRevLett.111.232002}},
\href{http://www.arXiv.org/abs/1307.6511}{\texttt{arXiv:1307.6511}}.

\bibitem{Abbott:1997fv}
\hrefCMSnoop {}{{\Dzero Collaboration}, ``{Measurement of the top quark mass
  using dilepton events}'',} \textit{ Phys. Rev. Lett.} \textbf{ 80} (1998)
  2063,
\href{http://www.arXiv.org/abs/hep-ex/9706014}{\texttt{arXiv:hep-ex/9706014}}.

\bibitem{Czarnecki:2010gb}
\hrefCMSnoop {}{A.~Czarnecki, J.~G. Korner, and J.~H. Piclum, ``{Helicity
  fractions of W bosons from top quark decays at NNLO in QCD}'',} \textit{
  Phys. Rev. D} \textbf{ 81} (2010) 111503,
\href{http://www.arXiv.org/abs/1005.2625}{\texttt{arXiv:1005.2625}}.

\bibitem{AguilarSaavedra:2006fy}
\hrefCMSnoop {}{J.~Aguilar-Saavedra {et~al.}, ``{Probing anomalous Wtb
  couplings in top pair decays}'',} \textit{ Eur. Phys. J. C} \textbf{ 50}
  (2007) 519,
\href{http://www.arXiv.org/abs/hep-ph/0605190}{\texttt{arXiv:hep-ph/0605190}}.

\bibitem{Zhang:2010dr}
\hrefCMSnoop {}{C.~Zhang and S.~Willenbrock, ``{Effective-Field-Theory Approach
  to Top-Quark Production and Decay}'',} \textit{ Phys. Rev. D} \textbf{ 83}
  (2011) 034006,
\href{http://www.arXiv.org/abs/1008.3869}{\texttt{arXiv:1008.3869}}.

\bibitem{Kane:1991bg}
\hrefCMSnoop {}{G.~L. Kane, G.~Ladinsky, and C.~Yuan, ``{Using the Top Quark
  for Testing Standard Model Polarization and CP Predictions}'',} \textit{
  Phys. Rev. D} \textbf{ 45} (1992)
124.

\bibitem{Aad:2012ky}
\hrefCMSnoop {}{{ATLAS Collaboration}, ``{Measurement of the W boson
  polarization in top quark decays with the ATLAS detector}'',} \textit{ JHEP}
  \textbf{ 06} (2012) 088,
\href{http://www.arXiv.org/abs/1205.2484}{\texttt{arXiv:1205.2484}}.

\bibitem{Valassi:2003mu}
\hrefCMSnoop {}{A.~Valassi, ``{Combining correlated measurements of several
  different physical quantities}'',} \textit{ Nucl. Instrum. Meth. A} \textbf{
  500} (2003)
391.

\bibitem{Chatrchyan:2013jna}
\hrefCMSnoop {}{{CMS Collaboration}, ``{Measurement of the W-boson helicity in
  top-quark decays from $t\bar{t}$ production in lepton+jets events in $pp$
  collisions at $\sqrt{s} =$ 7 TeV}'',} \textit{ JHEP} \textbf{ 10} (2013) 167,
\href{http://www.arXiv.org/abs/1308.3879}{\texttt{arXiv:1308.3879}}.

\bibitem{Khachatryan:2014vma}
\hrefCMSnoop {}{{CMS} Collaboration, ``{Measurement of the W boson helicity in
  events with a single reconstructed top quark in pp collisions at $ \sqrt{s}=8
  $ TeV}'',} \textit{ JHEP} \textbf{ 1501} (2015) 053,
  \href{http://dx.doi.org/10.1007/JHEP01(2015)053}{\doi{10.1007/JHEP01(2015)053}},
\href{http://www.arXiv.org/abs/1410.1154}{\texttt{arXiv:1410.1154}}.

\bibitem{Chatrchyan:2013qca}
\hrefCMSnoop {}{{CMS Collaboration}, ``{Measurement of associated production of
  vector bosons and top quark-antiquark pairs at $\sqrt(s) = 7$~TeV}'',}
  \textit{ Phys. Rev. Lett.} \textbf{ 110} (2013) 172002,
\href{http://www.arXiv.org/abs/1303.3239}{\texttt{arXiv:1303.3239}}.

\bibitem{Aaltonen:2011sp}
\hrefCMSnoop {}{{CDF Collaboration}, ``{Evidence for $t\bar{t}\gamma$
  Production and Measurement of $\sigma_t\bar{t}\gamma / \sigma_t\bar{t}$}'',}
  \textit{ Phys. Rev. D} \textbf{ 84} (2011) 031104,
\href{http://www.arXiv.org/abs/1106.3970}{\texttt{arXiv:1106.3970}}.

\bibitem{Khachatryan:2014ewa}
\hrefCMSnoop {}{{CMS} Collaboration, ``{Measurement of top quark-antiquark pair
  production in association with a W or Z boson in pp collisions at $\sqrt{s} =
  8$ $\,\text {TeV}$}'',} \textit{ Eur. Phys. J. C} \textbf{ 74} (2014) 3060,
  \href{http://dx.doi.org/10.1140/epjc/s10052-014-3060-7}{\doi{10.1140/epjc/s10052-014-3060-7}},
\href{http://www.arXiv.org/abs/1406.7830}{\texttt{arXiv:1406.7830}}.

\bibitem{Garzelli:2012bn}
\hrefCMSnoop {}{M.~Garzelli {et~al.}, ``{t $\bar{t}$ $W^{+-}$ and t $\bar{t}$ Z
  Hadroproduction at NLO accuracy in QCD with Parton Shower and Hadronization
  effects}'',} \textit{ JHEP} \textbf{ 11} (2012) 056,
\href{http://www.arXiv.org/abs/1208.2665}{\texttt{arXiv:1208.2665}}.

\bibitem{Campbell:2012dh}
\hrefCMSnoop {}{J.~M. Campbell and R.~K. Ellis, ``{$t \bar{t} W^{+-}$
  production and decay at NLO}'',} \textit{ JHEP} \textbf{ 07} (2012) 052,
\href{http://www.arXiv.org/abs/1204.5678}{\texttt{arXiv:1204.5678}}.

\bibitem{Campbell:2009ss}
\hrefCMSnoop {}{J.~M. Campbell {et~al.}, ``{Single-Top Production at Hadron
  Colliders}'',} \textit{ Phys. Rev. Lett.} \textbf{ 102} (2009) 182003,
\href{http://www.arXiv.org/abs/0903.0005}{\texttt{arXiv:0903.0005}}.

\bibitem{Campbell:2009gj}
\hrefCMSnoop {}{J.~Campbell {et~al.}, ``{NLO Predictions for $t$-Channel
  Production of Single Top and Fourth Generation Quarks at Hadron
  Colliders}'',} \textit{ JHEP} \textbf{ 10} (2009) 042,
  \href{http://www.arXiv.org/abs/0907.3933}{\texttt{arXiv:0907.3933}}.

\bibitem{Bordes:1994ki}
\hrefCMSnoop {}{G.~Bordes and B.~van Eijk, ``{Calculating QCD corrections to
  single top production in hadronic interactions}'',} \textit{ Nucl. Phys. B}
  \textbf{ 435} (1995)
23.

\bibitem{Stelzer:1997ns}
\hrefCMSnoop {}{T.~Stelzer, Z.~Sullivan, and S.~Willenbrock, ``{Single top
  quark production via $W$ - gluon fusion at next-to-leading order}'',}
  \textit{ Phys. Rev. D} \textbf{ 56} (1997) 5919,
\href{http://www.arXiv.org/abs/hep-ph/9705398}{\texttt{arXiv:hep-ph/9705398}}.

\bibitem{Stelzer:1998ni}
\hrefCMSnoop {}{T.~Stelzer, Z.~Sullivan, and S.~Willenbrock, ``{Single top
  quark production at hadron colliders}'',} \textit{ Phys. Rev. D} \textbf{ 58}
  (1998) 094021,
\href{http://www.arXiv.org/abs/hep-ph/9807340}{\texttt{arXiv:hep-ph/9807340}}.

\bibitem{Smith:1996ij}
\hrefCMSnoop {}{M.~C. Smith and S.~Willenbrock, ``{QCD and Yukawa corrections
  to single top quark production via $q \bar{q} \to t \bar{b}$}'',} \textit{
  Phys. Rev. D} \textbf{ 54} (1996) 6696,
\href{http://www.arXiv.org/abs/hep-ph/9604223}{\texttt{arXiv:hep-ph/9604223}}.

\bibitem{Giele:1995kr}
\hrefCMSnoop {}{W.~T. Giele, S.~Keller, and E.~Laenen, ``{QCD corrections to
  $W$ boson plus heavy quark production at the Tevatron}'',} \textit{ Phys.
  Lett. B} \textbf{ 372} (1996) 141,
\href{http://www.arXiv.org/abs/hep-ph/9511449}{\texttt{arXiv:hep-ph/9511449}}.

\bibitem{Zhu:2002uj}
\hrefCMSnoop {}{S.~Zhu, ``{Next-to-leading order QCD corrections to b g $\to$ t
  W- at the CERN Large Hadron Collider}'',} \textit{ Phys. Lett. B} \textbf{
  524} (2002)
283.

\bibitem{Brucherseifer:2014ama}
\hrefCMSnoop {}{M.~Brucherseifer, F.~Caola, and K.~Melnikov, ``{On the NNLO QCD
  corrections to single-top production at the LHC}'',} \textit{ Phys.Lett.}
  \textbf{ B736} (2014) 58--63,
  \href{http://dx.doi.org/10.1016/j.physletb.2014.06.075}{\doi{10.1016/j.physletb.2014.06.075}},
\href{http://www.arXiv.org/abs/1404.7116}{\texttt{arXiv:1404.7116}}.

\bibitem{Harris:2002md}
\hrefCMSnoop {}{B.~Harris {et~al.}, ``{The Fully differential single top quark
  cross-section in next to leading order QCD}'',} \textit{ Phys. Rev. D}
  \textbf{ 66} (2002) 054024,
\href{http://www.arXiv.org/abs/hep-ph/0207055}{\texttt{arXiv:hep-ph/0207055}}.

\bibitem{Sullivan:2004ie}
\hrefCMSnoop {}{Z.~Sullivan, ``{Understanding single-top-quark production and
  jets at hadron colliders}'',} \textit{ Phys. Rev. D} \textbf{ 70} (2004)
  114012,
\href{http://www.arXiv.org/abs/hep-ph/0408049}{\texttt{arXiv:hep-ph/0408049}}.

\bibitem{Sullivan:2005ar}
\hrefCMSnoop {}{Z.~Sullivan, ``{Angular correlations in single-top-quark and
  Wjj production at next-to-leading order}'',} \textit{ Phys. Rev. D} \textbf{
  72} (2005) 094034,
\href{http://www.arXiv.org/abs/hep-ph/0510224}{\texttt{arXiv:hep-ph/0510224}}.

\bibitem{Campbell:2004ch}
\hrefCMSnoop {}{J.~M. Campbell, R.~K. Ellis, and F.~Tramontano, ``{Single top
  production and decay at next-to-leading order}'',} \textit{ Phys. Rev. D}
  \textbf{ 70} (2004) 094012,
\href{http://www.arXiv.org/abs/hep-ph/0408158}{\texttt{arXiv:hep-ph/0408158}}.

\bibitem{Cao:2004ky}
\hrefCMSnoop {}{Q.-H. Cao and C.-P. Yuan, ``{Single top quark production and
  decay at next-to-leading order in hadron collision}'',} \textit{ Phys. Rev.
  D} \textbf{ 71} (2005) 054022,
\href{http://www.arXiv.org/abs/hep-ph/0408180}{\texttt{arXiv:hep-ph/0408180}}.

\bibitem{Cao:2005pq}
\hrefCMSnoop {}{Q.-H. Cao {et~al.}, ``{Next-to-leading order corrections to
  single top quark production and decay at the Tevatron: 2. $t^-$ channel
  process}'',} \textit{ Phys. Rev. D} \textbf{ 72} (2005) 094027,
\href{http://www.arXiv.org/abs/hep-ph/0504230}{\texttt{arXiv:hep-ph/0504230}}.

\bibitem{Campbell:2005bb}
\hrefCMSnoop {}{J.~M. Campbell and F.~Tramontano, ``{Next-to-leading order
  corrections to Wt production and decay}'',} \textit{ Nucl. Phys. B} \textbf{
  726} (2005) 109,
\href{http://www.arXiv.org/abs/hep-ph/0506289}{\texttt{arXiv:hep-ph/0506289}}.

\bibitem{Frixione:2005vw}
\hrefCMSnoop {}{S.~Frixione {et~al.}, ``{Single-top production in MC@NLO}'',}
  \textit{ JHEP} \textbf{ 03} (2006) 092,
\href{http://www.arXiv.org/abs/hep-ph/0512250}{\texttt{arXiv:hep-ph/0512250}}.

\bibitem{Frixione:2008yi}
\hrefCMSnoop {}{S.~Frixione {et~al.}, ``{Single-top hadroproduction in
  association with a W boson}'',} \textit{ JHEP} \textbf{ 07} (2008) 029,
\href{http://www.arXiv.org/abs/0805.3067}{\texttt{arXiv:0805.3067}}.

\bibitem{Alioli:2009je}
\hrefCMSnoop {}{S.~Alioli {et~al.}, ``{NLO single-top production matched with
  shower in POWHEG: s- and t-channel contributions}'',} \textit{ JHEP} \textbf{
  09} (2009) 111,
\href{http://www.arXiv.org/abs/0907.4076}{\texttt{arXiv:0907.4076}}.

\bibitem{Re:2010bp}
\hrefCMSnoop {}{E.~Re, ``{Single-top Wt-channel production matched with parton
  showers using the POWHEG method}'',} \textit{ Eur. Phys. J. C} \textbf{ 71}
  (2011) 1547,
\href{http://www.arXiv.org/abs/1009.2450}{\texttt{arXiv:1009.2450}}.

\bibitem{Mrenna:1997wp}
\hrefCMSnoop {}{S.~Mrenna and C.~Yuan, ``{Effects of QCD resummation on W+ h
  and t anti-b production at the Tevatron}'',} \textit{ Phys. Lett. B} \textbf{
  416} (1998) 200,
\href{http://www.arXiv.org/abs/hep-ph/9703224}{\texttt{arXiv:hep-ph/9703224}}.

\bibitem{Kidonakis:2006bu}
\hrefCMSnoop {}{N.~Kidonakis, ``{Single top production at the Tevatron:
  Threshold resummation and finite-order soft gluon corrections}'',} \textit{
  Phys. Rev. D} \textbf{ 74} (2006) 114012,
\href{http://www.arXiv.org/abs/hep-ph/0609287}{\texttt{arXiv:hep-ph/0609287}}.

\bibitem{Kidonakis:2007ej}
\hrefCMSnoop {}{N.~Kidonakis, ``{Higher-order soft gluon corrections in single
  top quark production at the LHC}'',} \textit{ Phys. Rev. D} \textbf{ 75}
  (2007) 071501,
\href{http://www.arXiv.org/abs/hep-ph/0701080}{\texttt{arXiv:hep-ph/0701080}}.

\bibitem{Kidonakis:2010ux}
\hrefCMSnoop {}{N.~Kidonakis, ``{Two-loop soft anomalous dimensions for single
  top quark associated production with a W- or H-}'',} \textit{ Phys. Rev. D}
  \textbf{ 82} (2010) 054018,
\href{http://www.arXiv.org/abs/1005.4451}{\texttt{arXiv:1005.4451}}.

\bibitem{Kidonakis:2010dk}
\hrefCMSnoop {}{N.~Kidonakis, ``{Next-to-next-to-leading soft-gluon corrections
  for the top quark cross section and transverse momentum distribution}'',}
  \textit{ Phys. Rev. D} \textbf{ 82} (2010) 114030,
\href{http://www.arXiv.org/abs/1009.4935}{\texttt{arXiv:1009.4935}}.

\bibitem{Kidonakis:2011wy}
\hrefCMSnoop {}{N.~Kidonakis, ``{Next-to-next-to-leading-order collinear and
  soft gluon corrections for t-channel single top quark production}'',}
  \textit{ Phys. Rev. D} \textbf{ 83} (2011) 091503,
\href{http://www.arXiv.org/abs/1103.2792}{\texttt{arXiv:1103.2792}}.

\bibitem{Aad:2012ux}
\hrefCMSnoop {}{{ATLAS Collaboration}, ``{Measurement of the $t$-channel single
  top-quark production cross section in $pp$ collisions at $\sqrt{s}=7$ TeV
  with the ATLAS detector}'',} \textit{ Phys. Lett. B} \textbf{ 717} (2012)
  330,
\href{http://www.arXiv.org/abs/1205.3130}{\texttt{arXiv:1205.3130}}.

\bibitem{Chatrchyan:2011vp}
\hrefCMSnoop {}{{CMS Collaboration}, ``{Measurement of the $t$-channel single
  top quark production cross section in $pp$ collisions at $\sqrt{s}=7$
  TeV}'',} \textit{ Phys. Rev. Lett.} \textbf{ 107} (2011) 091802,
\href{http://www.arXiv.org/abs/1106.3052}{\texttt{arXiv:1106.3052}}.

\bibitem{Chatrchyan:2012ep}
\hrefCMSnoop {}{{CMS Collaboration}, ``{Measurement of the single-top-quark
  $t$-channel cross section in $pp$ collisions at $\sqrt{s}=7$ TeV}'',}
  \textit{ JHEP} \textbf{ 12} (2012) 035,
\href{http://www.arXiv.org/abs/1209.4533}{\texttt{arXiv:1209.4533}}.

\bibitem{Khachatryan:2014iya}
\hrefCMSnoop {}{{CMS Collaboration}, ``{Measurement of the t-channel
  single-top-quark production cross section and of the $\mid V_{tb} \mid$ CKM
  matrix element in $pp$ collisions at $\sqrt{s}$= 8 TeV}'',} \textit{ JHEP}
  \textbf{ 06} (2014) 090,
  \href{http://dx.doi.org/10.1007/JHEP06(2014)090}{\doi{10.1007/JHEP06(2014)090}},
\href{http://www.arXiv.org/abs/1403.7366}{\texttt{arXiv:1403.7366}}.

\bibitem{Aad:2014fwa}
\hrefCMSnoop {}{{ATLAS} Collaboration, ``{Comprehensive measurements of
  $t$-channel single top-quark production cross sections at $\sqrt{s} = 7$ TeV
  with the ATLAS detector}'',} \textit{ Phys.Rev.} \textbf{ D90} (2014),
  no.~11, 112006,
  \href{http://dx.doi.org/10.1103/PhysRevD.90.112006}{\doi{10.1103/PhysRevD.90.112006}},
\href{http://www.arXiv.org/abs/1406.7844}{\texttt{arXiv:1406.7844}}.

\bibitem{Aad:2012xca}
\hrefCMSnoop {}{{ATLAS Collaboration}, ``{Evidence for the associated
  production of a $W$ boson and a top quark in ATLAS at $\sqrt{s}=7$ TeV}'',}
  \textit{ Phys. Lett. B} \textbf{ 716} (2012) 142,
\href{http://www.arXiv.org/abs/1205.5764}{\texttt{arXiv:1205.5764}}.

\bibitem{Chatrchyan:2012zca}
\hrefCMSnoop {}{{CMS Collaboration}, ``{Evidence for associated production of a
  single top quark and $W$ boson in $pp$ collisions at $\sqrt{s}$ = 7 TeV}'',}
  \textit{ Phys. Rev. Lett.} \textbf{ 110} (2013) 022003,
\href{http://www.arXiv.org/abs/1209.3489}{\texttt{arXiv:1209.3489}}.

\bibitem{Chatrchyan:2014tua}
\hrefCMSnoop {}{{CMS Collaboration}, ``{Observation of the associated
  production of a single top quark and a W boson in $pp$ collisions at
  $\sqrt(s) = 8$~TeV}'',} \textit{ Phys. Rev. Lett.} \textbf{ 112} (2014)
  231802,
  \href{http://dx.doi.org/10.1103/PhysRevLett.112.231802}{\doi{10.1103/PhysRevLett.112.231802}},
\href{http://www.arXiv.org/abs/1401.2942}{\texttt{arXiv:1401.2942}}.

\bibitem{Khachatryan:2014nda}
\hrefCMSnoop {}{{CMS Collaboration}, ``{Measurement of the ratio B(t to Wb)/B(t
  to Wq) in $pp$ collisions at $\sqrt(s) = 8$~TeV}'',} \textit{ Phys. Lett. B}
  \textbf{ 736} (2014) 33,
  \href{http://dx.doi.org/10.1016/j.physletb.2014.06.076}{\doi{10.1016/j.physletb.2014.06.076}},
\href{http://www.arXiv.org/abs/1404.2292}{\texttt{arXiv:1404.2292}}.

\bibitem{Abazov:2011zk}
\hrefCMSnoop {}{{\Dzero Collaboration}, ``{Precision measurement of the ratio
  ${\rm B}(t \to Wb)/{\rm B}(t \to Wq)$ and Extraction of $V_{tb}$}'',}
  \textit{ Phys. Rev. Lett.} \textbf{ 107} (2011) 121802,
  \href{http://dx.doi.org/10.1103/PhysRevLett.107.121802}{\doi{10.1103/PhysRevLett.107.121802}},
\href{http://www.arXiv.org/abs/1106.5436}{\texttt{arXiv:1106.5436}}.

\bibitem{Aaltonen:2013doa}
\hrefCMSnoop {}{{CDF Collaboration}, ``{Measurement of R=B(t→Wb)/B(t→Wq) in
  top-quark-pair decays using lepton+jets events and the full CDF run II
  dataset}'',} \textit{ Phys. Rev. D} \textbf{ 87} (2013) 111101,
  \href{http://dx.doi.org/10.1103/PhysRevD.87.111101}{\doi{10.1103/PhysRevD.87.111101}},
\href{http://www.arXiv.org/abs/1303.6142}{\texttt{arXiv:1303.6142}}.

\bibitem{Aaltonen:2014yua}
\hrefCMSnoop {}{{CDF Collaboration}, ``{Measurement of $R = {\mathcal{B}\left(t
  \rightarrow Wb \right)/\mathcal{B}\left(t \rightarrow Wq \right)} $ in
  Top--Quark--Pair Decays using Dilepton Events and the Full CDF Run II Data
  Set}'',} \textit{ Phys. Rev. Lett.} \textbf{ 112} (2014) 221801,
  \href{http://dx.doi.org/10.1103/PhysRevLett.112.221801}{\doi{10.1103/PhysRevLett.112.221801}},
\href{http://www.arXiv.org/abs/1404.3392}{\texttt{arXiv:1404.3392}}.

\end{thebibliography}\endgroup

\end{document}